\documentclass[conference]{IEEEtran}
\IEEEoverridecommandlockouts
\usepackage{cite}

\usepackage{amsthm,amssymb,amsmath}  
\usepackage{algorithmic}
\usepackage{url}
\usepackage{graphicx}
\usepackage{textcomp}
\usepackage{xcolor}
\usepackage{float}  
\usepackage{subfigure}  
\usepackage{balance}
\usepackage{epsfig,tabularx,subfigure,multirow, graphicx}
\usepackage[linesnumbered,ruled,vlined]{algorithm2e}
\SetKwRepeat{Do}{do}{while}
\SetCommentSty{mycommfont}

\long\def\comment#1{}
\usepackage{colortbl}
\usepackage{xcolor}
\usepackage{makecell}
\renewcommand{\baselinestretch}{1}

\definecolor{gray}{rgb}{0.5,0.5,0.5}

\newtheorem{definition}{Definition}

\newtheorem{theorem}{\bf Theorem}
\newtheorem{lemma}{\bf Lemma}

\newtheorem{example}{\bf Example}
\def\BibTeX{{\rm B\kern-.05em{\sc i\kern-.025em b}\kern-.08em
    T\kern-.1667em\lower.7ex\hbox{E}\kern-.125emX}}
\begin{document}

\title{Efficient $k$-clique Listing with Set Intersection Speedup \\
\thanks{*Zhirong Yuan and You Peng are joint first authors and contribute equally to this work. Li Han is the corresponding author.}
}

\author{
	{Zhirong Yuan{\small$~^{*}$}, You Peng{\small$~^{\#}$}, Peng Cheng{\small$~^{*}$}, Li Han{\small$~^{*}$},  Xuemin Lin{\small$~^{\#}$}, Lei Chen{\small$~^{\dagger}$}, Wenjie Zhang{\small$~^{\#}$}}\\
	\fontsize{10}{10}\itshape
	$~^{*}$East China Normal University, Shanghai, China\\
	\fontsize{9}{9}\upshape
	zryuan@stu.ecnu.edu.cn, pcheng@sei.ecnu.edu.cn, hanli@sei.ecnu.edu.cn\\
	\fontsize{10}{10}\itshape
	$~^{\dagger}$The Hong Kong University of Science and Technology, Hong Kong SAR, China\\
	\fontsize{9}{9}\upshape
	leichen@cse.ust.hk\\
	\fontsize{10}{10}\itshape
	$~^{\#}$The University of New South Wales, Australia\\
	\fontsize{9}{9}\upshape
	unswpy@gmail.com, lxue@cse.unsw.edu.au, wenjie.zhang@unsw.edu.au
}

\maketitle
\begin{abstract}
	Listing all $k$-cliques is a fundamental problem in graph mining, with applications in finance, biology, and social network analysis. However, owing to the exponential growth of the search space as $k$ increases, listing all $k$-cliques is algorithmically challenging. \textit{DDegree} and \textit{DDegCol} are the state-of-the-art algorithms that exploit ordering heuristics based on degree ordering and color ordering, respectively. Both \textit{DDegree} and \textit{DDegCol} induce high time and space overhead for set intersections cause they construct and maintain all induced subgraphs. Meanwhile, it is non-trivial to implement the data level parallelism to further accelerate on \textit{DDegree} and \textit{DDegCol}.
	
	In this paper, we propose two efficient algorithms \textit{SDegree} and \textit{BitCol} for $k$-clique listing. 
	We mainly focus on accelerating the set intersections for $k$-clique listing. Both \textit{SDegree} and \textit{BitCol} exploit the data level parallelism for further acceleration with single instruction multiple data (SIMD) or vector instruction sets. Furthermore, we propose two preprocessing techniques \textit{Pre-Core} and \textit{Pre-List}, which run in linear time. The preprocessing techniques significantly reduce the size of the original graph and prevent exploring a large number of invalid nodes. In the theoretical analysis, our algorithms have a comparable time complexity and a slightly lower space complexity than the state-of-the-art algorithms. The comprehensive experiments reveal that our algorithms outperform the state-of-the-art algorithms by $3.75$x for degree ordering and $5.67x$ for color ordering on average.
\end{abstract}

\section{Introduction}
Real-world graphs, such as social networks, road networks, world-wide-web networks, and IoT networks often consist of cohesive subgraph structures. Cohesive subgraph mining is a fundamental problem in network analysis, with applications in community detection~\cite{community_detection,densest_subgraph,densest_subgraph1,densest_subgraph2,densest_subgraph3}, real-time story identification~\cite{real-story}, frequent migration patterns mining in financial markets~\cite{financial}, and motif detection in biological networks~\cite{Biology}.

Clique is a cohesive subgraph structure par excellence with a variety of applications in network analysis~\cite{clique_app1,clique_app2,clique_app3}. A $k$-clique is a dense subgraph of a graph $G$ with $k$ nodes, and each pair of nodes are adjacent~\cite{baseline2}. 
The $k$-clique listing problem is a natural generalization of the triangle listing problem~\cite{tc2}.
In particular, the state-of-the-art triangle listing algorithm~\cite{tc} is capable of processing billion-scale graphs within $1,000$ seconds. However, the $k$-clique listing problem is often deemed not feasible even for million-scale graphs, since the number of $k$-cliques could be exponentially large for a relatively large $k$~\cite{baseline1,baseline2}.

\noindent\textbf{Applications.}~In recent years, the requirement of efficiently listing $k$-cliques has been raised by the data mining and database communities. We introduce the applications of $k$-clique listing as follows.

\vspace{0.5mm}
\noindent \textit{(1)
	\underline{Community Detection}}.
Detecting communities helps to reveal the structural organizations in real-world complex networks~\cite{kclique_community1}. 
Specifically, a $k$-clique community is the union of all the $k$-cliques that each $k$-clique is adjacent to another one~\cite{kclique_community1,kclique_community2}. Hui and Crowcroft~\cite{kclique_community_app} designed efficient forwarding algorithms for mobile networks with information in $k$-clique communities. The $k$-clique listing algorithms can be exploited to compute $k$-clique communities~\cite{kclique_community1,kclique_community2}.

\vspace{0.5mm}
\noindent \textit{(2)
	\underline{Spam Detection}}.
Link spam is an attempt to promote the ranking of websites by cheating the link-based ranking algorithm in search engines~\cite{linkspam}.
Clique identification in the network structure helps a lot in handling the search engine spam problems, especially the link spam~\cite{Spamdexing}. In particular, Jayanthi et al.~\cite{Spamdexing} proposed a  $k$-clique percolation method to detect the spam, which also needs to list the $k$-cliques.

\vspace{0.5mm}
\noindent \textit{(3)
	\underline{Biological Networks}}. 
Most cellular tasks are not performed by individual proteins, but by a group of functionally related proteins (often called modules). In gene association networks, Adamcsek et al.~\cite{cfinder} proposed CFinder to predict the function of a single protein and to discover novel modules. CFinder needs to locate the $k$-clique percolation clusters of the network interpreted as modules, in which a $k$-clique listing algorithm can be used for computing all $k$-cliques.

Motivated by the aforementioned studies, we investigate the problem of listing all the $k$-cliques in a graph. A $k$-clique can be expanded from a ($k$-$1$)-clique. Therefore, a basic idea of the $k$-clique listing algorithms is to recursively expand the clique, starting from a single node ($1$-clique).


\noindent \textbf{Existing Works.}~Based on a recursive framework, numerous practical algorithms have been developed for listing all the $k$-cliques in real-world graphs~\cite{ChibaNishizeki,baseline1,baseline2}. 
The state-of-the-art algorithms~\cite{baseline1,baseline2} are based on vertex ordering and mainly focus on the construction of each induced subgraph in the recursion.
Danisch et al.~\cite{baseline2} proposed an efficient ordering-based framework \textit{kClist} for listing all the $k$-cliques, which can be easily parallelized. 
Li et al.~\cite{baseline1} extended \textit{kClist} by proposing a new color ordering heuristics based on greedy graph coloring~\cite{greedy_coloring}, which can prune more unpromising search paths. 

To the best of our knowledge, all the state-of-the-art algorithms focus on selecting appropriate vertex ordering strategies to improve the theoretical upper bound of the time complexity. 



\noindent \textbf{Our Approaches.}~Existing algorithms~\cite{ChibaNishizeki,baseline1,baseline2} are implemented efficiently with well-designed hash tables. Furthermore, reordering nodes in the adjacency lists is sufficient for constructing the induced subgraphs, without generating a completely new one. 
Our approaches achieve even better performance with merge join, avoiding careful maintenance of induced subgraphs. 

In this paper, we focus on accelerating the set intersections in each recursion, thereby accelerating the $k$-clique listing.
Our algorithms are implemented based on the merge join, without additional operations on reordering nodes or recursively maintaining hash tables. 

Our algorithms exploit the data level parallelism, e.g., single instruction multiple data (SIMD) and vector instruction sets~\cite{autovec,autovec_gcc}, to further accelerate the set intersections~\cite{SIMD2,Roaring,SIGMOD18}.
We have to emphasize that, the state-of-the-art algorithms are based on hash join with the reordered nodes in each recursion, which is non-trivial to exploit the data level parallelism.





\noindent \textbf{Contributions.}
Our contributions in this paper are summarized as follows:
\begin{itemize}
	\item \textit{Pre-Core and Pre-List.}~We develop two preprocessing algorithms, named \textit{Pre-Core} and \textit{Pre-List}, that not only significantly reduce the search space of the original graph, but also finish in satisfactory time.
	
	\item \textit{\textit{SDegree} and \textit{BitCol}.}~We propose a simple but effective merge-based algorithm \textit{SDegree} for listing all the $k$-cliques, without reordering nodes or maintaining hash tables. \textit{BitCol} improves \textit{SDegree} by exploiting the color ordering and compressing the vertex set with a binary representation to compare more elements at a time. Both \textit{SDegree} and \textit{BitCol} are compatible with the data level parallelism to further accelerate the $k$-clique listing, which is non-trivial for hash-based algorithms.
	
	\item \textit{Efficiency.}~Our algorithms are efficient both theoretically and experimentally. On one hand, we achieve the same upper bound on the time complexity with less memory overhead compared to the state-of-the-art algorithms. On the other hand, we evaluate the serial and the parallel version of our algorithms on $10$ real-world datasets. Our algorithms outperform the state-of-the-art algorithms by $3.75$x for degree ordering and $5.67x$ for color ordering on average. 
\end{itemize}   
\noindent \textbf{Paper Organization.} The rest of the paper is organized as follows. Section~\ref{sec:relatedwork} surveys the important related works. Section~\ref{sec:preliminary} gives a formal definition of the problem and related concepts, and briefly introduces the existing solutions. Section~\ref{sec:framework} presents the overall framework briefly. The preprocessing techniques are proposed in Section~\ref{sec:preprocessing}. Our algorithms are formally introduced in Section~\ref{sec:SDegree} and Section~\ref{sec:BitCol}. The theoretical analysis is given in Section~\ref{sec:analysis}. Then the extensive experiments are conducted in Section~\ref{sec:experiment}. Finally, Section~\ref{sec:conclusion} concludes the paper.

\section{Related Work}
\label{sec:relatedwork}
\subsection{The Bron-Kerbosch Algorithm}
Relying on the fact that each $n$-clique ($n\ge k$) consists of $\tbinom{n}{k}$ $k$-cliques, the algorithms for maximal clique enumerating (MCE) can be applied to the problem of $k$-clique listing.
The classic \textit{Bron-Kerbosch} algorithm~\cite{BronKerbosch} solves the MCE problem by recursively maintaining the processed states of three sets $R$, $P$, $X$, to search the maximal cliques without redundant verification.
$R$ stands for the currently acquired growing clique;
$P$ stands for prospective nodes that may be adjacent to each node in $R$, with which $R$ can be further expanded;
$X$ stands for the nodes already processed.
Initially, $P=V$, $R=\emptyset$, and $X=\emptyset$. 
After exploring the vertex $v$, $v$ is transferred from $P$ to $X$. The \textit{Bron-Kerbosch} algorithm avoids reporting the maximal clique repeatedly by checking $X$. Specifically, a maximal clique is reported only when $P=\emptyset$ and $X=\emptyset$, since no more elements can be added to $R$.

Furthermore, Bron and Kerbosch proposed a variant version with a pivoting strategy, which reduces the number of unnecessary recursive calls. Eppstein et al.~\cite{degeneracy} improved the original \textit{Bron-Kerbosch} algorithm with vertex ordering, which reduces the exponential worst-case time complexity.

\subsection{Vertex Ordering Approach}
A common strategy to avoid redundancy and reduce the search space is the vertex ordering, which is a preprocessing step that transforms the original undirected graph into a directed acyclic graph (DAG). The orientation of each undirected edge is from the low order to the high order.
Initially, vertex ordering is a classical technique designed for the triangle listing problem. It is capable of accelerating the calculation both theoretically and experimentally, by properly selecting the orientation of each edge~\cite{tc}.  Therefore, it is critical in determining the appropriate order of each vertex.

In recent years, vertex ordering has been successfully applied to the $k$-clique listing problem with good performance~\cite{baseline1,baseline2}. The correctness is based on the fact that the $k$-clique can be discovered in any order, for the symmetrical structure of the clique.
Common vertex ordering approaches include degeneracy ordering~\cite{baseline2,degeneracy}, degree ordering~\cite{tc,tc2,baseline1}, and color ordering~\cite{baseline1}. 

The degeneracy ordering can be generated by the classic core-decomposition algorithm~\cite{coresdecomposition}, which repeatedly deletes the node with the minimum degree.
The degree ordering is simply generated by the degree of each vertex in descending order.
The color ordering is based on the greedy graph coloring~\cite{greedy_coloring}, which assigns each vertex a color value such that no adjacent vertices share the same color. 
In particular, the out-degree in the DAG based on degeneracy ordering is bounded by the degeneracy ($\beta$)~\cite{baseline2}, while the out-degree is bounded by the $h$-index ($\gamma$) for degree ordering~\cite{baseline1}.
\subsection{Set Intersection Acceleration}
As reported in ~\cite{SIGMOD18,set_intersect_2}, the set intersection is extensively involved in graph algorithms, such as triangle listing and maximal clique enumeration. Similarly, the $k$-clique listing problem also involves a large number of set intersections, which certainly causes a bottleneck.

Given two vertex sets $S_a$ and $S_b$, there are four major methods for set intersections.
Hash join has a time complexity of $O(|S_a|)$ with $S_b$ as the hash table. The additional cost is required to construct and maintain the hash tables.
If $S_a$ and $S_b$ are sorted in ascending order, we can use merge-based algorithms and galloping search for set intersections~\cite{set_intersect_3,set_intersect_4}.
The time complexity of the merge-based algorithm is $O(|S_a|)$ (or $O(|S_b|)$) in the best case while $O(|S_a|+|S_b|)$ in the worst case; the time complexity of the galloping search is $O(|S_a|\log{|S_b|})$, which is efficient in practice only for $|S_a|\ll |S_b|$.

Modern microprocessors are equipped with SIMD or vector instruction sets which allow compilers to exploit fine-grained data level parallelism~\cite{autovec}.
The application of SIMD instructions to accelerate set intersections is proposed by a bunch of algorithms~\cite{SIMD2,SIMD3,SIGMOD18}.
With SIMD instructions, the merge-based algorithm can be extended by reading and comparing multiple elements at a time. In particular, the state-of-the-art algorithms for accelerating the set intersections~\cite{set_intersect_2,SIGMOD18} are mostly developed based on merge join, instead of hash join or galloping search.

\section{Preliminary}
\label{sec:preliminary}
The formal definition of the $k$-clique listing problem is given in this section. The existing algorithms, \textit{Chiba-Nishizeki}~\cite{ChibaNishizeki}, \textit{kClist}~\cite{baseline2}, \textit{DDegree} and \textit{DDegCol}~\cite{baseline1}, are then introduced. TABLE~\ref{tab:notation} summarizes the most relevant notations.

\begin{table}[htb]
	\centering
	\caption{Notations}
	\begin{tabular}{ll}
		\hline
		\textbf{Notations} & \textbf{Description}\\
		\hline
		$G,\mathop{G}\limits ^{\rightarrow}$ & \makecell[l] {Undirected graph, directed graph}\\
		\hline
		$V$, $E$ & \makecell[l] {Graph vertex set, edge set}\\
		\hline
		$N_v,d_v$ & \makecell[l] {The neighbor set of $v\in G$ and its size }\\
		\hline
		$N^{+}_v,d^{+}_v$ & \makecell[l] {The out neighbors of $v\in \mathop{G}\limits ^{\rightarrow}$ and its size}\\
		\hline
		$G_u$ & \makecell[l] {The subgraph induced by the neighbors of $u$ in $G$}\\
		\hline
		$\Omega_k,C_k$ & \makecell[l] {$k$-core, $k$-clique}\\
		\hline
		$\Delta$ & \makecell[l] {The maximum out-degree in $\mathop{G}\limits ^{\rightarrow}$}\\
		\hline
		
	\end{tabular}
	\label{tab:notation}
	\vspace{-4mm}
\end{table}

\subsection{Problem Definition}
$G(V, E)$ is an undirected and unlabeled graph, where $V$ ($|V|=n$) and $E$ ($|E|=m$)
denote the set of nodes and the set of edges, respectively. The neighbor set of $v$ in $G$ is denoted by $N_v(G)$, and the degree of $v$ in $G$ is denoted by $d_v(G)=|N_v(G)|$. We use $N_v$ and $d_v$ to refer to $N_v(G)$ and $d_v(G)$ when the context is clear. If $V'\subseteq V$ and $E'=\{(u,v)|(u,v)\in E, u\in V'\land v\in V'\}$, a subgraph $G'(V',E')$ is termed an induced subgraph of $G$.

\begin{definition}[$k$-core]
	\label{def:kcore}
	Given a graph $G(V, E)$, an induced subgraph $\Omega_k(V_H, E_H)$ is a $k$-core of $G$, if $\Omega_k$ satisfies the following constraints: 
	\begin{enumerate}
		\item \textbf{Cohesive:}~For any $u \in V_H$, $d_u \geq k$. 
		\item \textbf{Maximal:}~$H$ is maximal, i.e., for any vertex set $H' \supset H$, the subgraph induced by $H'$ is not a $k$-core.
	\end{enumerate}
\end{definition}

\begin{definition}[$k$-clique]
	\label{def:klique}
	A $k$-clique $C_k(V_k,E_k)$ is an induced subgraph of $G$ where $|V_k|=k$ and $|E_k|=\tbinom{k}{2}$, i.e., every two nodes are adjacent in $C_k$.
\end{definition}

\begin{figure}[ht]\centering
	\scalebox{1}[1]{\includegraphics{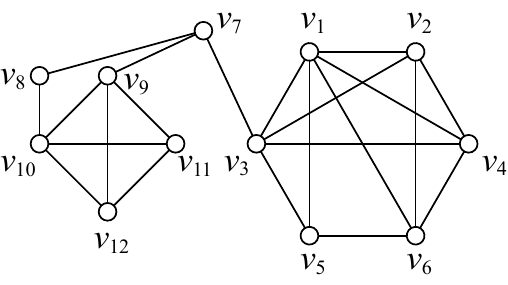}}
	\caption{\small The Input Graph $G$}
	\label{fig:graph}
\end{figure}

The $k$-clique listing problem is to find all the $k$-cliques in $G$, which is a natural generalization of the triangle listing problem~\cite{tc,tc2}. 
Since a $2$-clique is an edge, we only consider $k\ge 3$ by default in this paper.
\begin{example}
	The graph $G$ in Fig.\ref{fig:graph} has three $4$-cliques $\{v_9,v_{10},v_{11},v_{12}\},\{v_1,v_2,v_3,v_4\},$ and $\{v_1,v_2,v_4,v_6\}$.
\end{example}

\begin{figure}[ht]\centering
	\scalebox{1}[1]{\includegraphics{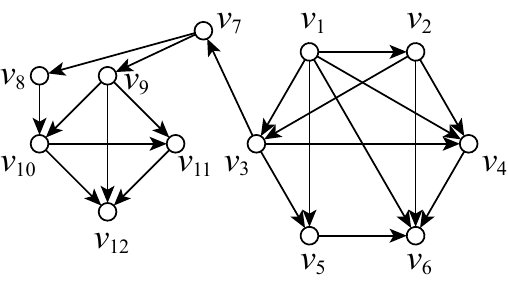}}
	\caption{\small The Directed Version of $G$}
	\label{fig:directed_graph}
\end{figure}

We use $\mathop{G}\limits ^{\rightarrow}$ to denote a directed graph.  The set of directed edges $\{\langle u,v\rangle \}$ in $\mathop{G}\limits ^{\rightarrow}$ is denoted as $E(\mathop{G}\limits ^{\rightarrow})$, where the direction of $\langle u,v\rangle$ is $u\rightarrow v$. The set of nodes in $\mathop{G}\limits ^{\rightarrow}$ is denoted as $V(\mathop{G}\limits ^{\rightarrow})$.
A directed acyclic graph (DAG) is a directed graph with no directed cycles. We denote with $N^{+}_v(\mathop{G}\limits ^{\rightarrow})$ the out-neighbor set of $v$ in $\mathop{G}\limits ^{\rightarrow}$, and $d^{+}_v(\mathop{G}\limits ^{\rightarrow})=|N^{+}_v(\mathop{G}\limits ^{\rightarrow})|$ denotes the out-degree of $v$.
If the context is clear, we use $N^{+}_v$ and $d^{+}_v$ to refer to $N^{+}_v(\mathop{G}\limits ^{\rightarrow})$ and $d^{+}_v(\mathop{G}\limits ^{\rightarrow})$.
The vertex ordering of the undirected graph $G(V,E)$ is determined by an assignment function $\eta:V\rightarrow \{1,2,\ldots,|V|\}$. An undirected edge $(u,v)$ can be converted into a directed edge $\langle u,v\rangle$ with $\eta$, where $\eta(u) < \eta(v)$. In graph $G$ of Fig.\ref{fig:graph}, we utilize the node ID as $\eta$ and the directed version $\mathop{G}\limits ^{\rightarrow}$ is shown in Fig.\ref{fig:directed_graph}.

Arboricity ($\alpha$)~\cite{ChibaNishizeki,arboricity,arboricity_hindex}, degeneracy ($\beta$)~\cite{degeneracy}, and $h$-index ($\gamma$)~\cite{arboricity_hindex,hindex1} are three important metrics in graph analysis that we will discuss. 

For algorithm design and complexity analysis, these metrics are commonly utilized in the $k$-clique listing problem. In particular, $\alpha$, $\beta$ and $\gamma$ are usually very small in real-world graphs\cite{degeneracy}.

\begin{itemize}
	\item \textit{Arboricity ($\alpha$)}.~The arboricity of an undirected graph $G(V, E)$, i.e., $\alpha$, is the minimum number of forests into which the edges of $G$ can be partitioned. The arboricity $\alpha$ can be used to measure the density of $G$, however, it is difficult to be calculated. 
	\item \textit{Degeneracy ($\beta$)}.~The degeneracy, denoted as $\beta$, is the maximum core number of $G$, where the core number of a vertex $u$ is the largest integer $l$ s.t. $u$ is contained by a $l$-core. $\beta$ can be calculated in linear time with the classic core-decomposition algorithm~\cite{coresdecomposition}. $\beta$ is frequently utilized to approximate the arboricity $\alpha$ since $\alpha \le \beta \le 2\alpha$ - $1$~\cite{baseline1}.
	\item \textit{$h$-index ($\gamma$)}.~The $h$-index of $G$, denoted as $\gamma$, can be written as $\gamma=\mathop{\arg\max}\limits_{h}{(|\{v|d_v(G)\ge h\}|\ge h, v\in V)}$, which is the maximum $h$ such that $G$ contains $h$ vertices with a degree at least $h$~\cite{arboricity_hindex}. Similarly, $\gamma$ also satisfies that $\alpha \le \gamma \le \sqrt{m}$.
\end{itemize}

\subsection{Existing Solutions}
\subsubsection{The Chiba-Nishizeki Algorithm}
We start by introducing the iconic \textit{Chiba-Nishizeki} algorithm~\cite{ChibaNishizeki}, which is the first practical algorithm for the $k$-clique listing problem.
As illustrated in Algorithm~\ref{algo:ChibaNishizeki}, the nodes are sorted by degree in descending order, i.e., $v_i < v_j$ for $d(v_i) > d(v_j)$ (Line \ref{line:chibanishizeki:reorder}).
For each vertex $v_i$, the induced subgraph $G_{v_i}$ is constructed by the neighbors of $v_i$ (Line \ref{line:chibanishizeki:inducedsubgraph}). Subsequently, the algorithm invokes the recursive procedure for each induced subgraph, with a recurrence depth of $k$-$1$ (Line \ref{line:chibanishizeki:recursion}). To prevent repeating computations, $v_i$ and its related edges are eliminated when $G_{v_i}$ is processed (Line \ref{line:chibanishizeki:remove}). 

The time complexity of the \textit{Chiba-Nishizeki} algorithm is $O(km\alpha^{k-2})$, which is closely associated with the arboricity. In most real-world graphs,
the arboricity is generally rather small. Therefore, the \textit{Chiba-Nishizeki} algorithm is proved to be effective in practice.
However, the parallelization of the \textit{Chiba-Nishizeki} algorithm is difficult, since each vertex $v_i$ is eliminated after $G_{v_i}$ has been processed, i.e., $G$ shrinks throughout each iteration.
\begin{algorithm}[t]
	\DontPrintSemicolon
	\KwIn{A graph $G$ and a positive integer $k$}
	\KwOut{All the $k$-cliques in $G$}
	List($G$, $\emptyset$, $k$)\;
	\textbf{Procedure} List($G$, $R$, $l$)\;
	\uIf{$l\neq2$}{
		Sort the nodes by degree in descending order\label{line:chibanishizeki:reorder}\;
		\For{$i=1$ to $|V(G)|$} {
			Compute the subgraph $G_{v_i}$ induced by $N_{v_i}$ \label{line:chibanishizeki:inducedsubgraph}\;
			List($G_{v_i}$, $R\cup\{v_i\}$, $l-1$)\label{line:chibanishizeki:recursion}\;
			Remove $v_i$ and the related edges in $G$\;\label{line:chibanishizeki:remove}
		}
	}
	\uElse{
		\For{each edge $(u,v)\in G$\label{line:chibanishizeki:output1}}
		{
			report a $k$-clique $R\cup \{u,v\}$\label{line:chibanishizeki:output2}\;
		}
	}
	
	\caption{\textsc{Chiba-Nishizeki}($G$, $k$)}
	\label{algo:ChibaNishizeki}
\end{algorithm}

\subsubsection{kClist}
Danisch et al.\cite{baseline2} proposed \textit{kClist} for listing all the $k$-cliques using the technique of vertex ordering. Algorithm 2 delves further into the framework of the \textit{kClist} algorithm. First, a total ordering $\eta$ on nodes is selected (Line \ref{line:kclist:order}). Based on the classic core-decomposition algorithm~\cite{coresdecomposition}, Danisch et al. specified the degeneracy ordering as the total ordering $\eta$.
\textit{kClist} creates a DAG $\mathop{G}\limits ^{\rightarrow}$ based on the total ordering (Line \ref{line:kclist:DAG}) and lists all the $k$-cliques on $\mathop{G}\limits ^{\rightarrow}$ without redundancy (Line \ref{line:kclist:main}). To prevent reporting the same $k$-clique repeatedly, \textit{kClist} lists each $k$-clique in the lexicographical order based on $\eta$. 
In the current DAG $\mathop{G}\limits ^{\rightarrow}$, \textit{kClist} recursively processes on the subgraph $\mathop{G_u}\limits ^{\rightarrow}$ induced by the out neighbors of each vertex $u$ (Lines \ref{line:kclist:foreach}-\ref{line:kclist:list}).

As indicated in the previous work, \textit{kClist} lists all the $k$-cliques in $O(km(\Delta/2)^{k-2})$ time, where $\Delta$ is the maximum out-degree in $\mathop{G}\limits ^{\rightarrow}$. 
Specifically, $\Delta=\beta$ (degeneracy) when the degeneracy ordering is specified.
Compared to the \textit{Chiba-Nishizeki} algorithm, \textit{kClist} is easy to be parallelized, with each thread processing on the different induced subgraphs.
\begin{algorithm}[t]
	\DontPrintSemicolon
	\KwIn{A graph $G$ and a positive integer $k$}
	\KwOut{All the $k$-cliques in $G$}
	Select a total ordering $\eta$ on nodes\;\label{line:kclist:order}
	Generate a DAG $\mathop{G}\limits ^{\rightarrow}$ on $\eta$\, where $u\rightarrow v$ if $\eta(u)<\eta(v)$\; \label{line:kclist:DAG}
	kCliqueList($k, \mathop{G}\limits ^{\rightarrow}, \emptyset$)\;\label{line:kclist:main}
	\textbf{Procedure} kCliqueList($l, \mathop{G}\limits ^{\rightarrow}, R$)\;
	\uIf{$l=2$}{
		\For{each edge $\langle u,v\rangle \in \mathop{G}\limits ^{\rightarrow}$} {
			report a $k$-clique $R\cup\{u,v\}$\;
		}
	}
	\uElse{
		\For{each node $u\in V(\mathop{G}\limits ^{\rightarrow})$\label{line:kclist:foreach}}{
			Let $\mathop{G_u}\limits ^{\rightarrow}$ be a subgraph of $\mathop{G}\limits ^{\rightarrow}$ induced by $N^{+}_u$\;
			kCliqueList($l-1, \mathop{G_u}\limits ^{\rightarrow}, R\cup \{u\}$)\;\label{line:kclist:list}
		}
	}
	\caption{\textsc{kClist}($G$, $k$)}
	\label{algo:kClist}
\end{algorithm}
\subsubsection{DDegree and DDegCol}~Li et al. proposed two state-of-the-art algorithms for listing $k$-cliques, namely \textit{DDegree} and \textit{DDegCol}~\cite{baseline1}.
Inspired by \textit{kClist}, \textit{DDegree} and \textit{DDegCol} both embrace the idea of vertex ordering. The difference is that \textit{DDegree} exploits degree ordering to generate the DAG $\mathop{G}\limits ^{\rightarrow}$, while \textit{DDegCol} utilizes the color ordering based on the greedy graph coloring~\cite{greedy_coloring}. 

Given an undirected graph $G(V, E)$, \textit{DDegree} and \textit{DDegCol} first build a DAG $\mathop{G}\limits ^{\rightarrow}$ based on the degeneracy ordering. Then, for each vertex $u$, a subgraph $G_u(V_u, E_u)$ induced by $N^{+}_u(\mathop{G}\limits ^{\rightarrow})$ is generated.
For \textit{DDegree} and \textit{DDegCol}, the \textit{kClist} algorithm is invoked on the subgraph $G_u(V_u,E_u)$, with degree ordering and color ordering, respectively. The time complexity is the same as \textit{kClist}. Both \textit{DDegree} and \textit{DDegCol} can be parallelized easily, as well.

\section{Framework Overview}
\label{sec:framework}
In this section, we present our overall approach to solving the $k$-clique listing problem.
Given an undirected graph $G$ and a positive integer $k$, the key steps of our approach are stated as follows:
\begin{enumerate}
	\item \textit{Preprocessing.}~Preprocess the undirected graph $G$  to eliminate the nodes and the associated edges that would not be contained in a $k$-clique, as illustrated in Section \ref{sec:preprocessing}. Then we perform our $k$-clique listing algorithms on the reduced graph.
	\item \textit{Set Intersection Acceleration.}~For listing $k$-cliques, we propose a recursive framework that uses SIMD instructions to speed up the set intersections. Furthermore, we exploit the technique of vertex ordering (for details, see Section \ref{sec:SDegree}).
	\item \textit{Additional Optimizations.}~We improve our algorithm based on color ordering with a more powerful pruning effect. We also compress the neighbor set with bitmaps to further accelerate the set intersections (for details, see Section \ref{sec:BitCol}).
	
\end{enumerate}

\section{Preprocessing Techniques}
\label{sec:preprocessing}
In this section, we propose preprocessing techniques to prune the invalid nodes that would not be contained in a $k$-clique, which run in linear time.

\subsection{Pre-Core}

Since the $k$-clique is a special case of the ($k$-$1$)-core, our first preprocessing technique \textit{Pre-Core} is based on the ($k$-$1$)-core.
Recall that ($k$-$1$)-cores of $G$ are connected components where the degree of each vertex is at least $k-1$.
During the \textit{Pre-Core} preprocessing, $v$ and its related edges are deleted for any vertex $v$ whose degree is less than $k-1$. Because $u$ is a neighbor of $v$, when $v$ is removed, $d_u$ will be updated to $d_u-1$. Furthermore, if $d_u<k-1$ after the update, $u$ will be removed as well, and the above process is repeated. Obviously, \textit{Pre-Core} runs in linear time.

As illustrated in Algorithm \ref{algo:precore}, the queue $Q$ is used to store the vertex $v$ that would not be contained in a ($k$-$1$)-core, and to update the degrees of $v$'s neighbors. The set $\mathcal{F}$ is used to hold the invalid nodes that are about to be removed. Both $Q$ and $\mathcal{F}$ are initialized as $\emptyset$ (Line \ref{line:pre-kcore:init}). $u$ is pushed into $Q$ and inserted into $\mathcal{F}$ for each node $u$ where $d_u<k-1$ (Lines \ref{line:pre-kcore:firstinsert1}-\ref{line:pre-kcore:firstinsert2}). Then, for each node $u\in Q$, we update the degree of each neighbor $v\in N_u$ to $d_v-1$, implying that $u$ is no longer a neighbor of $v$ (Line \ref{line:pre-kcore:updatedegree}). When $d_v<k-1$ after the update, $v$ is inserted into $\mathcal{F}$ and pushed into $Q$ (Lines \ref{line:pre-kcore:secondinsert1}-\ref{line:pre-kcore:secondinsert2}). 
Finally, each node $u\in  \mathcal{F}$ is eliminated from $G$ (Lines \ref{line:pre-kcore:remove1}-\ref{line:pre-kcore:remove2}).

\begin{figure}[ht]\centering
	\scalebox{1}[1]{\includegraphics{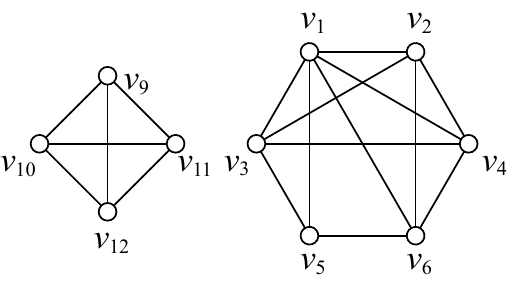}}
	\caption{\small The Pre-Core Preprocessing $(k=4)$}
	\label{fig:kcore}
\end{figure}

\begin{example}
	When $k=4$, the \textit{Pre-Core} preprocessing is performed on graph $G$ of Fig.\ref{fig:graph}, as shown in Fig.\ref{fig:kcore}. Since $d_{v_8}<3$, the vertex $v_8$ is eliminated. Then, $d_{v_7}$ is updated to $2$, and $v_7$ is removed. Finally, no more vertices can be eliminated since each remaining vertex has a degree of at least $3$.
\end{example}

\begin{algorithm}[t]
	\DontPrintSemicolon
	\KwIn{A graph $G$ and a positive integer $k$}
	$Q\gets \emptyset, \mathcal{F} \gets \emptyset$\label{line:pre-kcore:init}\;
	\For{each node $u$ in $G$\label{line:pre-kcore:firstinsert1}}{
		\uIf{$d_u< k-1$}{
			$Q.push(u)$\;
			$\mathcal{F}\gets \mathcal{F}\cup \{u\}$\label{line:pre-kcore:firstinsert2}\;
		}
	}
	\While{$Q\neq \emptyset$}{
		$u\gets Q.pop()$\;
		\For{each node $v\in N_u$}{
			$d_v\gets d_v-1$\label{line:pre-kcore:updatedegree}\;
			\uIf{$d_v < k-1 \wedge v\notin \mathcal{F}$\label{line:pre-kcore:secondinsert1}}{
				$\mathcal{F}\gets \mathcal{F}\cup \{v\}$\;
				$Q.push(v)$\label{line:pre-kcore:secondinsert2}\;
			}
		}
	}
	\For{each node $u\in \mathcal{F}$\label{line:pre-kcore:remove1}}{
		Remove $u$ from $G$\label{line:pre-kcore:remove2}\;
	}
	\caption{\textsc{Pre-Core}($G$, $k$)}
	\label{algo:precore}
\end{algorithm}

\subsection{Pre-List}
We propose the second preprocessing technique \textit{Pre-List} after performing \textit{Pre-Core}.
Since the \textit{Pre-Core} preprocessing leaves all the nodes with degrees of at least $k$, $|V(\mathcal{C})|\ge k$ will be satisfied for each connected component $\mathcal{C}$. If $\mathcal{C}$ is a clique, $\mathcal{C}$ contains $\tbinom{|V(\mathcal{C}_i)|}{k}$ $k$-cliques and we can directly report the $k$-cliques in $\mathcal{C}$. $\mathcal{C}$ is removed from $G$ after reporting the $k$-cliques in $\mathcal{C}$. Since we only need to perform a BFS to verify all the connected components, the time complexity of the \textit{Pre-List} preprocessing is $O(m)$.

Algorithm \ref{algo:prelist} briefly describes the \textit{Pre-List} preprocessing. The vertex size and the edge size of a connected component $\mathcal{C}$ are denoted as $m_c$ and $n_c$, respectively. If $\mathcal{C}$ is a clique, i.e., $m_c=n_c(n_c-1)$, we report $k$-cliques in $\mathcal{C}$ and remove $\mathcal{C}$ immediately.

\begin{figure}[ht]\centering
	\scalebox{1}[1]{\includegraphics{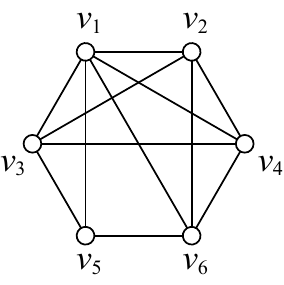}}
	\caption{\small The Pre-List Preprocessing}
	\label{fig:kcc}
\end{figure}

\begin{example}
	In the example of Fig.\ref{fig:kcore}, there are two remaining connected components after the \textit{Pre-Core} preprocessing. When performing \textit{Pre-List}, the $4$-clique $\{v_9,v_{10},v_{11},v_{12}\}$ can be reported and removed from $G$. Only one connected component is left, as illustrated in Fig.\ref{fig:kcc}.
\end{example}

\begin{algorithm}[t]
	\DontPrintSemicolon
	\KwIn{A graph $G$ and a positive integer $k$}
	\For{each connected components $\mathcal{C}\in G$}{
		$m_c\gets |E(\mathcal{C})|$, $n_c\gets |V(\mathcal{C})|$\;
		\uIf{$m_c=(n_c-1)n_c$}{
				output $k$-cliques in $\mathcal{C}$\;
			Remove $\mathcal{C}$ from $G$\;
		}
	}
	\caption{\textsc{Pre-List}($G$, $k$)}
	\label{algo:prelist}
\end{algorithm}

\section{The \textit{SDegree} Algorithm}
\label{sec:SDegree}
\subsection{Motivation}
The most efficient implementations of the state-of-the-art algorithms are based on the \textit{hash join}.
The out-neighbor set $N^{+}_u$ of each node $u\in \mathop{G}\limits ^{\rightarrow}$ is represented as
an adjacency list stored in an array. For listing all the $k$-cliques, each node $u\in \mathop{G}\limits ^{\rightarrow}$ is assigned a label $l$ ($l=k$ initially), which indicates that we want to find $l$-cliques rooted from $u$. For each node $v\in N^{+}_u$ whose label was $l$, $v$'s label is set to $l-1$ in order to further find $(l-1)$-cliques rooted from $v$ (i.e., $l$-cliques rooted from $\{u,v\}$). 
More specifically, the vertex set $V(\mathop{G}\limits ^{\rightarrow})$ in the current DAG $\mathop{G}\limits ^{\rightarrow}$ performs a \textit{hash join} with $u$'s out-neighbors, in order to 
obtain all the nodes with label $l-1$ and generate the new DAG in the next recursion.
For each $v\in N^{+}_u$, all the out-neighbors
of $v$ with label $l-1$ are moved in the first part of the adjacency list.

Since the out neighbors of each vertex are not sorted, the technique of data level parallelism is hard to be implemented. Meanwhile, for set intersections based on \textit{merge join}, SIMD instructions with data level parallelism are frequently exploited~\cite{SIMD2,SIMD3,SIGMOD18}, motivating us to accelerate the $k$-clique listing based on \textit{merge join}.

\begin{algorithm}[t]
	\DontPrintSemicolon
	\KwIn{A graph $G$ and a positive integer $k$}
	\KwOut{All the $k$-cliques in $G$}
	Perform \textit{Pre-Core} preprocessing on $G$\label{line:sdegree:pre-kcore}\;
	Perform \textit{Pre-List} preprocessing on $G$\label{line:sdegree:pre-list}\;
	Generate a DAG $\mathop{G}\limits ^{\rightarrow}$ based on degree ordering \label{line:sdegree:reorder}\;
	\For{each node $u\in V(\mathop {G}\limits ^ {\rightarrow})$\label{line:sdegree:main1}}{
		\uIf{$d^{+}_u \ge k-1$}{
			$R\gets \{u\}$\;
			SDegreeList($k-1$, $R$, $N^{+}_u$, $\mathop{G}\limits ^{\rightarrow}$)\;\label{line:sdegree:main2}
		}
	} 
	\textbf{Procedure} SDegreeList($l$, $R$, $C$, $\mathop{G}\limits ^{\rightarrow}$)\;
	\For{each node $u\in C$}{
		\uIf{$d^{+}_u\le l-2$\label{line:sdegree:degreeprune1}}{
			continue\;\label{line:sdegree:degreeprune2}
		}
		$C'=N^{+}_u\cap C$ \tcp*{ Merge Join with SIMD}\label{line:sdegree:intersect}
		\uIf{$l=2$\label{line:sdegree:list1}}{
			\For{each node $v\in C'$}{
				output $k$-cliques $R\cup \{u,v\}$\;\label{line:sdegree:list2}
			}
		}
		\uElse{
			\uIf{$|C'|>l-2$\label{line:sdegree:recursion1}}{
				SDegreeList($l-1$, $R\cup\{u\}$, $C'$, $\mathop{G}\limits ^{\rightarrow}$)\;\label{line:sdegree:recursion2}
			}
		}
	}
	
	\caption{\textsc{SDegree}($G$, $k$)}
	\label{algo:SDegreeList}
\end{algorithm}

\subsection{SDegree}
Based on \textit{merge join}, we propose a simple but effective framework \textit{SDegree} to list all the $k$-cliques. With data level parallelism, \textit{SDegree} can apply arbitrary vertex ordering, and we choose the degree ordering as the total ordering. As illustrated in Algorithm \ref{algo:SDegreeList}, \textit{SDegree} first performs the \textit{Pre-Core} preprocessing and the \textit{Pre-List} preprocessing on $G$ (Lines \ref{line:sdegree:pre-kcore}-\ref{line:sdegree:pre-list}). Then, based on degree ordering, a DAG $\mathop{G}\limits ^{\rightarrow}$ is generated (Line \ref{line:sdegree:reorder}). For each edge $e=(u,v)$, the orientation is $u\rightarrow v$ if $d_u< d_v$ (break ties by node ID).
We traverse each vertex $u$ and invoke the procedure \textit{SDegreeList} for $d^{+}_u\ge k-1$ (Lines \ref{line:sdegree:main1}-\ref{line:sdegree:main2}). Although \textit{Pre-Core} ensures $d_u\ge k-1$, it may still be the case that $d^{+}_u<k-1$ in the DAG $\mathop{G}\limits ^{\rightarrow}$, which can be pruned in advance. 

In our implementation of the procedure \textit{SDegreeList}, $R$ is the vertex set to form a clique, which is initialized as $\{u\}$. $C$ is the candidate set of vertices that would expand $R$ into a $(|R|+1)$-clique, and is initialized as $N^{+}_u$. The integer $l$ reflects the depth of the recursion, where $l=k-1$ initially and $l+|R|=k$. In other words, \textit{SDegreeList}($l$,$R$,$C$,$\mathop{G}\limits ^{\rightarrow}$) needs to find all the $l$-cliques rooted from $R$.

\textit{SDegreeList} processes nodes of $k$-cliques in the order of $\eta$, which ensures the correctness (See Section~\ref{sec:analysis}). Observe that for a $k$-clique $\{v_1,v_2,...,v_k\}$ in $\mathop{G}\limits ^{\rightarrow}$ where $\eta(v_i)<\eta(v_{i+1})$, $v_i$ has $(k-i)$ out-neighbors. 
In other words, \textit{SDegreeList} can safely prune the $i$-th node $v_i$ in the order of $\eta$ when $d^+_{v_i}<k-i$. 
Therefore, for each $u\in C$, \textit{SDegreeList} prunes $u$ for $d^{+}_u< l-1$ in advance (Lines \ref{line:sdegree:degreeprune1}-\ref{line:sdegree:degreeprune2}).

To obtain the new candidate set $C'$ for larger cliques, \textit{SDegreeList} executes a \textit{merge join} on $N^{+}_u$ and $C$ (Line \ref{line:sdegree:intersect}). 
For $l=2$, \textit{SDegreeList} outputs all the $k$-cliques in $R$, joined with $u\in C$ and $v\in C'$ (Lines \ref{line:sdegree:list1}-\ref{line:sdegree:list2}). Otherwise, we recursively invoke the procedure \textit{SDegreeList} at the $(l-1)$ level, with new parameters $R\cup \{u\}$ and $C'$. \textit{SDegreeList} also prunes the case where $|C'|\le l-2$, for the size constraint (Lines \ref{line:sdegree:recursion1}-\ref{line:sdegree:recursion2}). Note that \textit{SDegreeList} 
exploits \textit{merge join} for the set intersections, which can be accelerated with SIMD instructions~\cite{SIGMOD18,SIMD2,SIMD3}.

\begin{figure}[ht]\centering
	\scalebox{1}[1]{\includegraphics{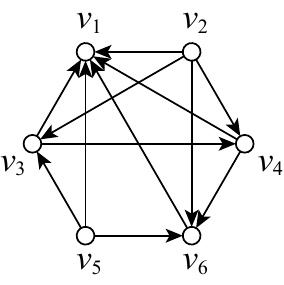}}
	\caption{\small The DAG after Preprocessing}
	\label{fig:directed_kcc}
\end{figure}

\begin{example}
	For $k=4$ and $G(V,E)$ in Fig.\ref{fig:graph}, $G$ is reduced as shown in Fig.\ref{fig:kcc} after \textit{Pre-Core} and \textit{Pre-List}. We obtain the DAG $\mathop{G}\limits ^{\rightarrow}$ as shown in Fig.\ref{fig:directed_kcc}. First, \textit{SDegree} prunes the other vertices except $v_2$ and $v_5$ since only $d^{+}_{v_2}\ge k-1$ and $d^{+}_{v_5}\ge k-1$. Starting from $v_2$, \textit{SDegree} invokes the procedure \textit{SDegreeList} with $N^{+}_{v_2}$. Only $v_3$ and $v_4$ can be further expanded due to the size constraint. Take $v_3$ as an example, at the second level $(l=2)$, $R=\{v_2,v_3\}$, $C=\{v_1,v_4\}$. For $v_4\in C$, $C'=N^{+}_{v_4}\cap C=\{v_1\}$. Therefore, we report the $4$-clique $R\cup\{v_1,v_4\}=\{v_2,v_3,v_4,v_1\}$. Another $4$-clique $\{v_2,v_4,v_6,v_1\}$ can be found similarly.
\end{example}

\subsection{Data Parallelism with SIMD Instructions}
\begin{figure}[ht]\centering
	\scalebox{1}[1]{\includegraphics[width=0.75\linewidth]{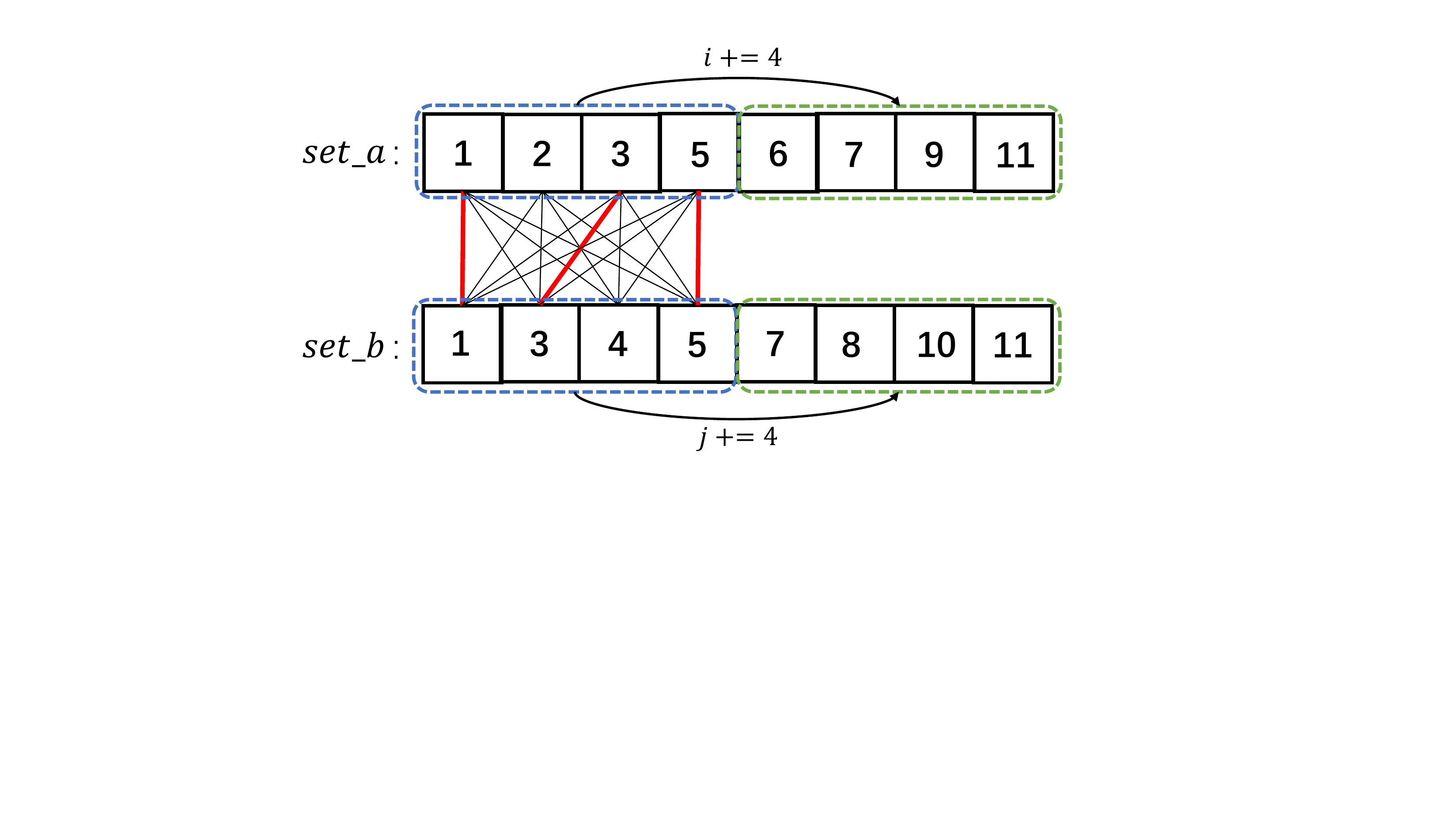}}
	\caption{\small Merge-based Intersection with SIMD Instructions}
	\label{fig:SIMD_set_intersection}
\end{figure}
We describe the main steps of the merge-based set intersection with SIMD instructions (Line \ref{line:sdegree:intersect}) as follows~\cite{SIGMOD18,SIMD3}.
\begin{itemize}
	\item\textit{Load vectors.}~Both two vectors of four $32$-bit integers are loaded into $128$-bit registers by SIMD load instructions such as $\_mm\_lddqu\_si128()$.
	\item\textit{Fully compare vectors.}~Make an all-pairs comparison between two vectors. First, compare four $32$-bit integers in both two vectors by SIMD compare instructions  ($\_mm\_cmpeq\_epi32()$). Then shuffle one vector  ($\_mm\_shuffle\_epi32()$) and repeat the comparison. Finally, store the intersection result ($\_mm\_storeu\_si128()$).
	\item\textit{Forward comparison.}~Compare the last elements of the two vectors. Advance both pointers to the next block when they are equal. Otherwise, the pointer of the smaller one is moved forward.
\end{itemize}

The intrinsic $\_mm\_lddqu\_si128()$ loads consecutive $128$-bit data from memory to a $128$-bit SIMD register; $\_mm\_cmpeq\_epi32(a,b)$ compares four $32$-bit integers in registers $a$ and $b$ for equality; $\_mm\_shuffle\_epi32(a,m)$ shuffles the four $32$-bit integers in the register $a$ with the mask $m$; $\_mm\_storeu\_si128()$ writes the $128$-bit data from the register to the result array.
\begin{example}
	In Fig.\ref{fig:SIMD_set_intersection},
	two vectors $V_a=set\_a[i:i+3]$ and $V_b=set\_b[j:j+3]$ are loaded into two $128$-bit registers with $\_mm\_lddqu\_si128()$. We compare the vectors with $\_mm\_cmpeq\_epi32()$ and get two common values $1$ and $5$. Then $V_b$ is shuffled as $[3,4,5,1]$, $[4,5,1,3]$ and $[5,1,3,4]$ with $\_mm\_shuffle\_epi32()$. We compare $V_a$ with each shuffled $V_b$ and get the common value $3$. Finally, the common values $[1,3,5,null]$ are written back to the result array with $\_mm\_storeu\_si128()$. We advance $i$ and $j$ to the next block (i.e., $i+=4,j+=4$) since $set\_a[i+3]=set\_b[j+3]$.
\end{example}

\section{Optimization Strategies}
\label{sec:BitCol}
\subsection{Color Ordering}
The main defect of other ordering heuristics is that the pruning effect is limited. They are based only on the size constraint that a $k$-clique must have at least k nodes.
Color ordering~\cite{baseline1} exploits the technique of greedy coloring~\cite{greedy_coloring}, and prunes more unpromising search paths in the $k$-clique listing  procedure. It is based on the following observation.

\begin{lemma}
	\label{lemma:color}
	If $G_k$ contains a $k$-clique, then at least $k$ colors are needed to color $G_k$.
\end{lemma}

The greedy coloring colors the nodes following a fixed order, which is specified as the \textit{inverse degree ordering} here. 
When coloring a vertex $v$, it always selects the minimum color value which has not been used by $v$'s neighbors.
The greedy coloring colors each node in descending order of the degree, which tends to assign small color values to the high-degree nodes.

The color ordering first assigns a color value $c_u$ to each node $u$ with the greedy coloring.  To construct the DAG $\mathop{G}\limits ^{\rightarrow}$ by color ordering, the vertices are reordered based on the color values. Specifically, the orientation of $e=(u,v)$ is $u\rightarrow v$ if $c_u>c_v$. For the node $u\in\mathop{G}\limits ^{\rightarrow}$ with $c_u<l$, the out-neighbors of $u$ have color values strictly smaller than $l-1$. Therefore, $u$ does not have $l-1$ out-neighbors with different colors, indicating that no $l$-clique rooted from $u$ exists.

\begin{figure}[ht]\centering
	\subfigure[][{\scriptsize Graph coloring}]{
		\scalebox{1}[1]{\includegraphics{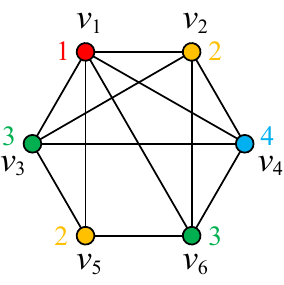}}
		\label{subfig:graph_coloring}}
	\subfigure[][{\scriptsize DAG on color-ordering}]{
		\scalebox{1}[1]{\includegraphics{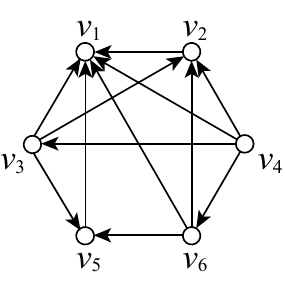}}
		\label{subfig:color_ordering}}
	\caption{\small The Example of Color Ordering}
	\label{fig:ColorOrder}
\end{figure}

Compared to the size constraint, the constraint of color values is stronger, which provides more pruning power.  For example in Fig.\ref{subfig:color_ordering}, since $v_4$ has $4$ out-neighbors, we can not prune $v_4$ based on the size constraint when finding a $5$-clique. However, the color value of $v_4$ is $4$, indicating that $v_4$ does not have $4$ out-neighbors with different colors. As a result, we can not find a $5$-clique rooted from $v_4$ and safely prune $v_4$. 
\begin{example}
	Assume the graph $G$ in Fig.\ref{subfig:graph_coloring} is an induced subgraph, we generate the DAG $\mathop{G}\limits ^{\rightarrow}$ based on the color ordering. Following the inverse degree ordering, the order for coloring the nodes is $(v_1,v_2,v_3,v_4,v_6,v_5)$. Firstly, $v_1$ is assigned with the smallest color value $1$, $v_2$ is colored with value $2$, then $v_3$ is colored with value $3$. Other nodes are colored similarly. As shown in Fig.\ref{subfig:color_ordering}, the DAG $\mathop{G}\limits ^{\rightarrow}$ is generated based on the color ordering.
\end{example}

\subsection{Out-neighbor Reduction}
In \textit{SDegree}, we perform the set intersection between a candidate set $C$ and an out-neighbor set $N^{+}_v(\mathop{G}\limits ^{\rightarrow})$ on each recursion. Initially, $C=N^{+}_u(\mathop{G}\limits ^{\rightarrow})$ is the out neighbors of a certain vertex $u$. Some vertices of $N^{+}_v(\mathop{G}\limits ^{\rightarrow})$ may not be in $N^{+}_u(\mathop{G}\limits ^{\rightarrow})$, which causes unnecessary comparisons. 

Motivated by the idea of induced subgraphs in~\cite{baseline1,baseline2}, we first construct the undirected subgraph $G_u$ induced by $N^{+}_u(\mathop{G}\limits ^{\rightarrow})$ and list the ($k$-$1$)-cliques $C_{k-1}$ on $G_u$. 
All the $k$-cliques $C_k$ are listed as $\{u\}\cup C_{k-1}$. Therefore, we can efficiently prune the invalid nodes that will not be contained in $N^{+}_u(\mathop{G}\limits ^{\rightarrow})$.

For the color-ordering based algorithms, the strategy of induced subgraphs can also be applied to reduce the worst-case time complexity~\cite{baseline1}. We can first generate a DAG $\mathop{G}\limits ^{\rightarrow}$ based on the degeneracy ordering. For each node $u$, we perform the ($k$-$1$)-clique listing on the undirected subgraph $G_u$ with color ordering.

\subsection{More Efficient Set Intersection Strategies}
We can efficiently reduce the set size for the intersection in two ways.
On one hand, we perform the $k$-clique listing procedure on the undirected subgraph $G_u$ induced by $N^{+}_u(\mathop{G}\limits ^{\rightarrow})$ for each node $u$.
On the other, we exploit vertex ordering to reduce the maximum degree in $\mathop{G_u}\limits ^{\rightarrow}$.
After the neighbor set is greatly reduced, we can further accelerate the set intersections with the idea of bitmaps. 

Firstly, we fix the size of nodes that a number can represent as $\mathcal{L}$. The vertex set $N^{+}_u(\mathop{G}\limits ^{\rightarrow})$ is encoded as a vector $B_u$ with $\lceil d^{+}_u(\mathop{G}\limits ^{\rightarrow})/\mathcal{L}\rceil$ numbers, and the out-neighbors of any node $v\in \mathop{G_u}\limits ^{\rightarrow}$ are encoded as $B_u(v)$ with $\lceil d^{+}_u(\mathop{G_u}\limits ^{\rightarrow})/\mathcal{L} \rceil$ numbers. The $i$-th bit of $B_u(v)$ is $1$, when the $i$-th out-neighbor of $u$ is also the out neighbor of $v$ in $\mathop{G_u}\limits ^{\rightarrow}$.
Meanwhile, a mask of $\mathcal{L}$-bits is calculated in advance, with which the neighbor sets can be recovered from a vector. Therefore, we can compress the neighbors of each node with a bitmap vector, instead of recording each specific neighbor in an adjacency list. Meanwhile, we only need to perform the bitwise AND operation based on the compressed neighbors, to obtain the intersection set that can further expand the current clique.
\begin{example}
	Let the graph in Fig.\ref{subfig:graph_coloring} be an induced subgraph $G_u$ and Fig.\ref{subfig:color_ordering} be the DAG $\mathop{G_u}\limits ^{\rightarrow}$ generated by color ordering. For $\mathcal{L}=3$, the compressed out-neighbor set of each vertex is shown in TABLE.\ref{tab:encode}, and the corresponding mask is shown in TABLE.\ref{tab:decode}.  In this example, two numbers are needed to store each neighbor set in $\mathop{G_u}\limits ^{\rightarrow}$. The intersection of $B_u(v_3)$ and $B_u(v_4)$ can be obtained with a bitwise AND operation $(\langle 3,2\rangle \& \langle 7,4\rangle = \langle 3,0\rangle)$, which can be recovered as $\{v_1,v_2\}$ with the mask.
\end{example}

\begin{table}
	\caption{Bitmaps of Neighbor Sets}
	\centering
	\subtable[Encoding]{
		
		\begin{tabular}{ccc}
			\hline
			Bitmap & $v_1 - v_3$ & $v_4 - v_6$ \\
			\hline
			$B_u(u)$      & $111(7)$ & $111(7)$\\
			$B_u(v_1)$ & $000(0)$ & $000(0)$\\
			$B_u(v_2)$ & $001(1)$ & $000(0)$\\
			$B_u(v_3)$ & $011(3)$ & $010(2)$\\
			$B_u(v_4)$ & $111(7)$ & $100(4)$\\
			$B_u(v_5)$ & $001(1)$ & $000(0)$\\
			$B_u(v_6)$ & $011(3)$ & $010(2)$\\
			\hline
		\end{tabular}
		\vspace{2mm}
		\label{tab:encode}
	}
	\subtable[Decoding]{
		\begin{tabular}{ccc}
			\hline
			Mask &Neighbors\\
			\hline
			$1(001)$ & $\{v_1\}$\\
			$2(010)$ & $\{v_2\}$\\
			$3(011)$ & $\{v_1,v_2\}$\\
			$4(100)$ & $\{v_3\}$\\
			$5(101)$ & $\{v_1,v_3\}$\\
			$6(110)$ & $\{v_2,v_3\}$\\
			$7(111)$ & $\{v_1,v_2,v_3\}$\\
			\hline
		\end{tabular}
		\vspace{2mm}
		\label{tab:decode}
	}
\end{table}

\subsection{BitCol}
Based on the above optimizations, we propose our improved algorithm \textit{BitCol}.
First, a DAG $\mathop{G}\limits ^{\rightarrow}$ is generated from $G(V,E)$, based on the degeneracy ordering. For each node $u$, \textit{BitCol} constructs an undirected subgraph $G_u(V_u,E_u)$ induced by $N^{+}_u(\mathop{G}\limits ^{\rightarrow})$. Specifically, $V_u=N^{+}_u(\mathop{G}\limits ^{\rightarrow})$ and $E_u=\{(v_i,v_j)|(v_i,v_j)\in E,v_i\in V_u,v_j\in V_u\}$. After that, a DAG $\mathop{G_u}\limits ^{\rightarrow}$ is generated based on color ordering. Finally, \textit{BitCol} iteratively processes on each induced subgraph $\mathop{G_u}\limits ^{\rightarrow}$.

An appealing feature is that the size of vertices in each induced subgraph is restricted within the degeneracy $\beta$. Notice that the degeneracy is often very small in real-world graphs~\cite{degeneracy}. Therefore, we take the idea of bitmaps and propose a simple but effective strategy to accelerate the set intersections. 

\begin{algorithm}[t]
	\DontPrintSemicolon
	\KwIn{A graph $G$ and a positive integer $k$}
	\KwOut{All the $k$-cliques in $G$}
	Perform \textit{Pre-Core} preprocessing on $G$\;\label{line:bitcol:pre-kcore}
	Perform \textit{Pre-List} preprocessing on $G$\;\label{line:bitcol:pre-list}
	Generate a DAG $\mathop{G}\limits ^{\rightarrow}$ based on  degeneracy ordering\;
	Fix $\mathcal{L}$ for bitmaps\;
	\For{each node $u\in V(\mathop {G}\limits ^ {\rightarrow})$}{
		\uIf{$d^{+}_u \ge k-1$}{
			$R\gets \{u\}$\;
			$(\mathop{G_u}\limits ^{\rightarrow}, color)\gets ColorOrdering(G_u)$\;\label{line:bitcol:colororder}
			$B_u\gets BitEncode(\mathop{G_u}\limits ^{\rightarrow}, \mathcal{L})$\;\label{line:bitcol:encode}
			BitColList($k-1$, $color$, $R$, $B_u$, $B_u(u)$)\;\label{line:bitcol:main}
		}
	} 
	\textbf{Procedure} BitColList($l$, $color$, $R$, $B$, $C_B$)\;
	\For{each node $v\in BitDecode(C_B)$\label{line:bitcol:decode}}{
		\uIf{$color(v) < l$\label{line:bitcol:colorprune1}}{
			continue\;\label{line:bitcol:colorprune2}
		}
		$C'_B\gets BitJoin(B(v), C_B)$\;\label{line:bitcol:bitjoin}
		\uIf{$l=2$\label{line:bitcol:recursion1}}{
			\For{each node $w\in BitDecode(C'_B)$}{
				output $k$-cliques $R\cup \{v, w\}$\;
			}
		}
		\uElse{
			\uIf{$|BitDecode(C'_B)|)\ge l-1$}{         BitColList($l-1$, $color$, $R\cup\{v\}$, $B$, $C'_B$)\;\label{line:bitcol:recursion2}
			}
		}
	}
	\textbf{Procedure} BitJoin($B$, $C_B$)\;
	\For{i = $1,2,\cdots,|B|$\label{line:bitcol:bitjoin_start}}{
		$C'_B[i]\gets B[i] \& C_B[i]$ \tcp*{ Bitwise AND Operation}\label{line:bitcol:bitjoin_end}
	}
	\Return{$C'_B$}
	\caption{\textsc{BitCol}($G$, $k$)}
	\label{algo:BitCol}
\end{algorithm}

As illustrated in Algorithm \ref{algo:BitCol}, \textit{BitCol} first reduces the original graph by preprocessing (Lines \ref{line:bitcol:pre-kcore}-\ref{line:bitcol:pre-list}). A DAG $\mathop{G}\limits ^{\rightarrow}$ is generated based on degeneracy ordering. Then \textit{BitCol} obtains the color values $color$ and the induced DAG $\mathop {G_u}\limits ^ {\rightarrow}$ by reordering the induced subgraph $G_u$ on color ordering (Line \ref{line:bitcol:colororder}). After that, \textit{BitCol} encodes the adjacency lists in $\mathop {G_u}\limits ^ {\rightarrow}$ with bitmaps, and invokes the procedure \textit{BitColList} (Lines \ref{line:bitcol:encode}-\ref{line:bitcol:main}). 

For the procedure \textit{BitColList}, $R$ represents the current clique, $B$ is the bitmap, and $C_B$ is the encoded bitmap of the candidate set with which $R$ can be expanded into a $(|R|+1)$-clique. Before processing on each vertex $v$, \textit{BitColList} decodes the bitmap of $C_B$ into the candidate set (Line \ref{line:bitcol:decode}).
According to Lemma \ref{lemma:color}, \textit{BitColList} prunes the search space for $color(v)<l$ (Lines \ref{line:bitcol:colorprune1}-\ref{line:bitcol:colorprune2}). Since $color(v)<l$, $v$ does not have $l-1$ out neighbors with different colors, which means $v$ and its out neighbors can not form any $l$-clique. 

\textit{BitColList} accelerates the set intersections with the procedure \textit{BitJoin} (Line \ref{line:bitcol:bitjoin}), which performs bitwise AND operation on the candidate vector with bitmap $C_B$ and the encoded out-neighbor vector $B(v)$ (Lines \ref{line:bitcol:bitjoin_start}-\ref{line:bitcol:bitjoin_end}). 
Note that the procedure \textit{BitJoin} exploits the data level parallelism with the compiler auto-vectorization~\cite{autovec,autovec_gcc}, which can obtain further acceleration.

\subsection{Data Parallelism with Auto-vectorization}
Automatic vectorization is supported on Intel® 64 architectures~\cite{autovec,autovec_gcc}. If vectorization is enabled (compiled using O2 or higher options), the compiler may use the additional registers to perform four bitwise operations in a single instruction. The obstacles to auto-vectorization are non-contiguous memory access and data dependencies, both of which are avoided by the Procedure \textit{BitJoin} (Lines \ref{line:bitcol:bitjoin_start}-\ref{line:bitcol:bitjoin_end}). However, the hash join accesses non-contiguous memory so that auto-vectorization can not be exploited.

For the original serial loop, each instruction can only handle single data. Instead, single instruction processes on a block of elements for the vectorized loop. For example, the loop in \textit{BitJoin} can be vectorized as follows:
\begin{equation}
	C_B'[i:i+3]\gets B[i:i+3] \& C_B[i:i+3], \nonumber
\end{equation}
where the loop bound is $\lfloor|B|/4\rfloor$ and single instruction can handle four results of the bitwise AND operation. The remaining ($|B|$ mod $4$) elements at the tail will be processed in a serial loop.

\section{Theoretical Analysis}
\label{sec:analysis}
In this section, we give a theoretical analysis of the correctness, time complexity, and space complexity of our algorithms. Let $m$ be the number of edges, $\Delta$ be the upper bound of the out-degree and $N$ be the number of threads. 
The state-of-the-art algorithms list all the $k$-cliques in $O(km(\frac{\Delta}{2})^{k-2})$ time, using $O(m+N\Delta^2)$ memory.

The correctness is guaranteed by the unique listing order of the vertices which represent a $k$-clique.
\begin{theorem}[Correctness]
	\label{theorem:correcteness}
	Both \textit{SDegree} and \textit{BitCol} list every $k$-clique in $G$ without repetition.
	\begin{proof}
		Obviously, \textit{Pre-Core} and \textit{Pre-List} will not affect the final results of $k$-clique listing. Let $\{v_1,\dots,v_k\}$ be the nodes of a $k$-clique. There is the only ordering such that $\eta(v_1)<\eta(v_2)<\dots<\eta(v_k)$. 
		$\forall i\in[2,k]$, vertex $v_i$ will be detected after $v_{i-1}$ since $v_i\in N^{+}_{v_{i-1}}$. Therefore, the $k$-clique will only be listed in the order $(v_1,v_2,\dots,v_k)$, without repetition.
	\end{proof}
\end{theorem}

\begin{theorem}
	\label{theorem:space}
	\textit{SDegree} lists all the $k$-cliques with $O(m+kN\Delta)$ space and \textit{BitCol} uses
	$O(m+N\Delta^2/\mathcal{L})$ space, where $\mathcal{L}$ is the size of nodes that each number can represent.
	\begin{proof}
		The space overhead is mainly divided into two parts, the input original graph $G$ and each subgraph $G_u$ induced by $N^{+}_u$. Both \textit{SDegree} and \textit{BitCol} require $O(m)$ memory for storing the input graph $G$. Then we perform the analysis on the induced subgraph $G_u$.
		
		\textit{SDegree} only maintains the vertex set of $G_u$ in each recursion, which is the candidate set $C$ ($|C|\le \Delta$). Therefore, \textit{SDegree} requires additional $O(k\Delta)$ space for each thread ($k\le \Delta$). 
		For \textit{BitCol}, each neighbor set of induced subgraph $G_u$ is compressed with a binary representation, where each number can represent $\mathcal{L}$ nodes. Therefore, \textit{BitCol} requires additional $O(\Delta^2/\mathcal{L})$ space for each thread.
	\end{proof}
\end{theorem}

To present a formal analysis of the time complexity, several necessary lemmas are given in the following. The $k$-clique listing problem can be solved in linear time for $\Delta < 2 $ or $k<3$. Therefore, we only consider $\Delta\ge 2$ and $k\ge3$ in this paper.
\begin{lemma} 
	\label{lemma:CliqueListing}
	Let $C$ be the candidate set for expanding the $k$-cliques. The time complexity of the procedure \textit{SDegreeList} in Algorithm \ref{algo:SDegreeList} can be upper bounded by $T(l,C)$ written as the following recurrence:
	\begin{equation*}
		\left\{\begin{array}{ll}
			T(2,C) = 2\sum\limits_{u\in C}(|C|+d^{+}_{u}) \\
			T(l,C) = \sum\limits_{u\in C}(|C|+d^{+}_{u})+\sum\limits_{u\in C}T(l-1,N^{+}_{u}\cap C)
		\end{array}\right.
	\end{equation*} 
	
	\begin{proof}
		For each node $u\in C$, \textit{SDegreeList} first calculates the intersection $C'$ of $N^{+}_u$ and $C$, which runs in $O(\sum\limits_{u\in C}(|C|+d^{+}_{u}))$. If $l=2$, $k$-cliques of $R\cup \{u,v\}$ are reported for each node $v\in C'$ ($O(\sum\limits_{u\in C}(|C|+d^{+}_{u}))$). Otherwise, \textit{SDegreeList} is recursively executed with the new parameters $l-1$ and $C'=N^{+}_u\cap C$.
	\end{proof}
\end{lemma}

\begin{lemma}
	\label{lemma:sum_intersection}
	Let $C$ be the candidate set in Algorithm \ref{algo:SDegreeList}. For each node $u$, the following equation holds, where $\Delta$ is the upper bound of the out-degree in $\mathop{G}\limits ^{\rightarrow}$.
	\begin{equation*}
		\sum\limits_{u\in C}|N^{+}_{u}\cap C|\le \frac{|C|(\Delta-1)}{2}
	\end{equation*}
	\begin{proof}
		Consider the subgraph $\mathop{G_c}\limits ^{\rightarrow}(C,E_c)$ induced by $C$ in $\mathop{G}\limits ^{\rightarrow}$, where $E_c=\{\langle u,v\rangle|\langle u,v\rangle \in E(\mathop{G}\limits ^{\rightarrow}), u\in C \wedge v\in C\}$. For each $u\in C$, $N^{+}_u(\mathop{G_c}\limits ^{\rightarrow})=N^{+}_u(\mathop{G}\limits ^{\rightarrow})\cap C$. Therefore, we have the following derivation.
		\begin{align}
			\label{inequ:lemma3_1}\sum\limits_{u\in C}|N^{+}_{u}(\mathop{G}\limits ^{\rightarrow})\cap C| &= \sum\limits_{u\in C}|N^{+}_u(G_c)| \\
			\label{inequ:lemma3_2}                                      &\le \frac{|C|(|C|-1)}{2}\\
			\label{inequ:lemma3_3}                                      &\le \frac{|C|(\Delta-1)}{2}
		\end{align}
		Equation (\ref{inequ:lemma3_2}) follows from the fact that $\sum\limits_{u\in C}|N^{+}_u(G_c)|=|E_c|$, and $|E_c|$ is at most the number of edges in a $|C|$-clique.
	\end{proof}
	
\end{lemma}

\begin{lemma}
	\label{lemma:Sum}
	Let $\Delta$ be the upper bound of the out-degree in $\mathop{G}\limits ^{\rightarrow}$ and $C$ be the candidate set in Algorithm \ref{algo:SDegreeList}, the following equation holds.
	\begin{equation*}
		\begin{array}{ll}
			T(l,C)\le 2\Delta (k+\frac{l}{2})(\frac{\Delta}{2})^{l-2}|C|
		\end{array}
	\end{equation*} 
	
	\begin{proof}
		We prove by the induction on $l$, where $2\le l\le k-1$.
		For $l=2$ and $k\ge 3$, $T(2,C)=2\sum\limits_{u\in C}(|C|+d^{+}_{u})$. Obviously, we have $|C|\le \Delta$ and $d^{+}_u\le \Delta$ for each $u\in C$. Therefore, Lemma \ref{lemma:Sum} holds for $l=2$. For $l>2$, we have the following derivation.
		\begin{align}
			\label{inequ:lemma4_1}T(l,C)&\le \sum\limits_{u\in C}(|C|+d^{+}_{u})+\sum\limits_{u\in C}T(l-1,N^{+}_{u}\cap C)\\ 
			\label{inequ:lemma4_2} &\le 2\Delta |C|+\sum\limits_{u\in C}2\Delta(k+\frac{l-1}{2})(\frac{\Delta}{2})^{l-3}|N^{+}_{u}\cap C|\\
			\label{inequ:lemma4_3} &= 2\Delta |C|+2\Delta(k+\frac{l-1}{2})(\frac{\Delta}{2})^{l-3}\sum\limits_{u\in C}|N^{+}_{u}\cap C|\\
			\label{inequ:lemma4_4} &\le 2\Delta |C|+2\Delta(k+\frac{l-1}{2})(\frac{\Delta}{2})^{l-2}|C|\\
			&\ \ \ -\Delta(k+\frac{l-1}{2})(\frac{\Delta}{2})^{l-3}|C| \nonumber\\
			\label{inequ:lemma4_5} &\le 2\Delta(k+\frac{l}{2})(\frac{\Delta}{2})^{l-2}|C|
		\end{align}
		Equation (\ref{inequ:lemma4_1}) follows from Lemma \ref{lemma:CliqueListing}, Equation (\ref{inequ:lemma4_2}) follows from the inductive hypothesis, Equation (\ref{inequ:lemma4_4}) follows from Lemma \ref{lemma:sum_intersection}, and Equation (\ref{inequ:lemma4_5}) follows from the fact that $2\Delta|C|\le \Delta(k+\frac{l-1}{2})(\frac{\Delta}{2})^{l-3}|C|$ for $l> 2$, $k\ge 3$, and $\Delta \ge 2$.
	\end{proof}
\end{lemma}

Derived from Lemma \ref{lemma:CliqueListing} and Lemma \ref{lemma:Sum}, we can formally give the main theorems on the time complexity.
\begin{theorem}
	\label{theorem:sdegree}
	\textit{SDegree} lists all the $k$-cliques in $O(km(\frac{\Delta}{2})^{k-2})$ time.
	\begin{proof}
		According to Lemma \ref{lemma:CliqueListing}, the time complexity of \textit{SDegree} can be formulated as follows.
		\begin{equation*}
			\begin{array}{ll}
				\mathcal{T}(k,G)=\sum\limits_{u\in V }T(k-1,N^{+}_u)+O(m+n)
			\end{array}
		\end{equation*} 
		First, \textit{Pre-Core} and \textit{Pre-List} both run in linear time. 
		According to Lemma \ref{lemma:Sum}, we have the following derivation.
		\begin{align*}
			\mathcal{T}(k,G)&=\sum\limits_{u\in V}T(k-1,N^{+}_u)+O(m+n)\\
			&\le \sum\limits_{u\in V}2\Delta(k+\frac{k-1}{2})(\frac{\Delta}{2})^{k-3}d^{+}_u+O(m+n)\\ &\le6 km(\frac{\Delta}{2})^{k-2}+O(m+n)
		\end{align*}
		The last equation above follows from the fact that $\sum\limits_{u\in V}d^{+}_u = m$. 
	\end{proof}
\end{theorem}

\begin{theorem}
	\label{theorem:bitcol}
	\textit{BitCol} lists all the $k$-cliques in $O(km(\frac{\Delta}{2})^{k-2})$ time. 
	\begin{proof}
		This theorem can be proved similarly as the Theorem \ref{theorem:sdegree}.
	\end{proof}
\end{theorem}

\section{Experiments}
\label{sec:experiment}
\begin{table}
	\caption{Statistics of Datasets}
	\centering
	\begin{tabular}{c c c c c c c}
		\hline
		Dataset &Name   & $|V|$ & $|E|$ & $d_{avg}$ & $\omega$  & $N_{max}$
		\\ \hline
		BerkStan &BS & 685K & 7M & 19.41 & 201 & 4
		\\ 
		Pokec &PK & 1.6M & 22.3M & 27.32 & 29 & 6
		\\
		DBLP & DB & 317k & 1M & 6.62 & 114 & 1
		\\
		CitPatents & CP & 6M & 16.5M & 5.49 & 11 & 2
		\\
		Linkedin & LK & 6.7M & 19.4M & 5.76 & 11 & 33
		\\
		Stanford & BB & 282K & 2M & 14.14 & 61 & 10
		\\
		WebUK05 & UK05 & 129K & 11.7M & 181.19 & 500 & 2
		\\
		ClueWeb09 & CW & 428M & 446M & 2.09 & 56 & 160
		\\
		Wikipedia13 & WP & 27.1M & 543M & 40.01 & 428 & 4
		\\
		AllWebUK02 & UK02 & 18.5M & 262M & 28.27 & 944 & 1
		\\ \hline
	\end{tabular}
	\vspace{2mm}
	\label{tb:dataset}
\end{table}
In this section, we conduct extensive experiments to evaluate the efficiency of our algorithms \textit{SDegree} and \textit{BitCol}. 

\subsection{Experimental Setup}
\noindent \textbf{Settings.} 
All experiments are carried on a Linux machine, equipped with 1TB disk, 128GB memory, and 4 Intel Xeon CPUs (4210R @2.40GHz, 10 cores). All algorithms are implemented in C/C++ and compiled with -O3 option. The source codes of \textit{DDegree} and \textit{DDegCol}~\footnote{https://github.com/gawssin/kcliquelisting} are publicly available in~\cite{baseline1}. For all the algorithms, we set the time limit to 24 hours, and the reported running time is the total CPU time excluding only the I/O time of loading graph from disk.

\noindent \textbf{Datasets.} All datasets are downloaded from the public website NetworkRepository~\footnote{https://networkrepository.com}. 
The detailed data descriptions are demonstrated in TABLE~\ref{tb:dataset}, where $d_{avg}$ denotes the average degree, $\omega$ denotes the maximum clique size, and $N_{max}$ denotes the number of maximum cliques. 

If $\omega$ is large, the number of $k$-cliques is exponentially large with a relatively large $k$. For instance, BerkStan has $4$ $201$-cliques and each of them has around $1.8 \times 10^{27}$ $20$-cliques, which can not be listed in a reasonable time for all the algorithms.

\noindent \textbf{Algorithms.} We compare two state-of-the-art algorithms \textit{DDegree} and \textit{DDegCol} for $k$-clique listing with our two proposed algorithms. We fix $\mathcal{L}=24$ for \textit{BitCol}.
\begin{itemize}
	\item \textit{DDegree} is the state-of-the-art algorithm for degree ordering.
	\item \textit{DDegCol} is the state-of-the-art algorithm for color ordering. For general $k$-clique listing algorithms, there does not exist a clear winner between \textit{DDegree} and \textit{DDegCol}~\cite{baseline1}, thus we compare both of them with our algorithms.
	\item \textit{SDegree} is our proposed algorithm based on degree ordering.
	\item \textit{BitCol} is our proposed algorithm based on color ordering.
\end{itemize}
To be more specific, we compare \textit{SDegree} with \textit{DDegree} for degree ordering, and compare \textit{BitCol} with \textit{DDegCol} for color ordering, respectively.

\subsection{$k$-clique Listing Time in Serial}
\begin{figure*}[htbp]
	\vspace{-0.3cm}
	
	\centering
	\subfigure[AllWebUK02]{
		\includegraphics[width=0.23\linewidth]{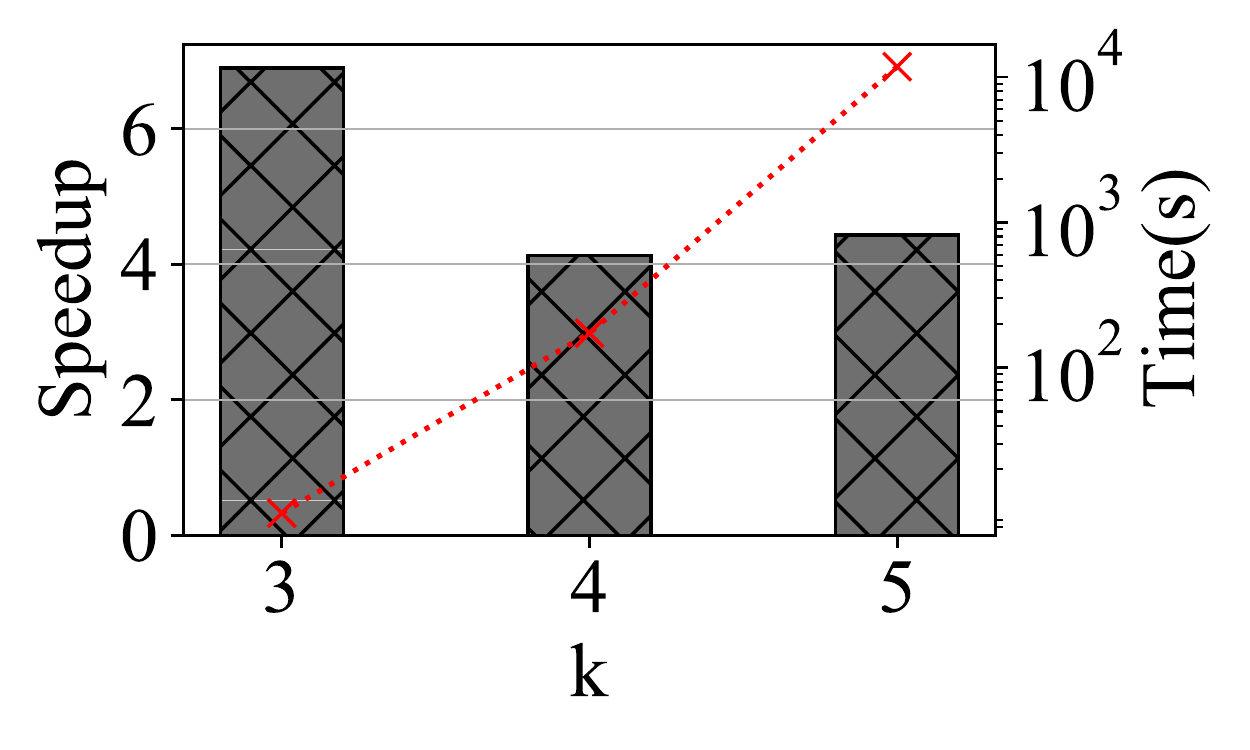}
		\label{fig:AllWebUK02_deg}
	}
	\subfigure[Stanford]{
		\includegraphics[width=0.23\linewidth]{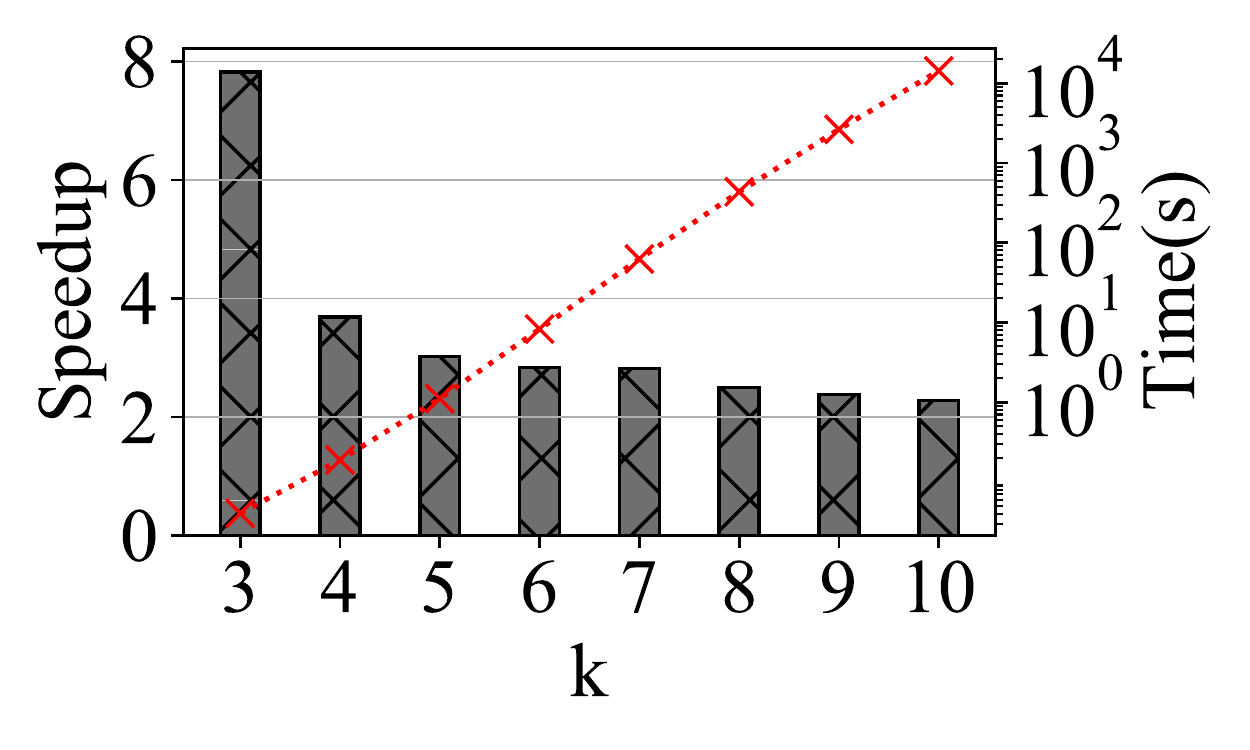}%
		\label{fig:Stanford_deg}
	}
	\subfigure[BerkStan]{
		\includegraphics[width=0.23\linewidth]{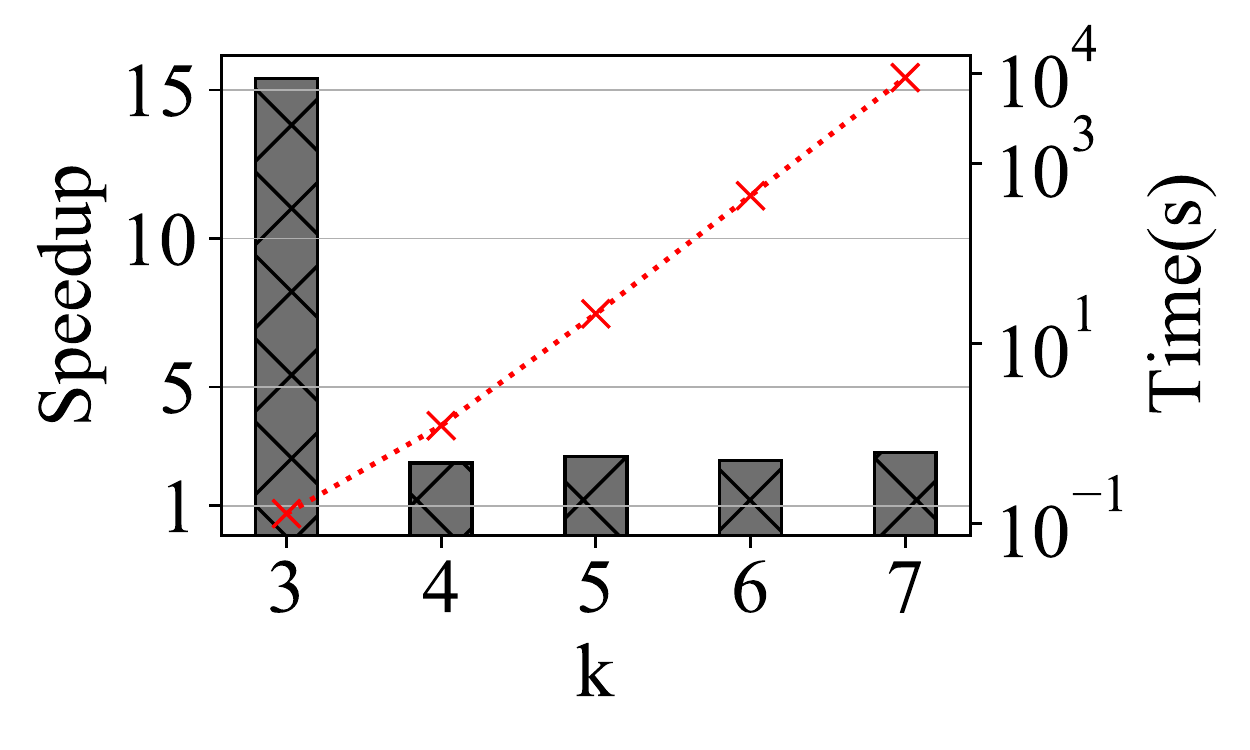}%
		\label{fig:BerkStan_deg}
	}
	\subfigure[CitPatents]{
		\includegraphics[width=0.23\linewidth]{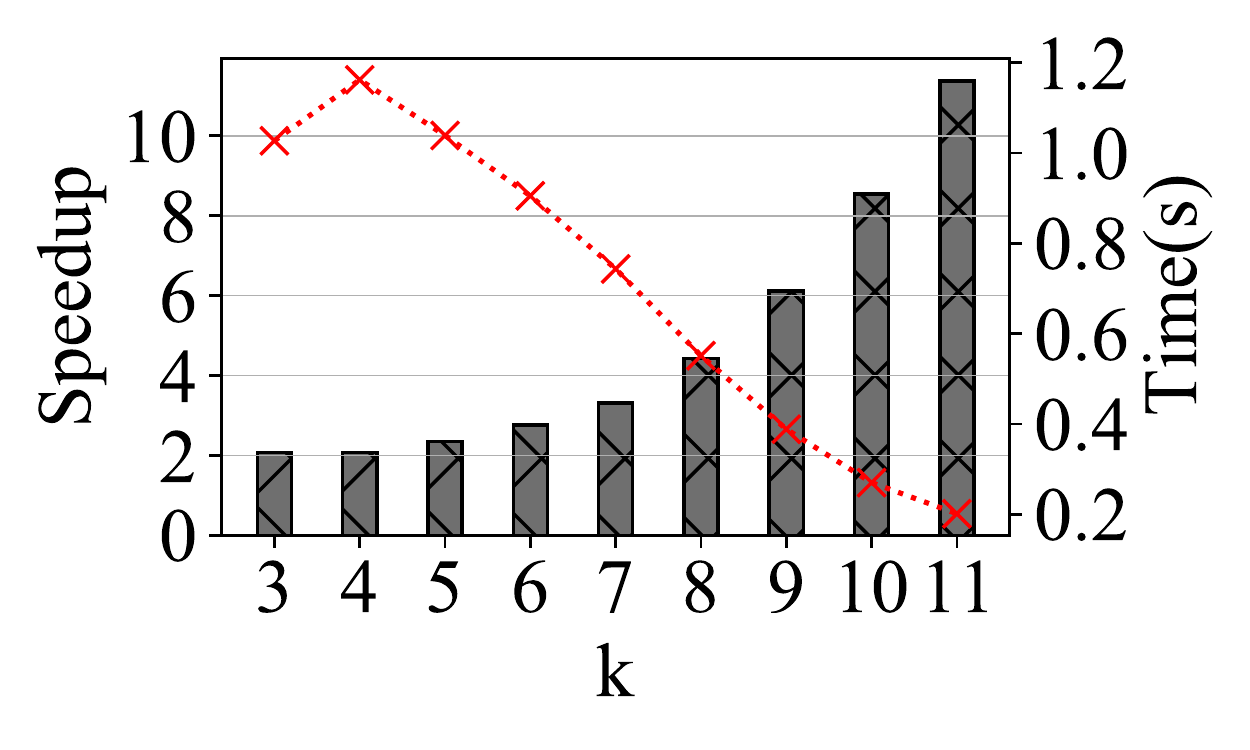}%
		\label{fig:CitParents_deg} 
	}

	\subfigure[DBLP]{
		\includegraphics[width=0.23\linewidth]{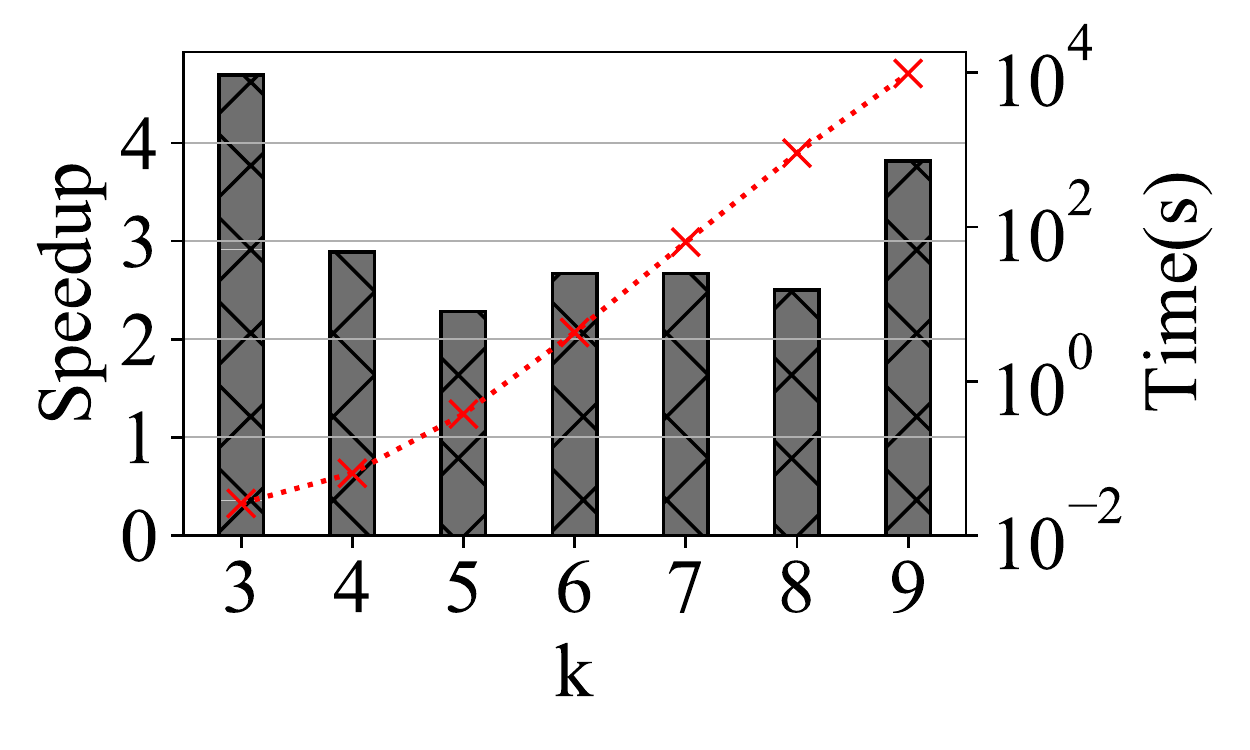}%
		\label{fig:DBLP_deg}
	}
	\subfigure[Linkedin]{
		\includegraphics[width=0.23\linewidth]{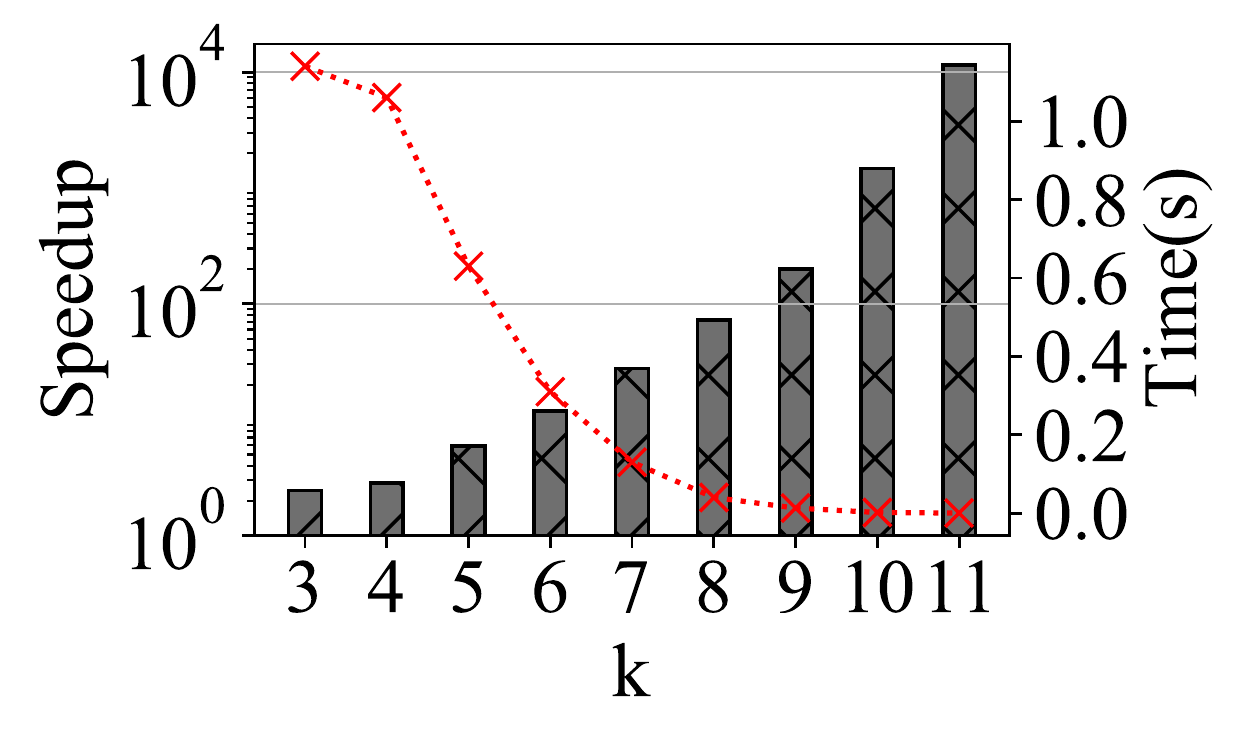}%
		\label{fig:Linkedin_deg}
	}
	\subfigure[Pokec]{
		\includegraphics[width=0.23\linewidth]{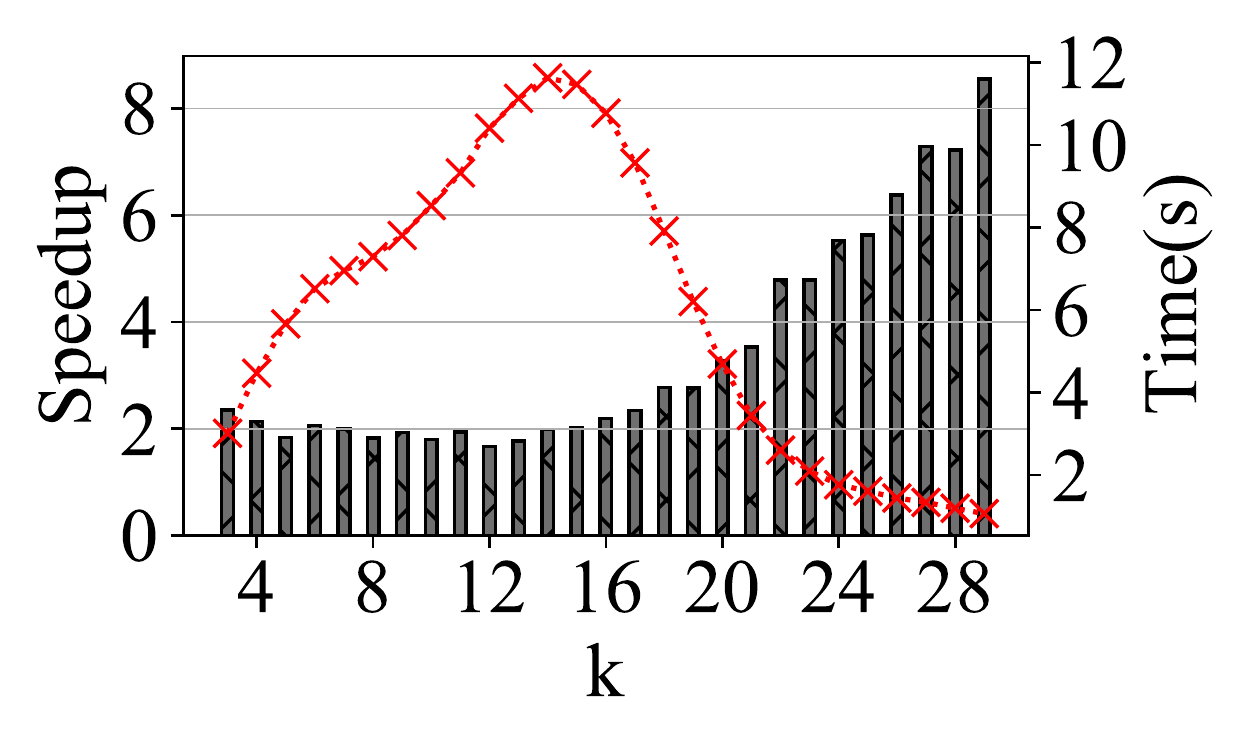}%
		\label{fig:Pokec_deg} 
	}
	\subfigure[WebUK05]{
		\includegraphics[width=0.23\linewidth]{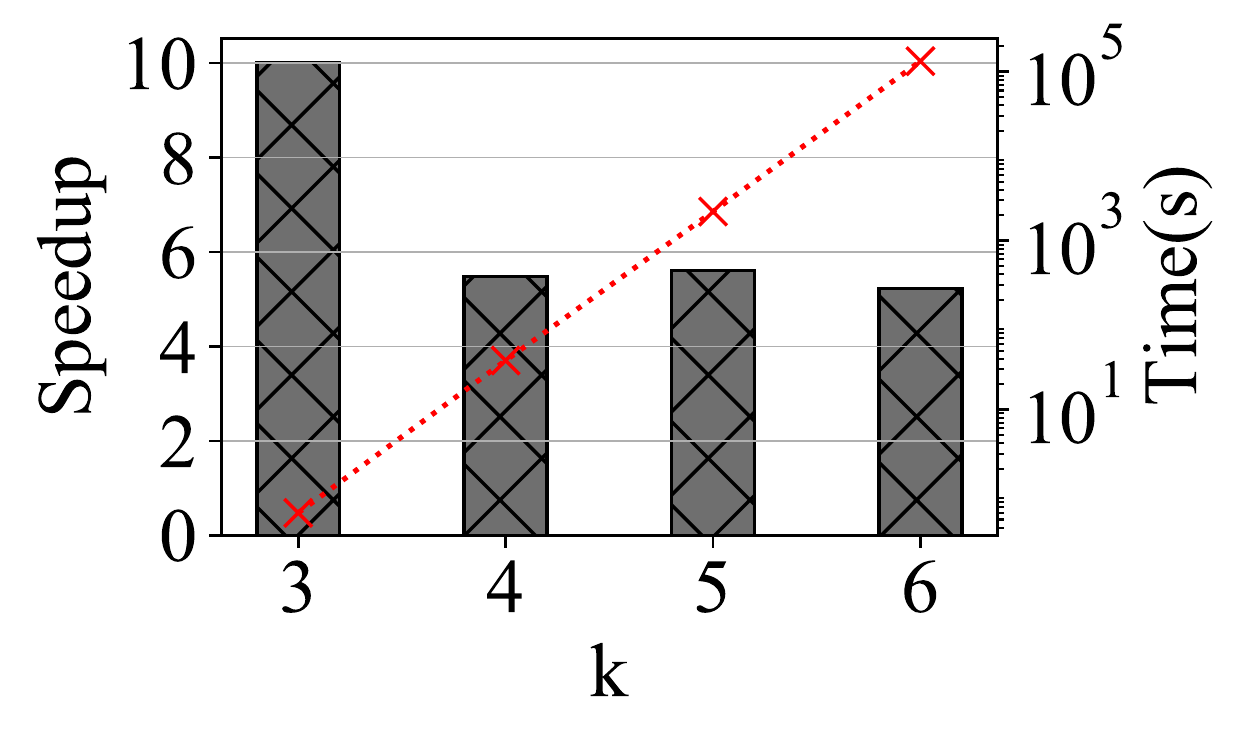}
		\label{fig:WebUK05_deg}
	}
	
	\vspace{-3mm}
	\caption{Running Time on Degree Ordering. Red lines represent  running  time of  \textit{SDegree}; Histograms represent  speedups  of \textit{SDegree} over \textit{DDegree}.}
	\label{fig:kclique-degree}
\end{figure*}

\begin{figure*}[!htb]
	\vspace{-0.2cm}
	
	\centering
	\subfigure[AllWebUK02]{
		\includegraphics[width=0.23\linewidth]{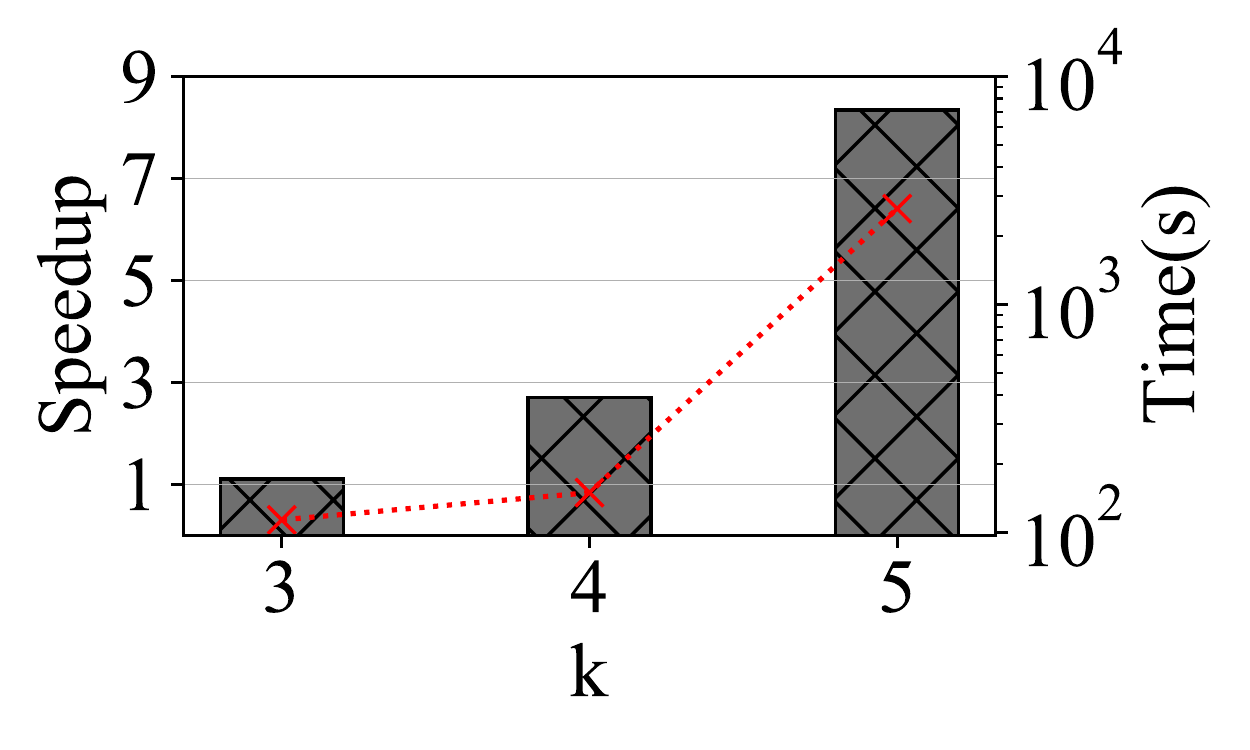}
		\label{fig:AllWebUK02_col}
	}
	\subfigure[BerkStan]{
		\includegraphics[width=0.23\linewidth]{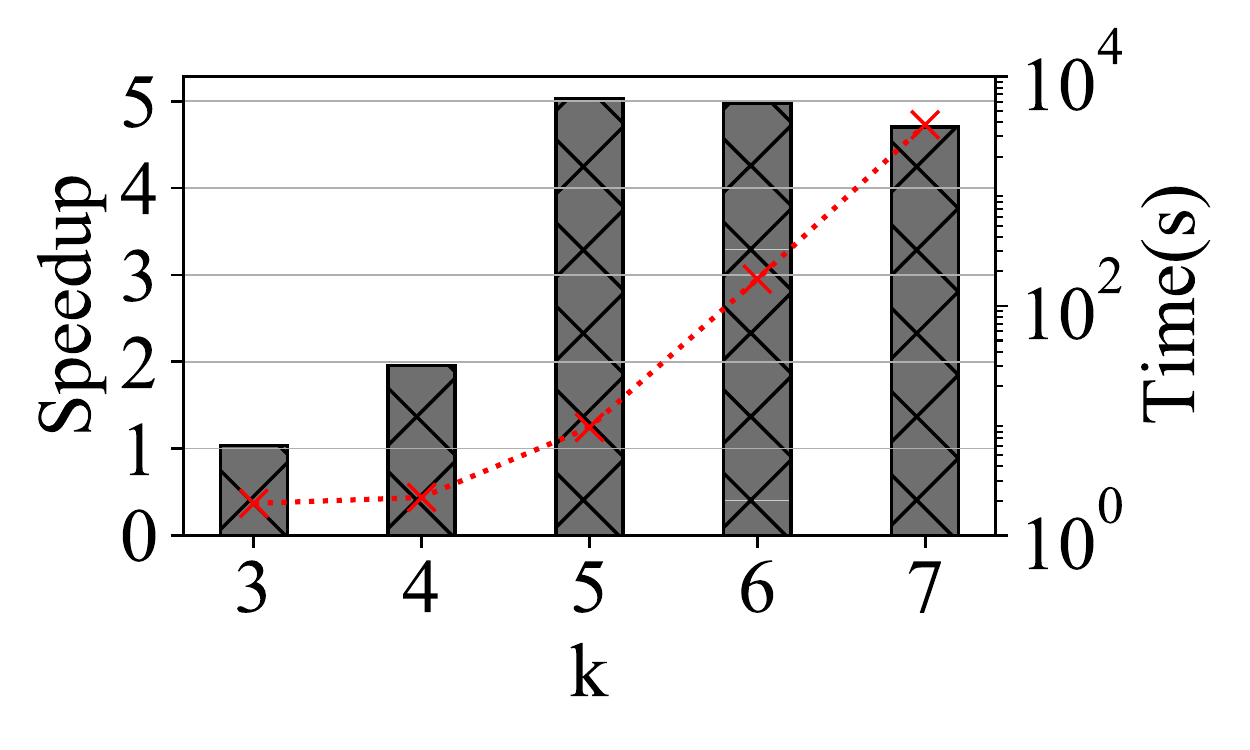}%
		\label{fig:BerkStan_col}
	}
	\subfigure[CitPatents]{
		\includegraphics[width=0.23\linewidth]{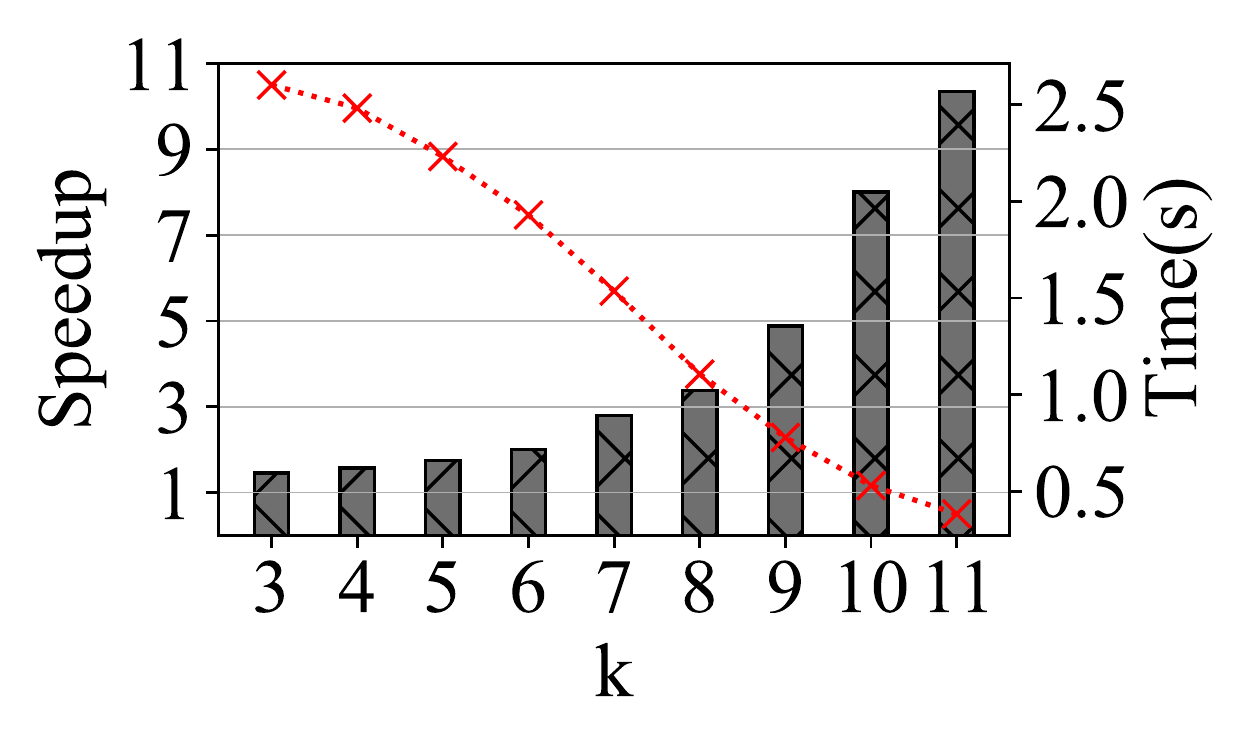}%
		\label{fig:CitParents_col} 
	}
	\subfigure[Stanford]{
		\includegraphics[width=0.23\linewidth]{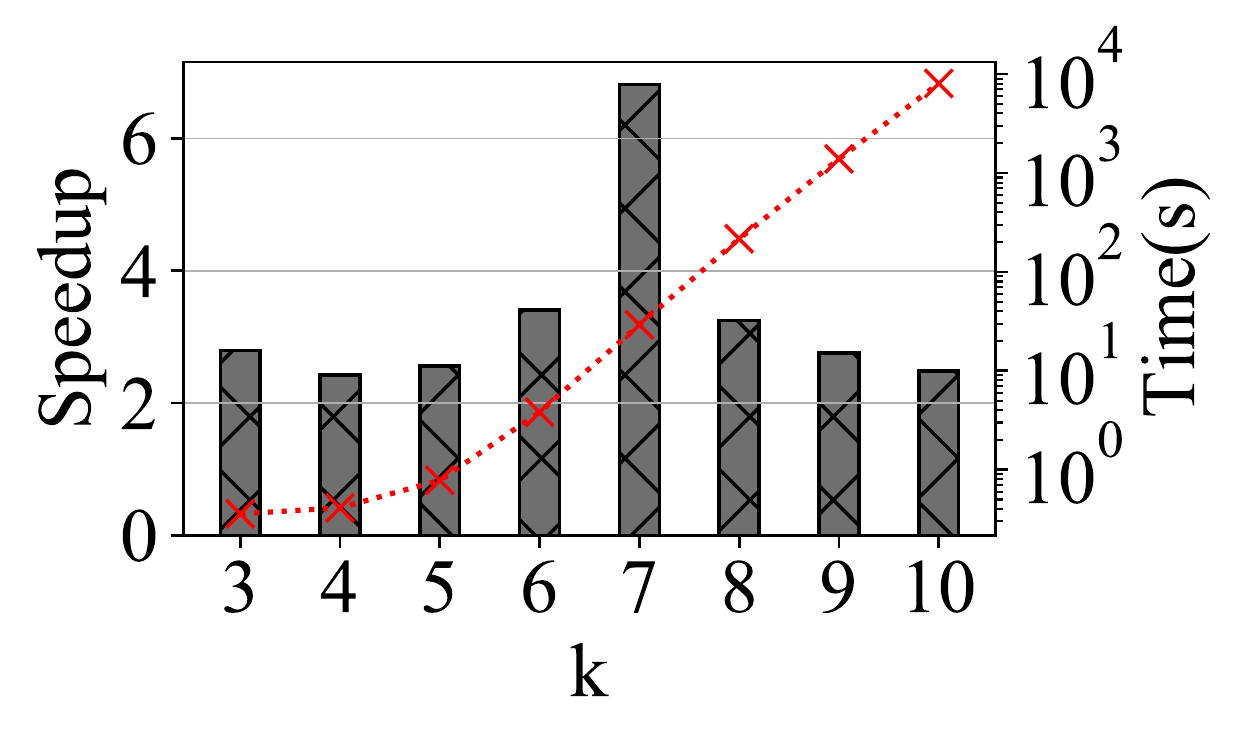}
		\label{fig:Stanford_col}
	}
	
	\subfigure[DBLP]{
		\includegraphics[width=0.23\linewidth]{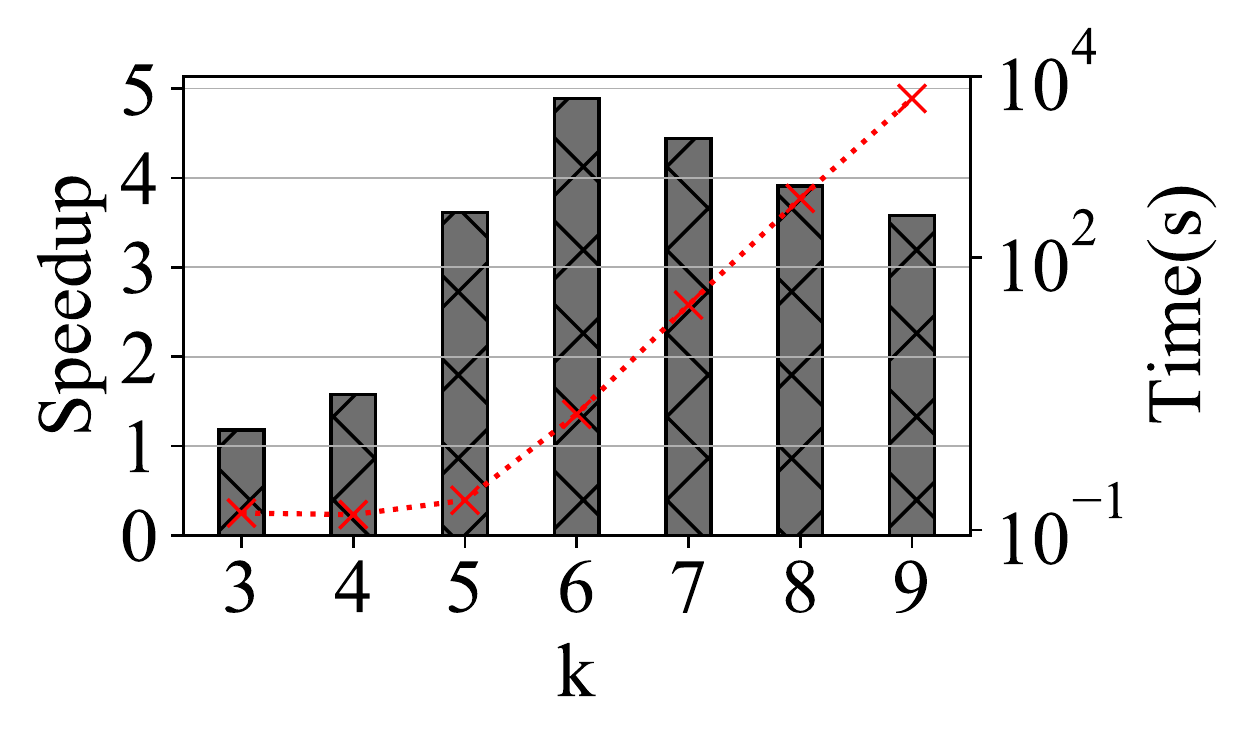}%
		\label{fig:DBLP_col}
	}
	\subfigure[Linkedin]{
		\includegraphics[width=0.23\linewidth]{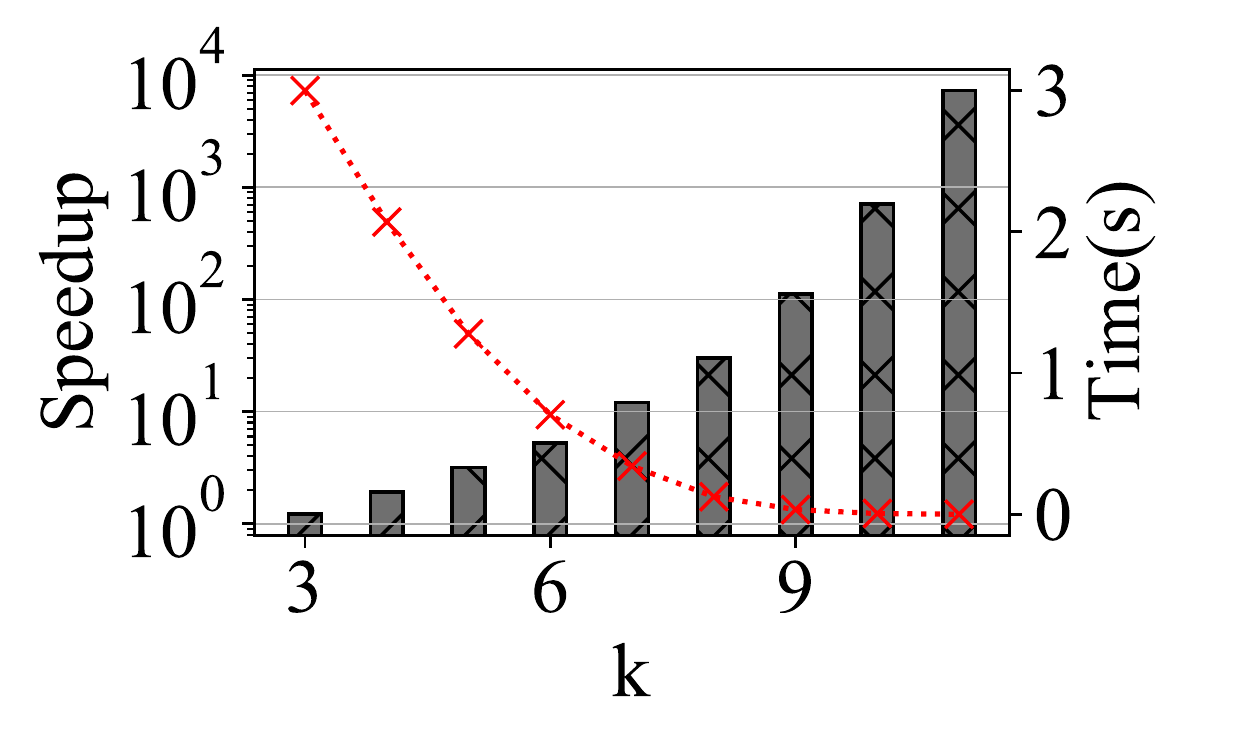}%
		\label{fig:Linkedin_col}
	}
	\subfigure[Pokec]{
		\includegraphics[width=0.23\linewidth]{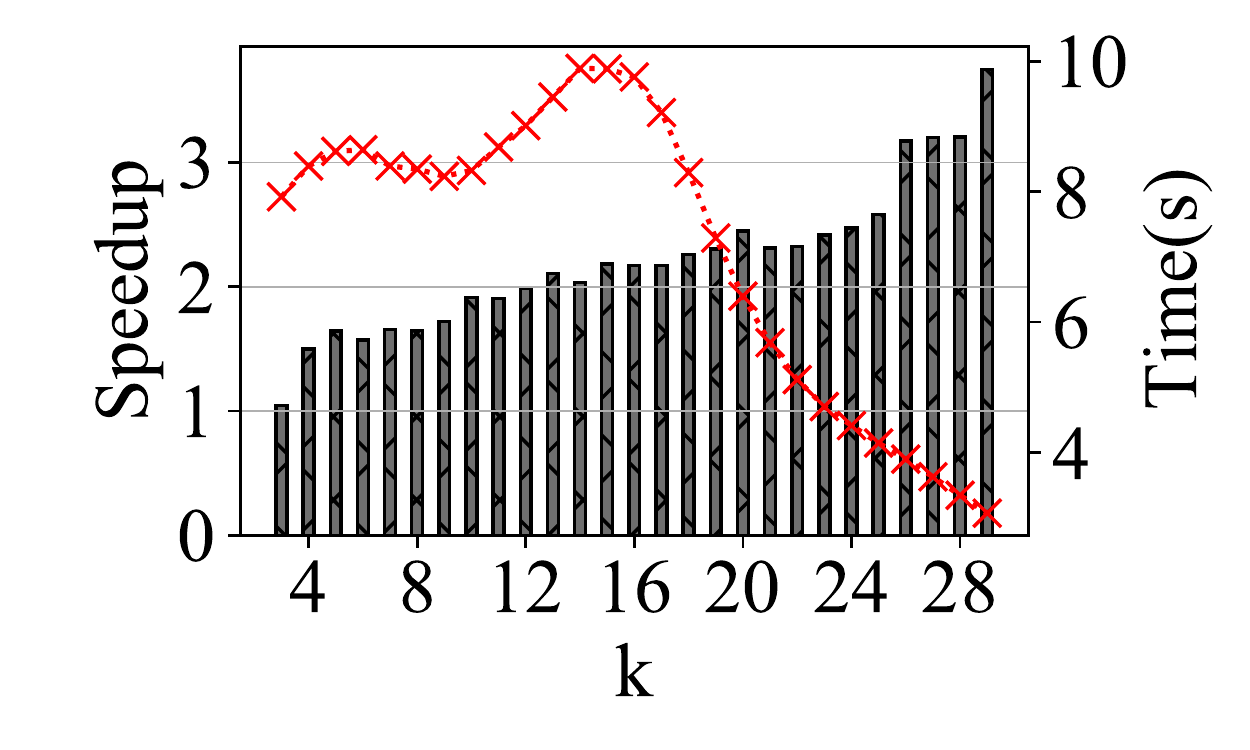}%
		\label{fig:Pokec_col} 
	}
	\subfigure[WebUK05]{
		\includegraphics[width=0.23\linewidth]{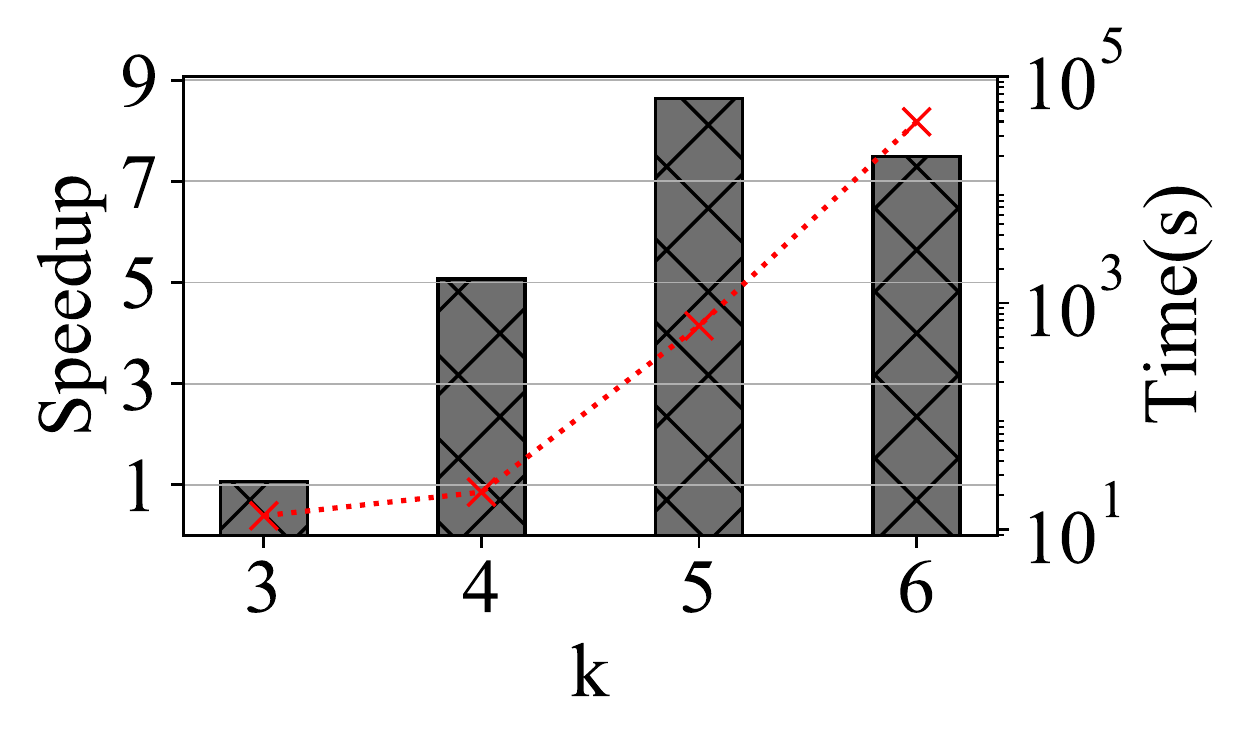}
		\label{fig:WebUK05_col}
	}
	
	\vspace{-2mm}
	\caption{Running Time on Color Ordering. Red lines represent  running  time of  \textit{BitCol}; Histograms represent  speedups  of \textit{BitCol} over \textit{DDegCol}.}
	\label{fig:kclique-color}
\end{figure*}

In this experiment, we evaluate the $k$-clique listing time in serial with varying the size $k$. The experiments are conducted based on degree ordering and color ordering respectively, which are illustrated in Fig.\ref{fig:kclique-degree} and Fig.\ref{fig:kclique-color}.
The red dotted line in Fig.\ref{fig:kclique-degree} represents the $k$-clique listing time of \textit{SDegree}, and the histogram represents the speedup of \textit{SDegree} over \textit{DDegree}. Similarly, the red dotted line in Fig.\ref{fig:kclique-color} represents the $k$-clique listing time of \textit{BitCol}, and the histogram represents the speedup of \textit{BitCol} over \textit{DDegCol}. 
The results shown in Fig.\ref{fig:kclique-degree} and Fig.\ref{fig:kclique-color} indicate that our algorithms outperform \textit{DDegree} and \textit{DDegCol} for fixed $k$, no matter based on degree ordering or color ordering.

From the perspective of the acceleration ratio w.r.t $k$, there are some interesting findings. In Fig.\ref{fig:kclique-degree}, \textit{SDegree} has a significant effect when dealing with triangle listing $(k=3)$. We explain that \textit{SDegree} does not need to construct induced subgraphs, which requires additional time overhead.
Correspondingly, \textit{BitCol} only has a tiny speedup over \textit{DDegCol} for triangle listing, since the listing process is dominated by constructing and reordering induced subgraphs when $k$ is small. The advantage of \textit{BitCol} with bitmaps to accelerate the set intersections is gradually obvious as $k$ increases. This further demonstrates that the set intersection is an essential step in the process of $k$-clique listing.

Within the time limit, we compare the total time required for all data sets and for all $k$ in serial as a metric to calculate the speedups of our algorithms. In a word, \textit{SDegree} outperforms \textit{DDegree} by $3.75$x, and \textit{BitCol} outperforms \textit{DDegCol} by $5.67$x for the total time.

\begin{figure}[htp]
	\subfigure[DBLP]{
		\includegraphics[width=0.45\linewidth]{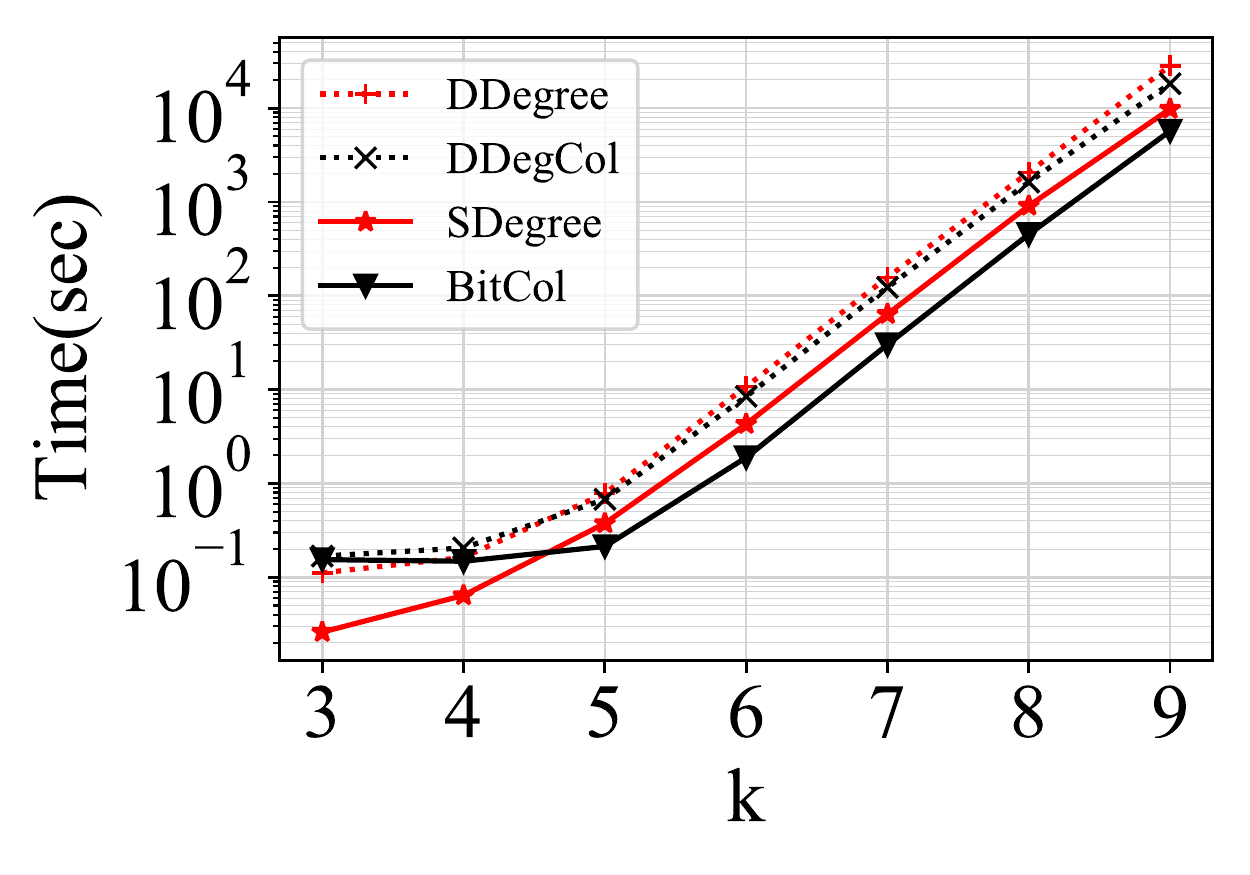}
		\label{fig:DBLP_Effect_of_k}
	}
	\subfigure[ClueWeb09]{
		\includegraphics[width=0.45\linewidth]{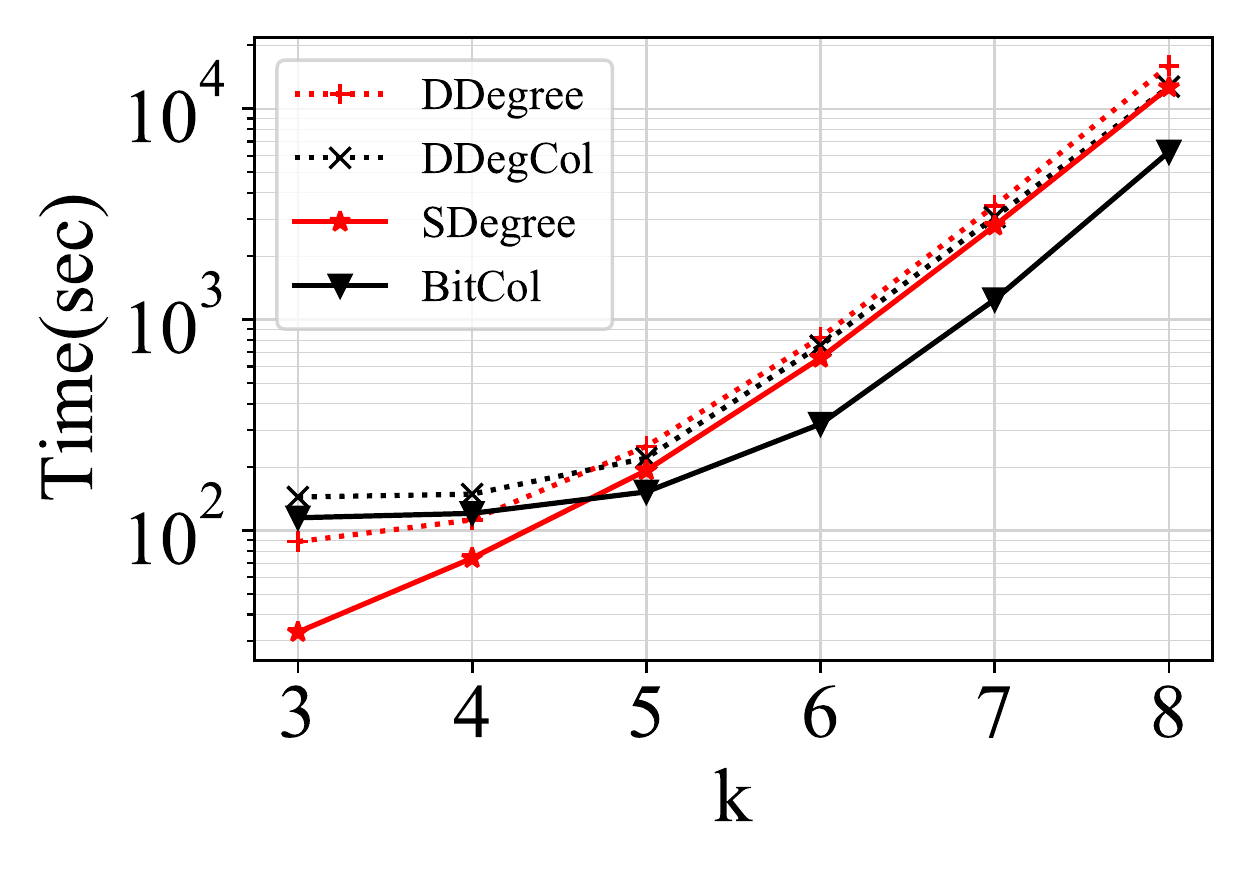}%
		\label{fig:ClueWeb09_Effect_of_k}
	}
	
	\subfigure[Pokec]{
		\includegraphics[width=0.45\linewidth]{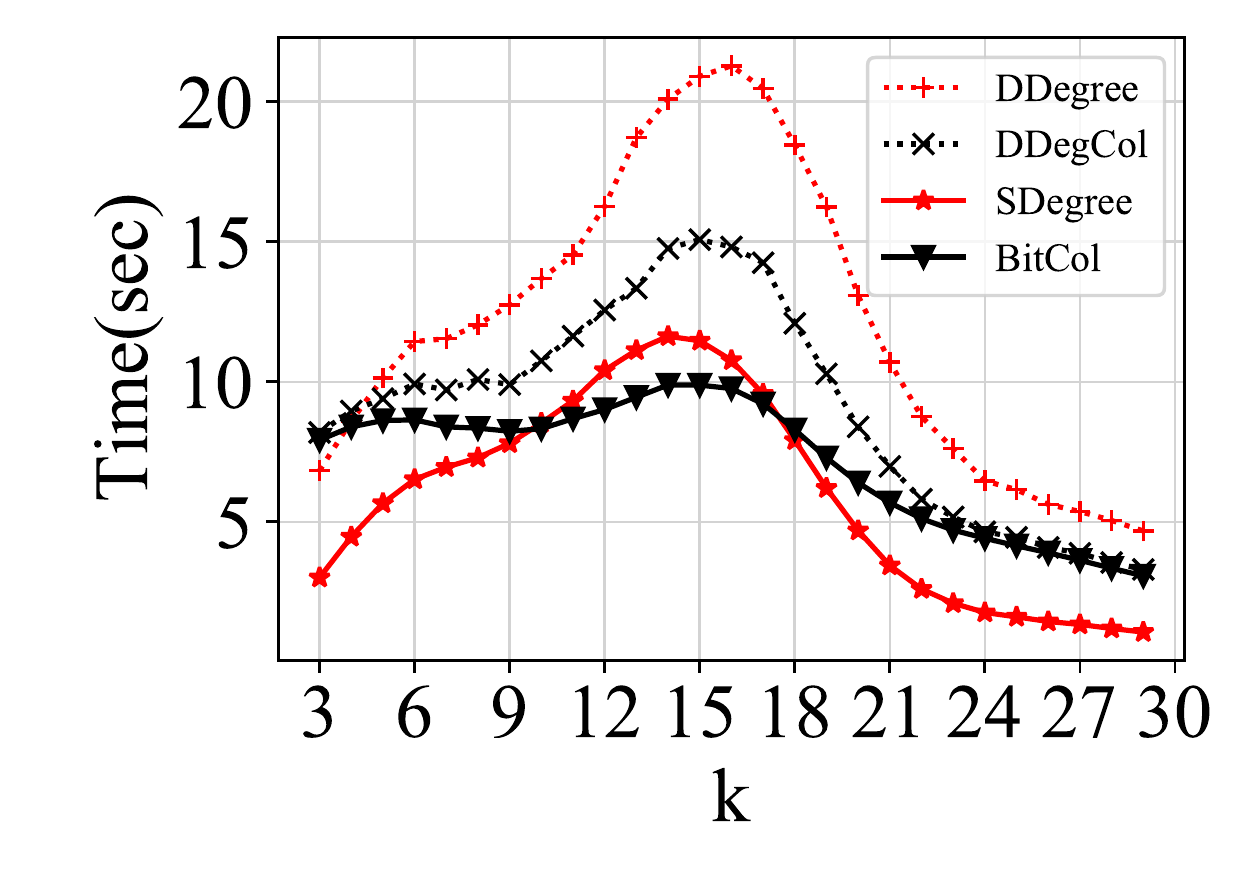}%
		\label{fig:Pokec_Effect_of_k}
	}
	\subfigure[BerkStan]{
		\includegraphics[width=0.45\linewidth]{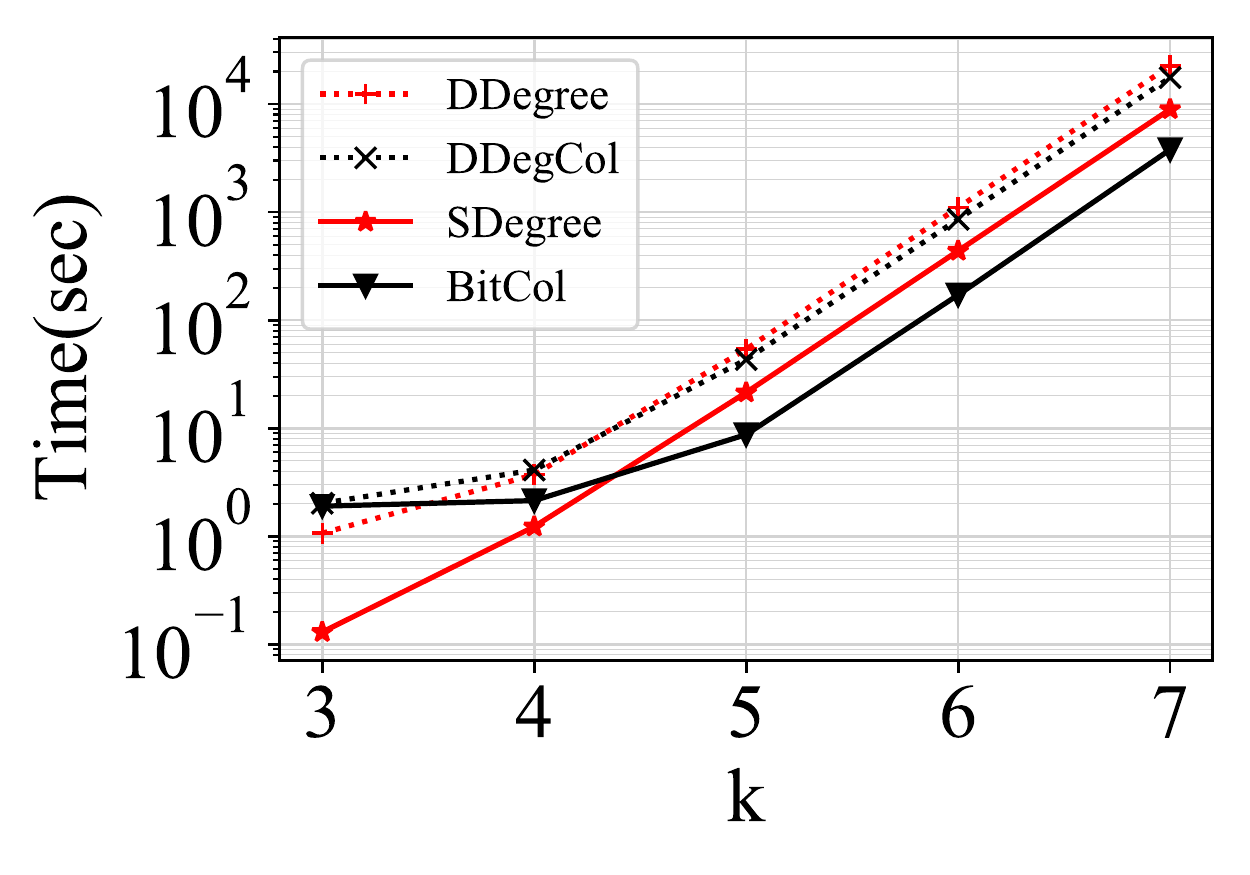}%
		\label{fig:Patents_Effect_of_k}
	}
	\vspace{-2mm}
	\caption{Effect of $k$ after our preprocessing techniques}
	\label{fig:Effect_of_k}
\end{figure}

\begin{figure}[htp]
	\subfigure[Preprocessing (Pokec)]{
		\includegraphics[width=0.45\linewidth]{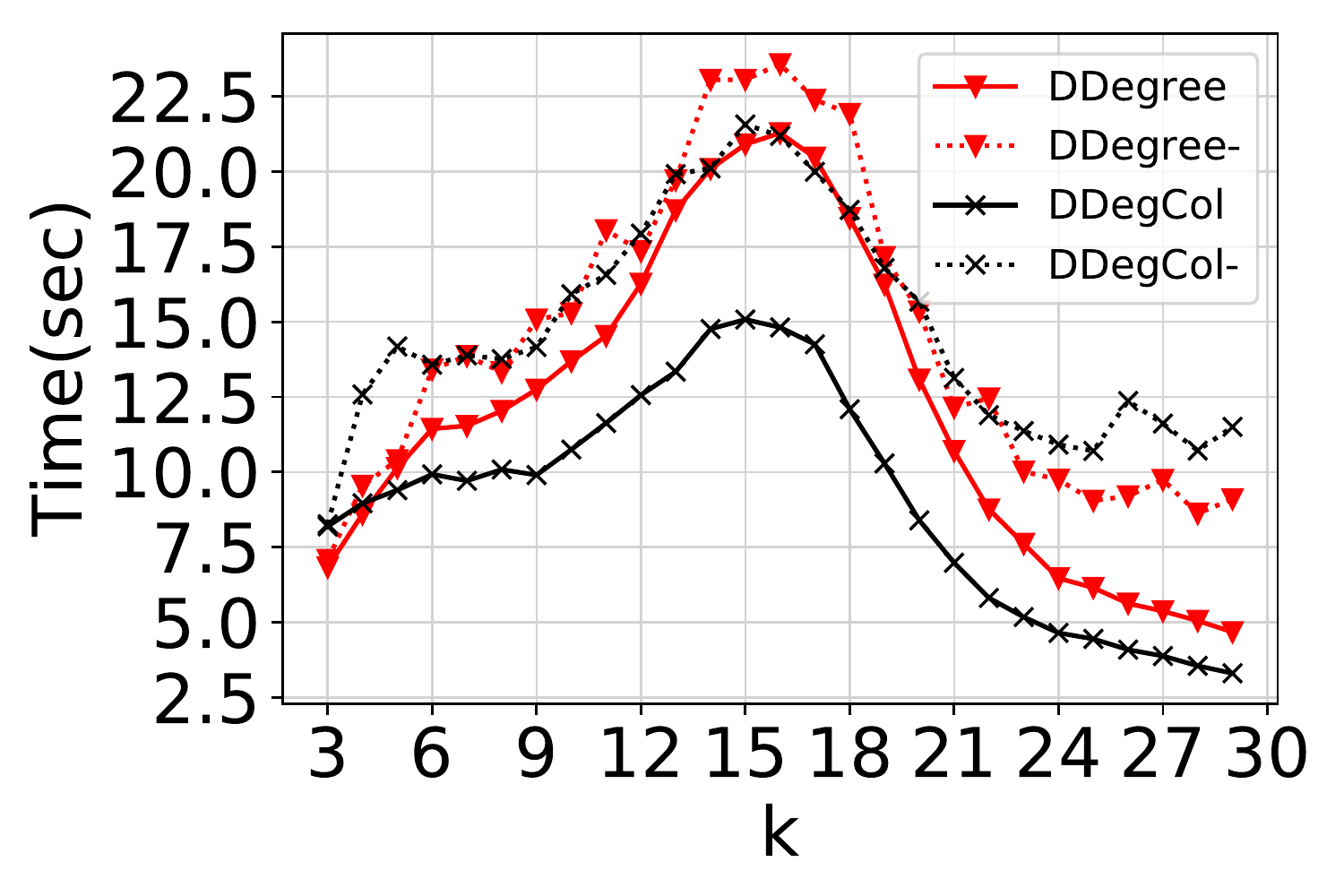}
		\label{fig:pre_pokec}
	}
	\subfigure[Pruning (Pokec)]{
		\includegraphics[width=0.45\linewidth]{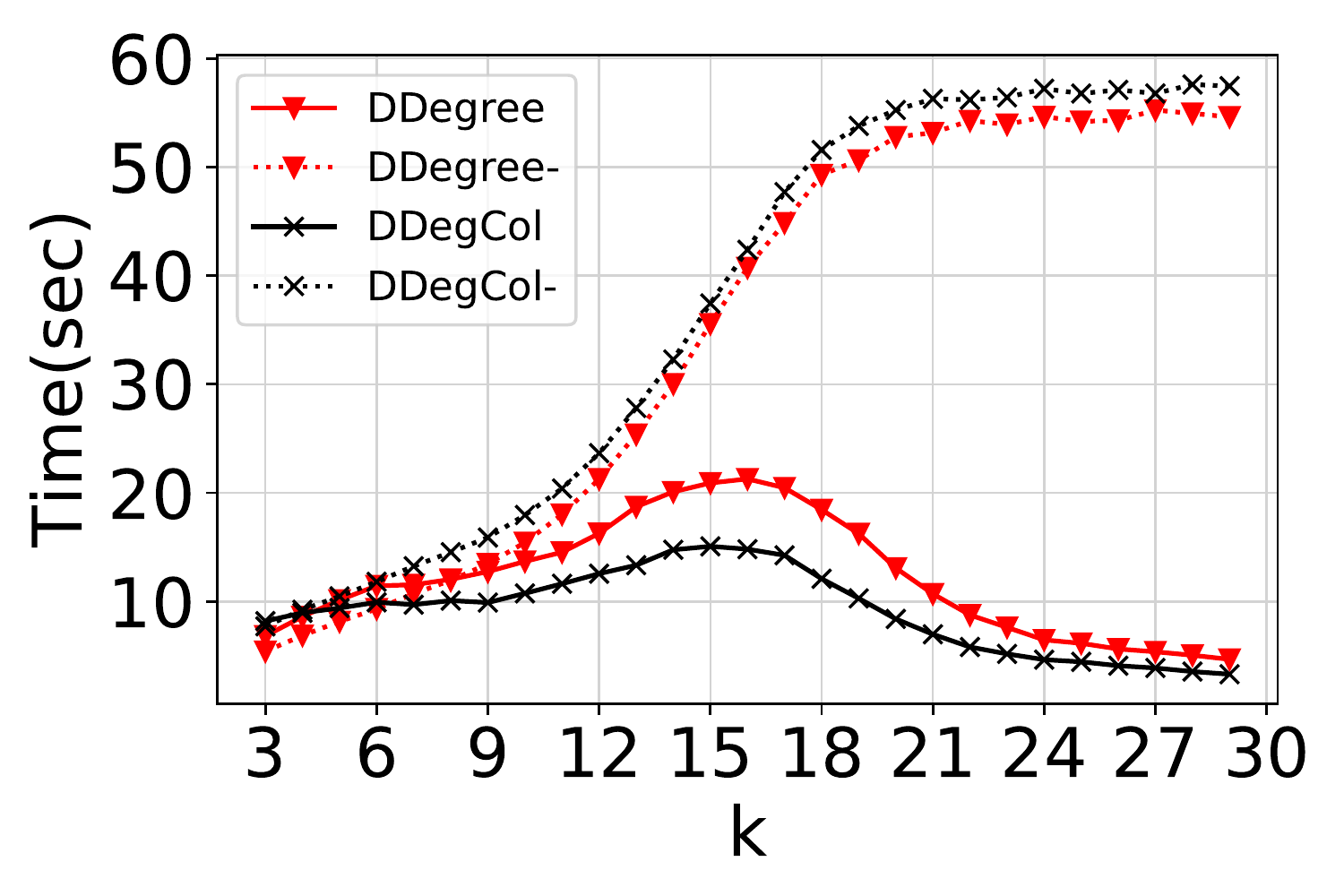}%
		\label{fig:prune_pokec}
	}
	\vspace{-2mm}
	\caption{Comparison of preprocessing and pruning techniques on Pokec. \textit{DDegree-} means \textit{DDegree} without our preprocessing ($a$) or pruning techniques ($b$), so does \textit{DDegCol-}.}
	\label{fig:no_pre}
\end{figure}

\noindent 
\textbf{Effect of $k$.}
\textit{DDegree} and \textit{DDegCol} can also exploit \textit{Pre-Core} and \textit{Pre-List}. Fig.\ref{fig:Effect_of_k} illustrates \textit{SDegree}, \textit{BitCol}, \textit{DDegree}, and \textit{DDegCol} after our proposed preprocessing techniques. For DBLP, ClueWeb09, and BerkStan, it is in line with expectations that the listing time grows exponentially w.r.t $k$ since the number of $k$-cliques is exponential in the size $k$. However, it does not hold for Pokec where the time grows marginally or even decreases w.r.t $k$. An explanation would be that the maximum clique size $\omega$ is small, and most of the search space is pruned in the preprocessing stage. As shown in Fig.\ref{fig:pre_pokec}, both \textit{DDegree} and \textit{DDegCol} run faster after our preprocessing techniques on Pokec. What's more, the pruning effects of the color constraint and the size constraint become more and more obvious as $k$ increases, which is illustrated in Fig.\ref{fig:prune_pokec}. Similar results can be obtained on Pokec for \textit{SDegree} and \textit{BitCol}.

When $k$ is small, degree-based algorithms run faster than color-based algorithms. For example, \textit{DDegree} runs in $88$s and \textit{DDegCol} runs in $144$s for ClueWeb09 and $k=3$ in Fig.\ref{fig:ClueWeb09_Effect_of_k}. Similar results can be observed for \textit{SDegree} and \textit{BitCol}. One explanation could be that the running time is dominated by greedy coloring.

\noindent \textbf{Effect of Dataset.}
For a given $k$, the $k$-clique listing time varies in datasets with different graph topologies. The maximum clique size $\omega$ is closely related to the running time, determining the lower bound of time overhead.
For example, the running time on DBLP is significantly higher than Pokec as $k$ increases for all the algorithms, even though the scale of DBLP is much smaller. This is because the maximum clique size of DBLP ($114$) is much larger than that of Pokec ($29$). 

In Pokec, the running time for both the degree-ordering based and the color-ordering based algorithms first increases as $k$ increases to around $\omega/2$, and then drops as $k$ increases to $\omega$. The reason could be in two ways. 
First, our preprocessing and pruning techniques can prune more invalid nodes as $k$ increases in advance. Furthermore, the pruning performance becomes more effective for a larger $k$. For the sparse graph Linkedin in Fig.\ref{fig:kclique-degree} and Fig.\ref{fig:kclique-color}, most of the search space is pruned after the preprocessing, and we obtain up to three magnitudes of acceleration.

\subsection{$k$-clique Listing Time in Parallel}
\begin{figure*}[!htbp]
	\vspace{-0.3cm}
	
	\centering
	\subfigure[AllWebUK02]{
		\includegraphics[width=0.23\linewidth]{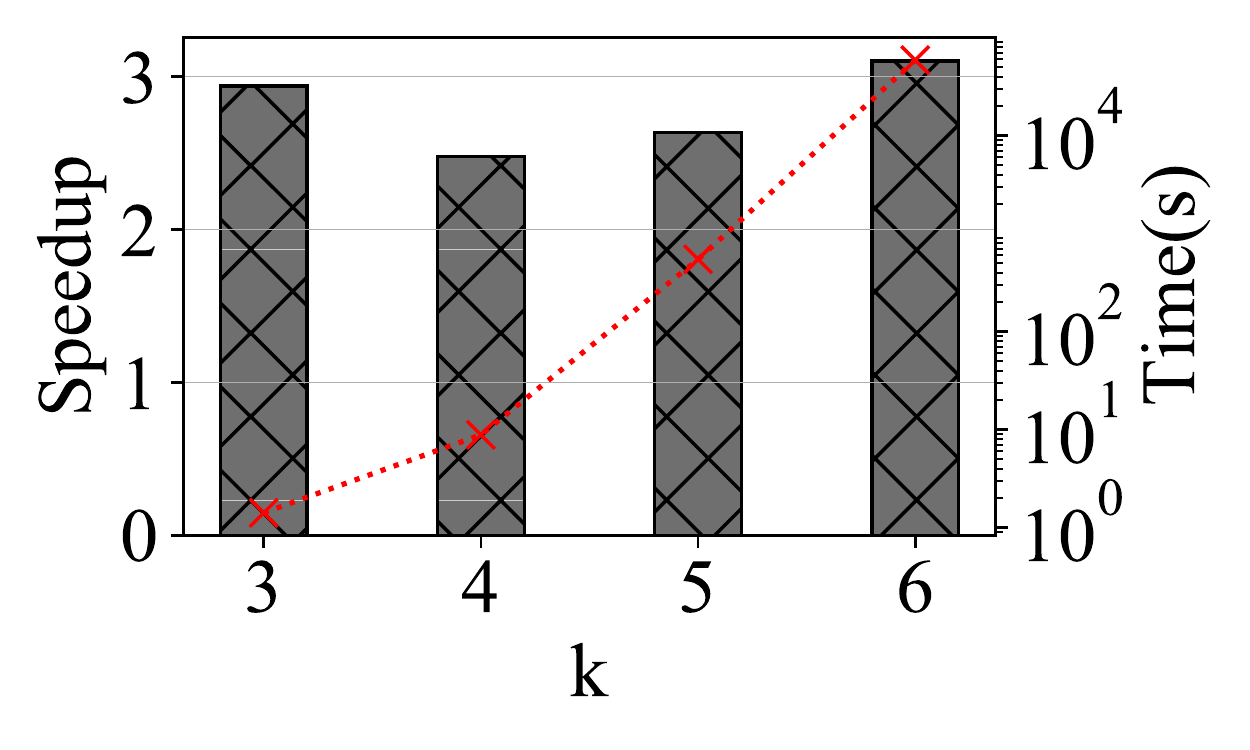}
		\label{fig:AllWebUK02_deg_32}
	}
	\subfigure[Stanford]{
		\includegraphics[width=0.23\linewidth]{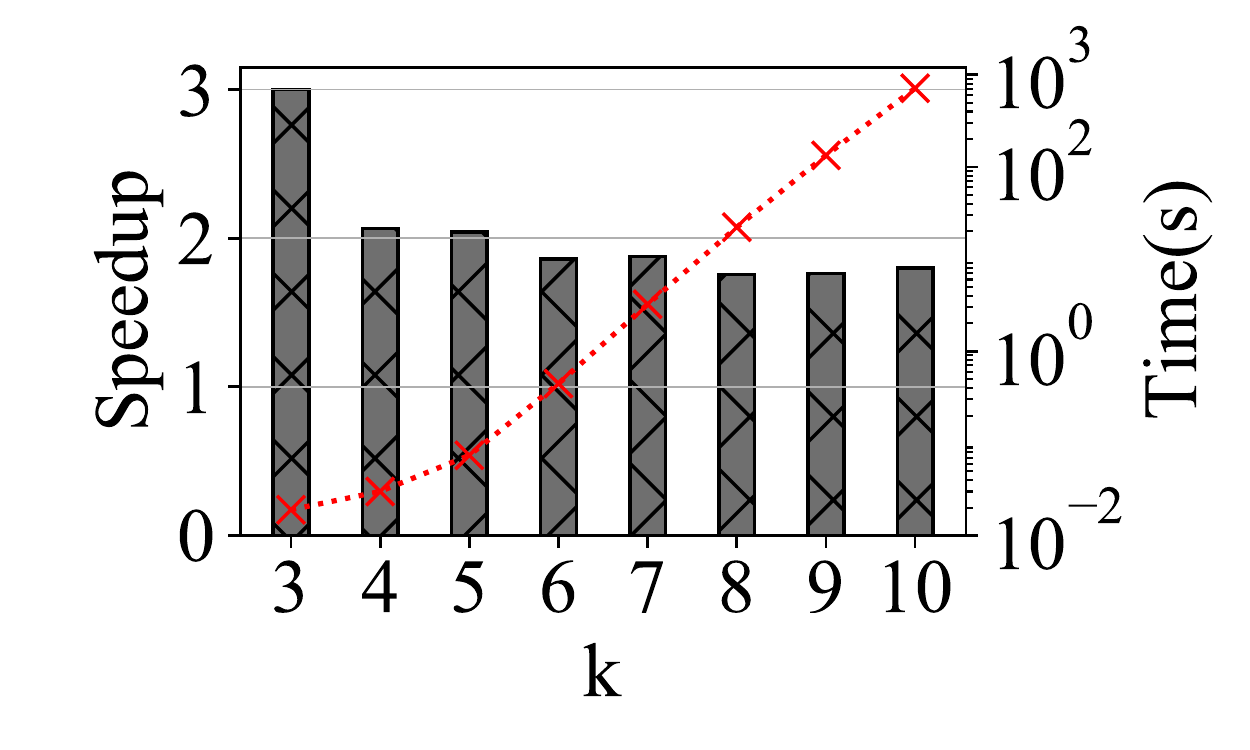}%
		\label{fig:Stanford_deg_32}
	}
	\subfigure[BerkStan]{
		\includegraphics[width=0.23\linewidth]{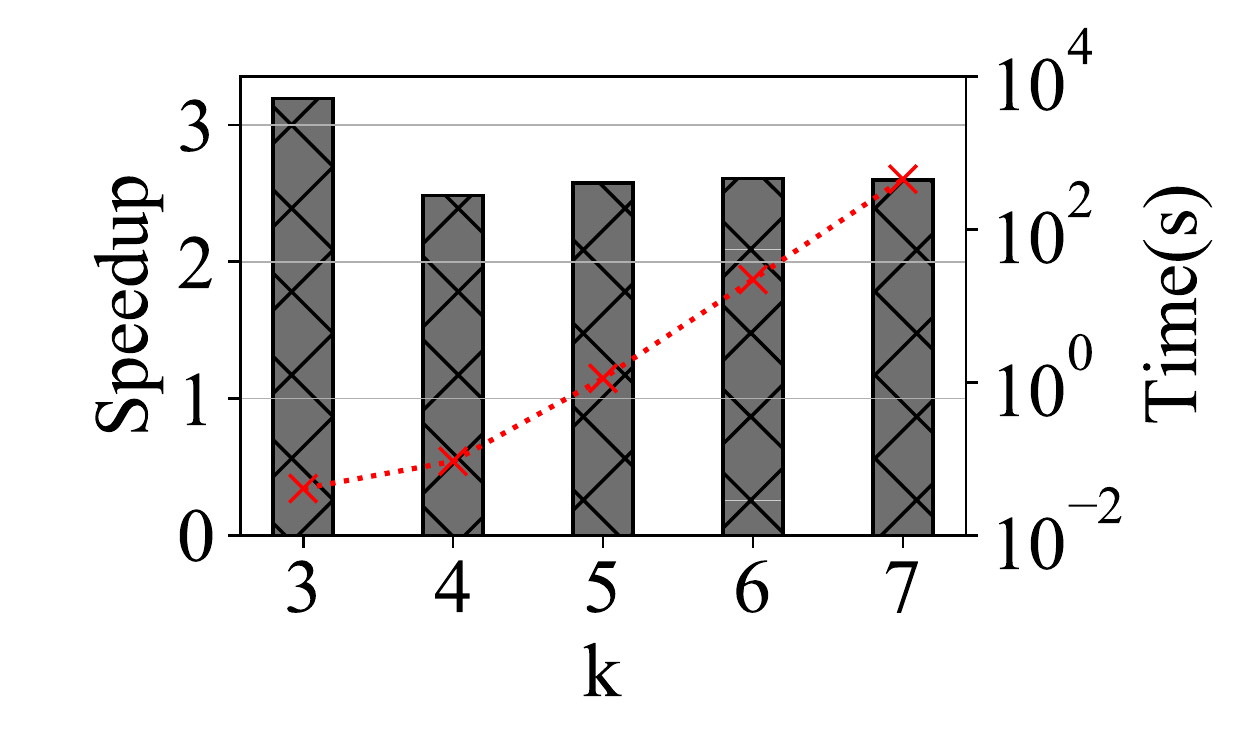}%
		\label{fig:BerkStan_deg_32}
	}
	\subfigure[DBLP]{
		\includegraphics[width=0.23\linewidth]{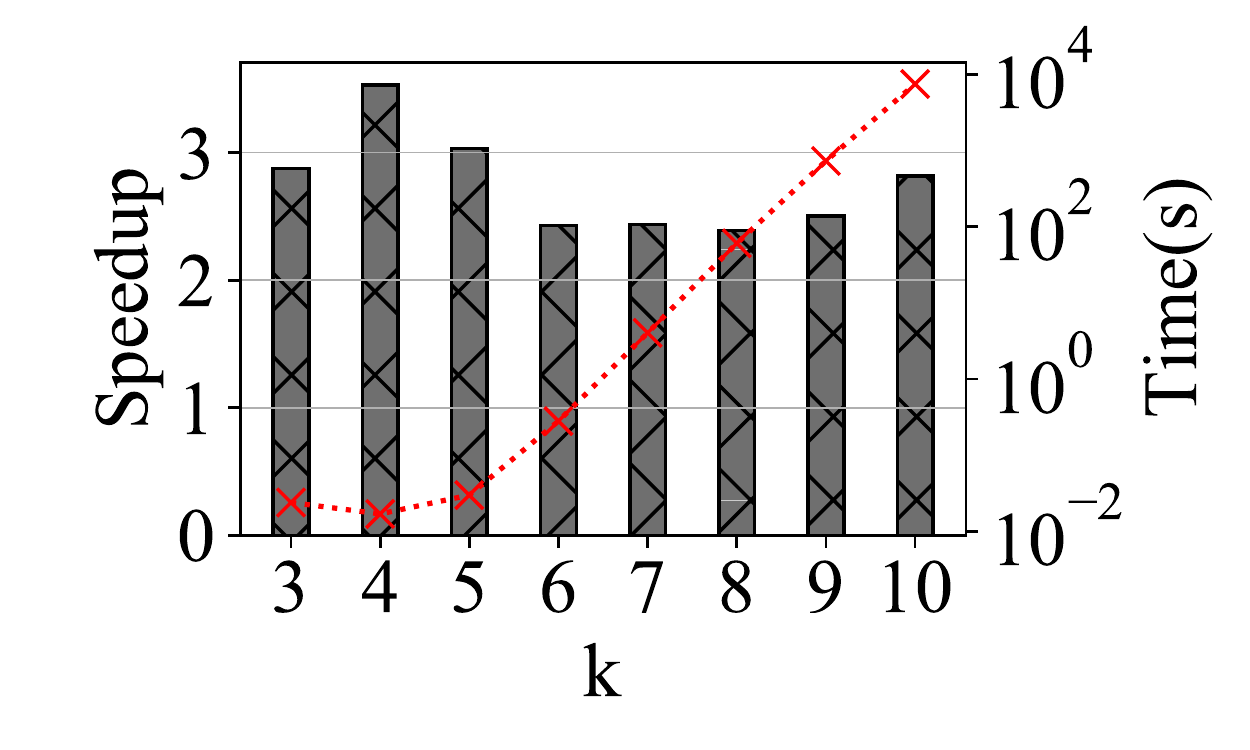}%
		\label{fig:DBLP_deg_32}
	}

	\vspace{-3mm}
	\caption{Running Time on Degree Ordering (NodeParallel). Red lines represent  running  time of  $\textit{SDegree}_{32}$; Histograms represent  speedups of $\textit{SDegree}_{32}$.}
	\label{fig:kclique-degree-32}
\end{figure*}

\begin{figure*}[!htbp]

	\centering
	\subfigure[AllWebUK02]{
		\includegraphics[width=0.23\linewidth]{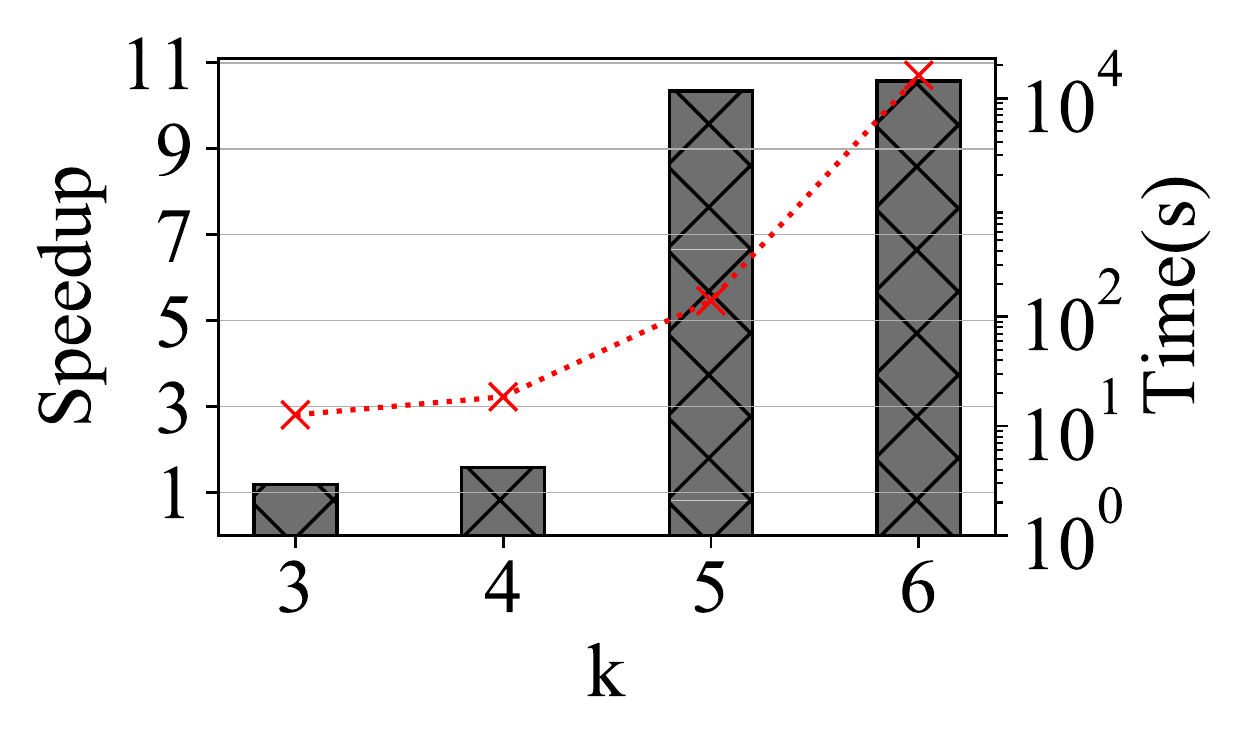}
		\label{fig:AllWebUK02_col_32}
	}
	\subfigure[Stanford]{
		\includegraphics[width=0.23\linewidth]{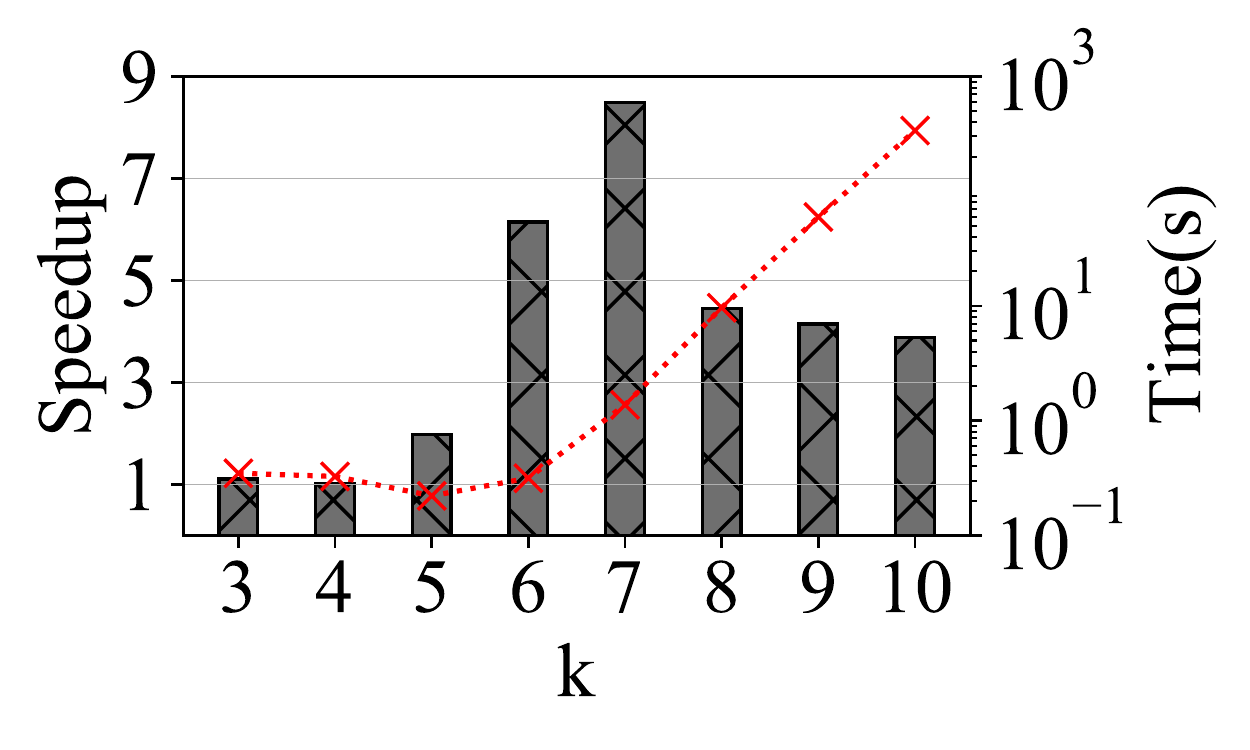}%
		\label{fig:Stanford_col_32}
	}
	\subfigure[BerkStan]{
		\includegraphics[width=0.23\linewidth]{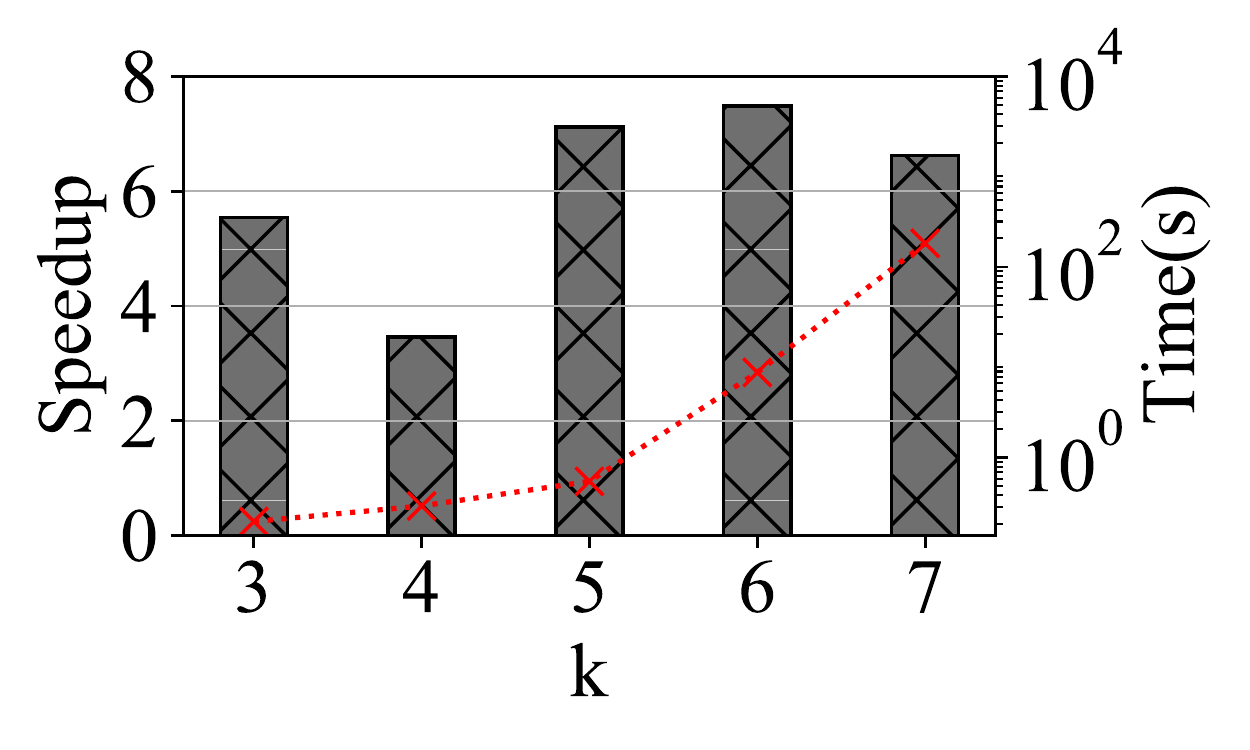}%
		\label{fig:BerkStan_col_32}
	}
	\subfigure[DBLP]{
		\includegraphics[width=0.23\linewidth]{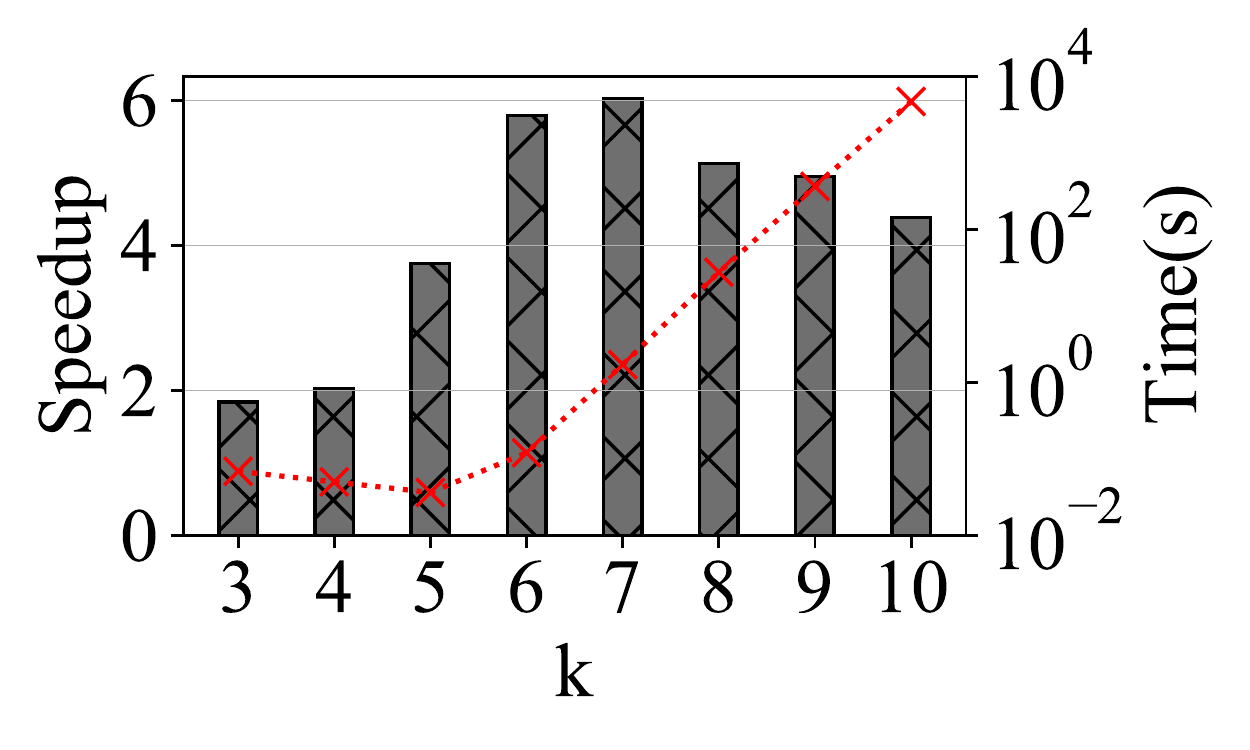}%
		\label{fig:DBLP_col_32}
	}

	\caption{Running Time on Color Ordering (NodeParallel). Red lines represent  running  time of  $\textit{BitCol}_{32}$; Histograms represent  speedups  of $\textit{BitCol}_{32}$.}
	\label{fig:kclique-color-32}
\end{figure*}

\begin{figure*}[!htbp]

	\centering
	\subfigure[BerkStan]{
		\includegraphics[width=0.23\linewidth]{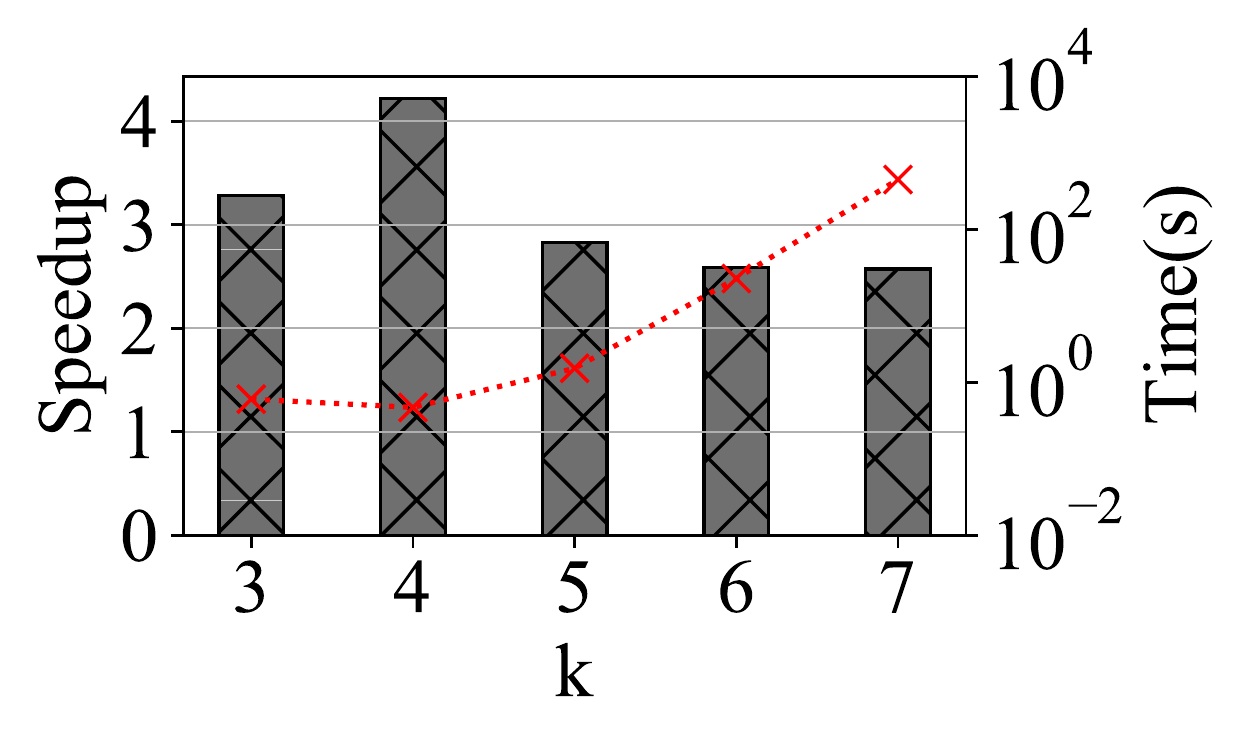}
		\label{fig:BerkStan_deg_32e}
	}
	\subfigure[Stanford]{
		\includegraphics[width=0.23\linewidth]{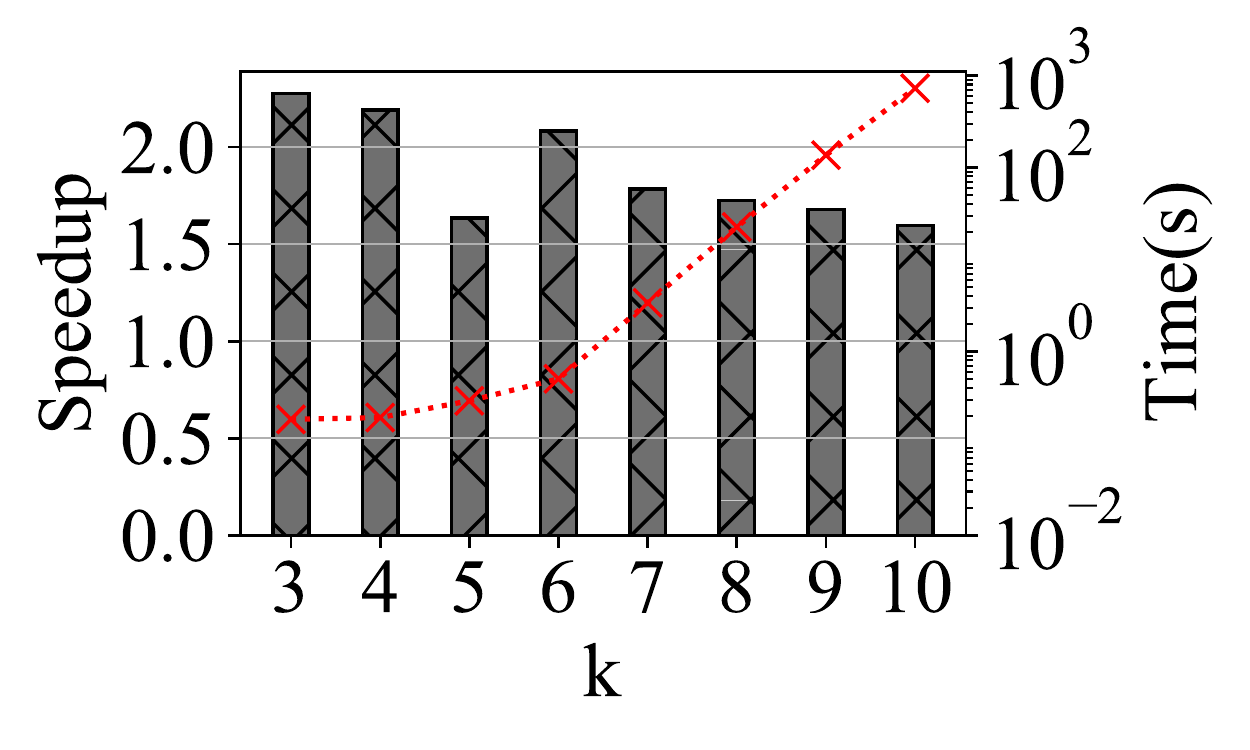}%
		\label{fig:Stanford_deg_32e}
	}
	\subfigure[Pokec]{
		\includegraphics[width=0.23\linewidth]{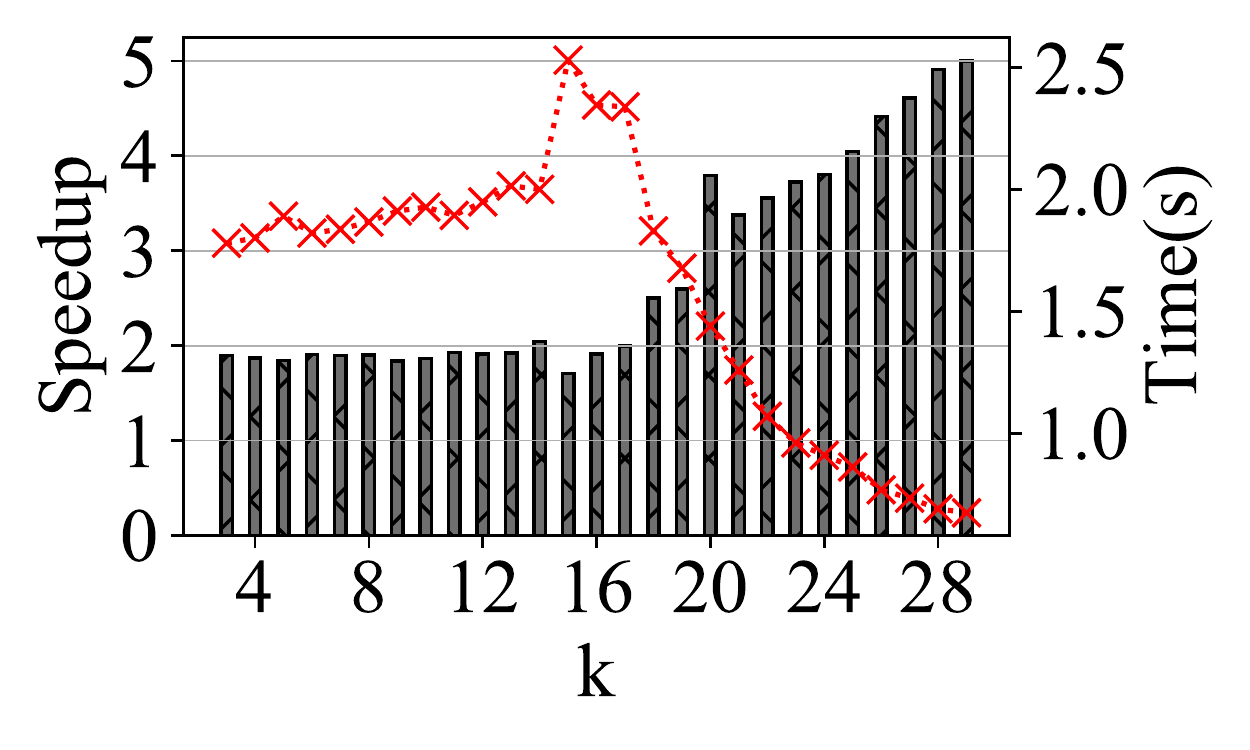}%
		\label{fig:Pokec_deg_32e}
	}
	\subfigure[DBLP]{
		\includegraphics[width=0.23\linewidth]{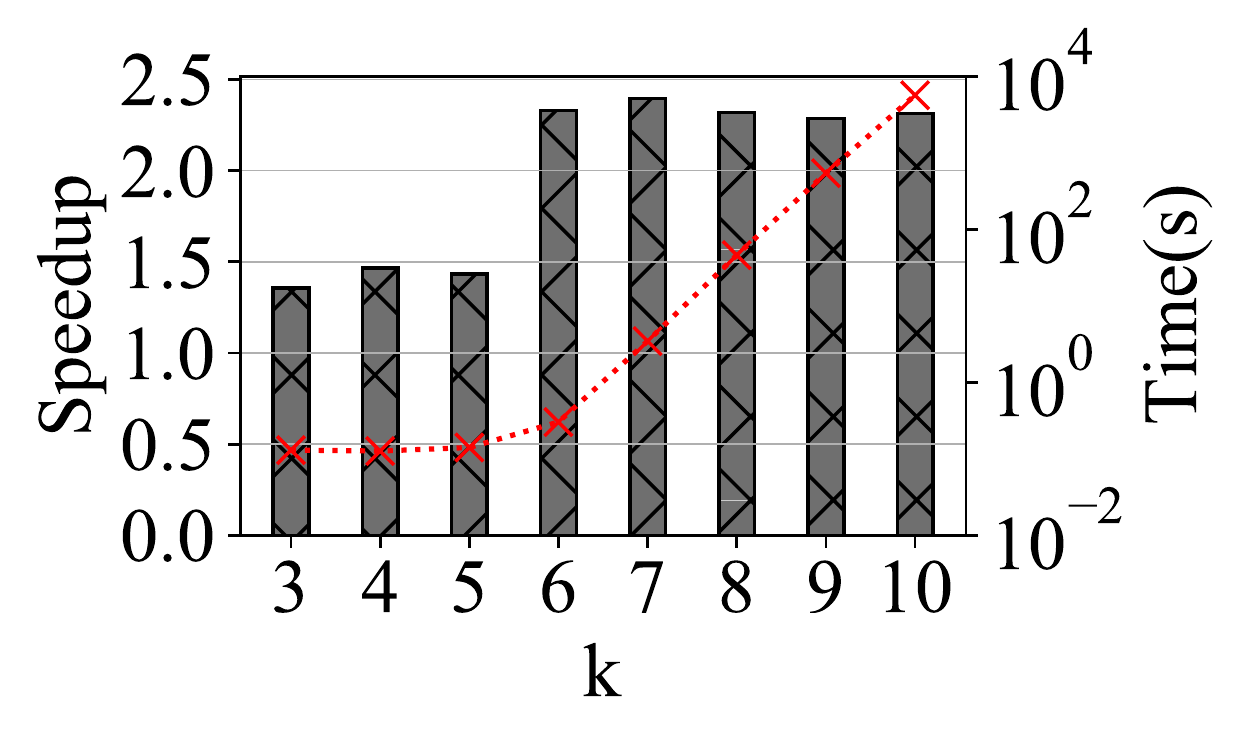}%
		\label{fig:DBLP_deg_32e}
	}
	\caption{Running Time on Degree Ordering (EdgeParallel). Red lines represent  running  time of  $\textit{SDegree}_{32}$; Histograms represent  speedups  of $\textit{SDegree}_{32}$.}
	\label{fig:kclique-degree-32e}
\end{figure*}

\begin{figure*}[!htbp]
	
	\centering
	\subfigure[BerkStan]{
		\includegraphics[width=0.23\linewidth]{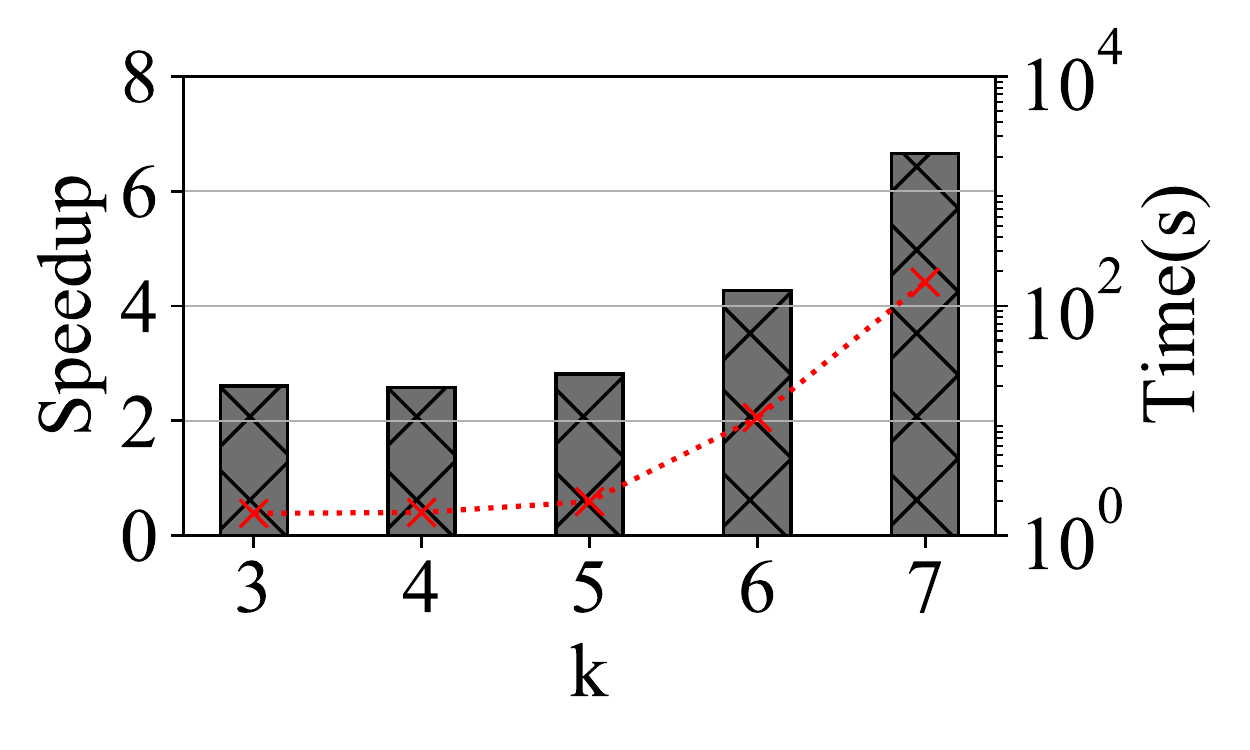}
		\label{fig:BerkStan_col_32e}
	}
	\subfigure[Stanford]{
		\includegraphics[width=0.23\linewidth]{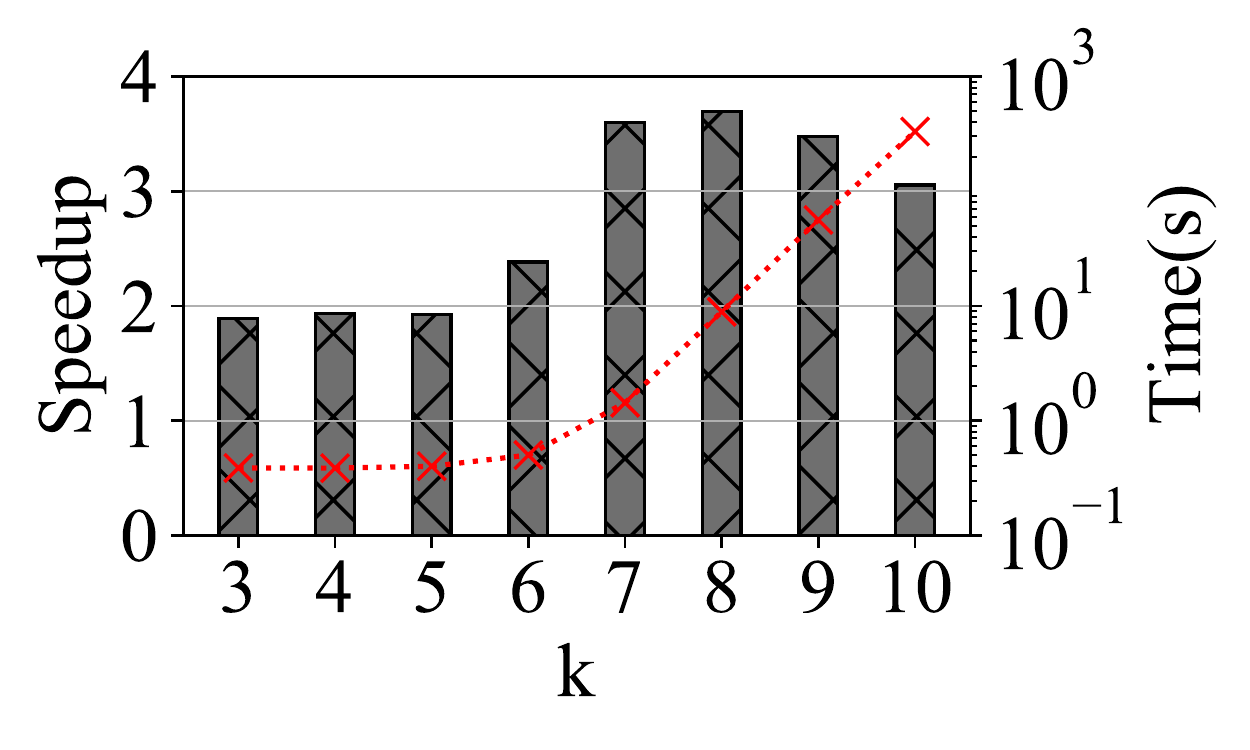}%
		\label{fig:Stanford_col_32e}
	}
	\subfigure[Pokec]{
		\includegraphics[width=0.23\linewidth]{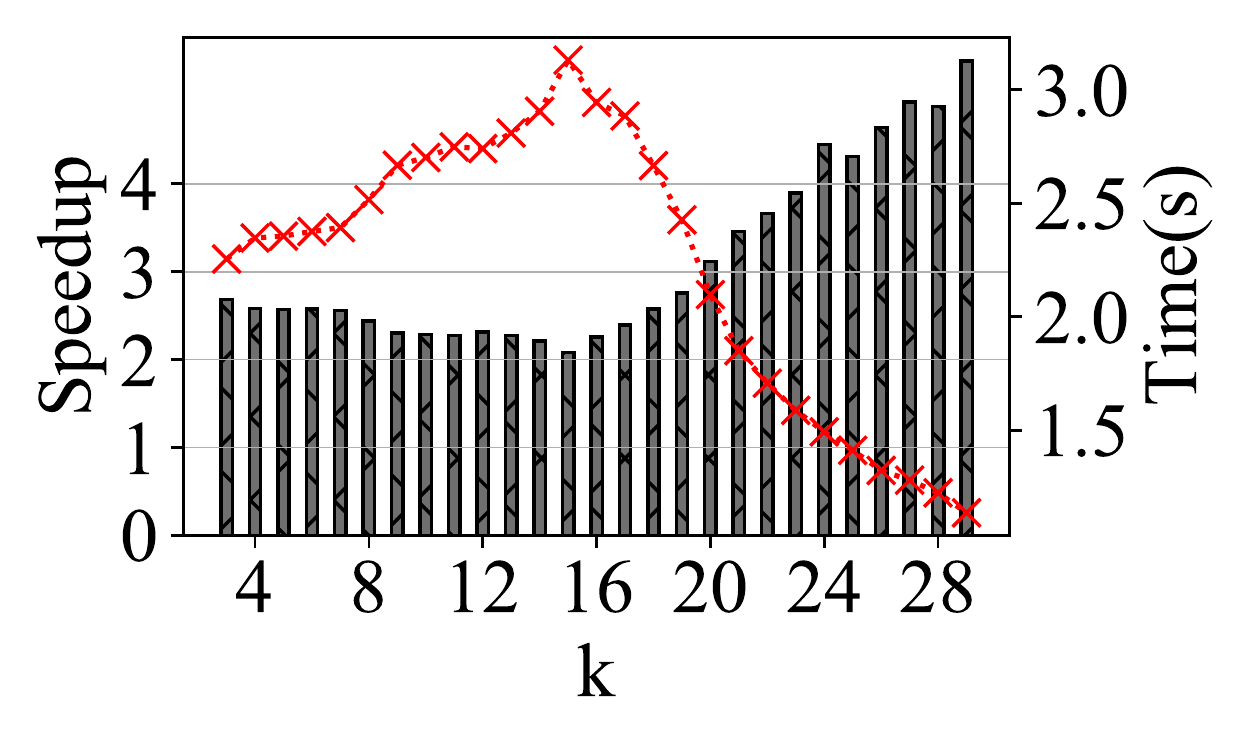}%
		\label{fig:Pokec_col_32e}
	}
	\subfigure[DBLP]{
		\includegraphics[width=0.23\linewidth]{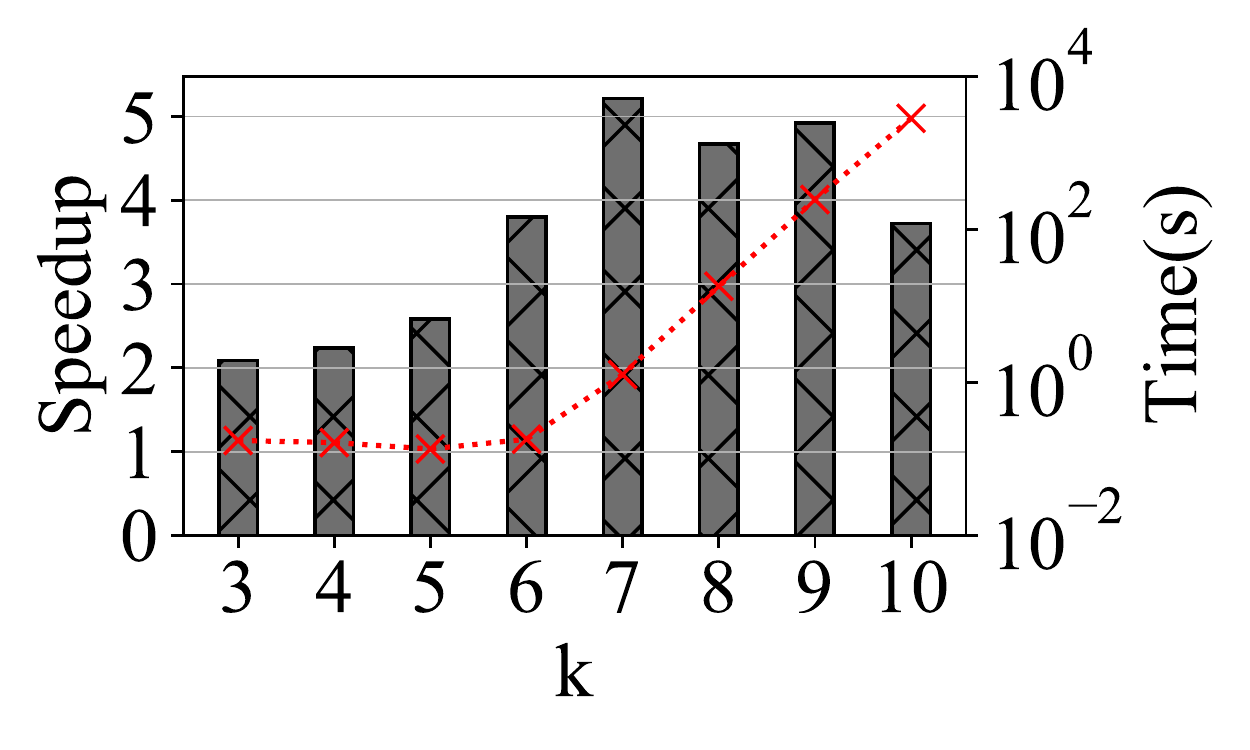}%
		\label{fig:DBLP_col_32e}
	}
	\caption{Running Time on Color Ordering (EdgeParallel). Red lines represent  running  time of  $\textit{BitCol}_{32}$; Histograms represent  speedups  of $\textit{BitCol}_{32}$.}
	\label{fig:kclique-color-32e}
\end{figure*}

In this section, we conduct several experiments to evaluate the performance of \textit{SDegree} and \textit{BitCol} in the scenario of $32$ threads. We denote the parallel version of \textit{DDegree}, \textit{SDegree}, \textit{DDegCol}, and \textit{BitCol} as $\textit{DDegree}_{32}$, $\textit{SDegree}_{32}$, $\textit{DDegCol}_{32}$, and $\textit{BitCol}_{32}$.

\noindent \textbf{Exp-I: NodeParallel with varying $k$.}
In this subsection, we evaluate the performance of our algorithms with the strategy of \textit{NodeParallel}. More specifically, for each vertex $u\in G$, \textit{SDegree} processes each  candidate set $N^{+}_u$ in parallel, and each thread processes on each induced subgraph $G_u$ for \textit{BitCol}. The 
experimental results are illustrated in Fig.\ref{fig:kclique-degree-32} and Fig.\ref{fig:kclique-color-32}.

Similarly, the running time grows exponentially w.r.t $k$ for most graphs. In parallel, all the algorithms further accelerate  $k$-clique listing and are capable of listing larger cliques within the time limit. For example, we can list all the $6$-cliques in AllWebUK02 and $10$-cliques in DBLP, which is infeasible in serial. In general, $\textit{SDegree}_{32}$ outperforms $\textit{DDegree}_{32}$, and $\textit{BitCol}_{32}$  outperforms $\textit{DDegCol}_{32}$, respectively. In particular, $\textit{BitCol}_{32}$ achieves more speedup with multiple threads than \textit{BitCol}, which implies that \textit{BitCol} can make better use of parallelism. For example, $\textit{BitCol}_{32}$ achieves around $8$x speedup while \textit{BitCol} achieves around $5$x speedup, for BerkStan and $k=6$.

\begin{figure}[htp]
	\newskip\subfigtoppskip \subfigtopskip = -0.02cm
	\newskip\subfigcapskip \subfigcapskip = -0.1cm
	\begin{minipage}[b]{\columnwidth}
		\centering
		\includegraphics[width=0.5\columnwidth]{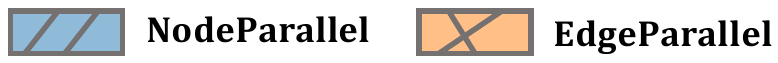}%
	\end{minipage}
	
	\subfigure[DBLP ($k=8$)]{
		\includegraphics[width=0.45\linewidth]{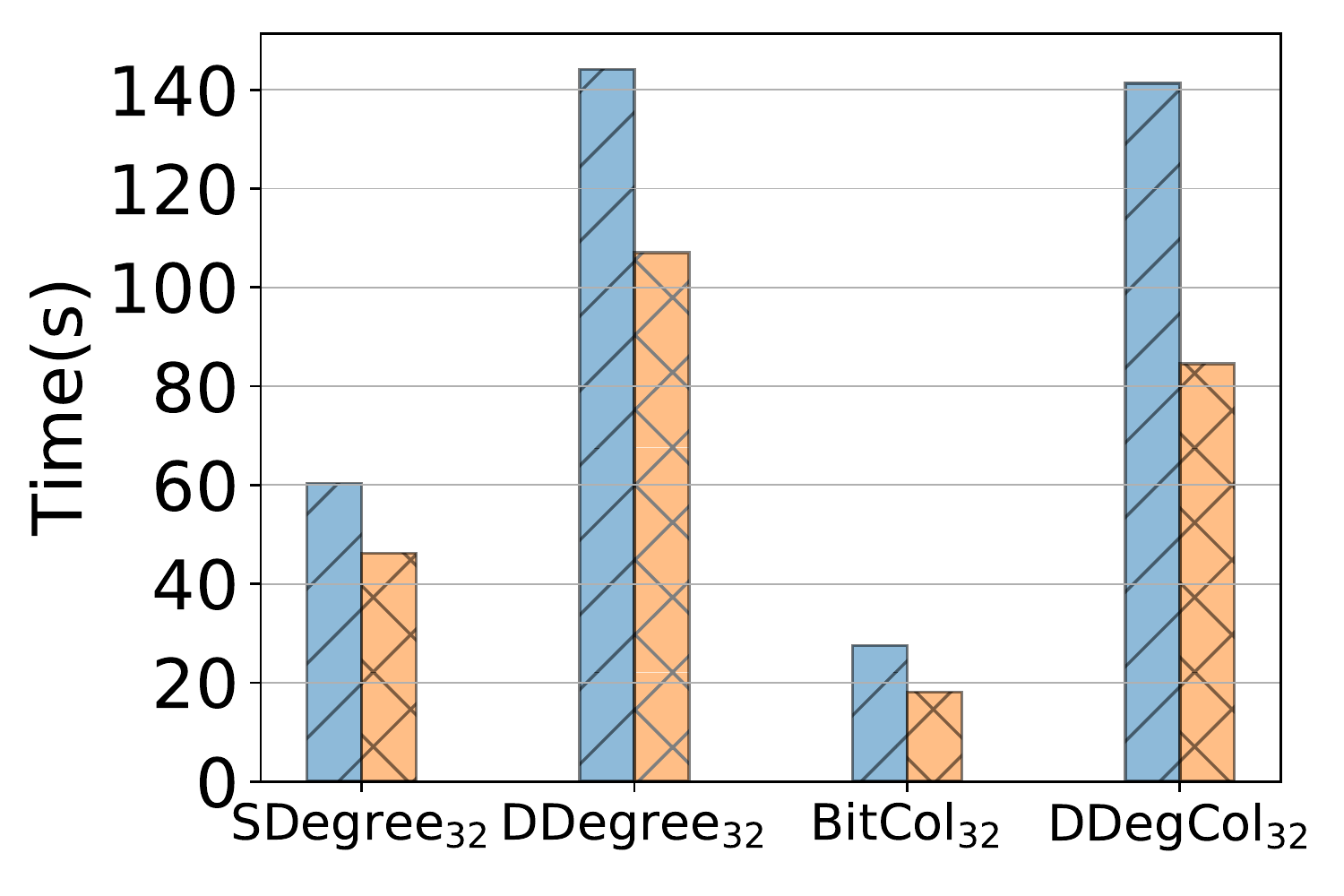}
		\label{fig:DBLP_EdgeParallel}
	}
	\subfigure[BerkStan ($k=7$)]{
		\includegraphics[width=0.45\linewidth]{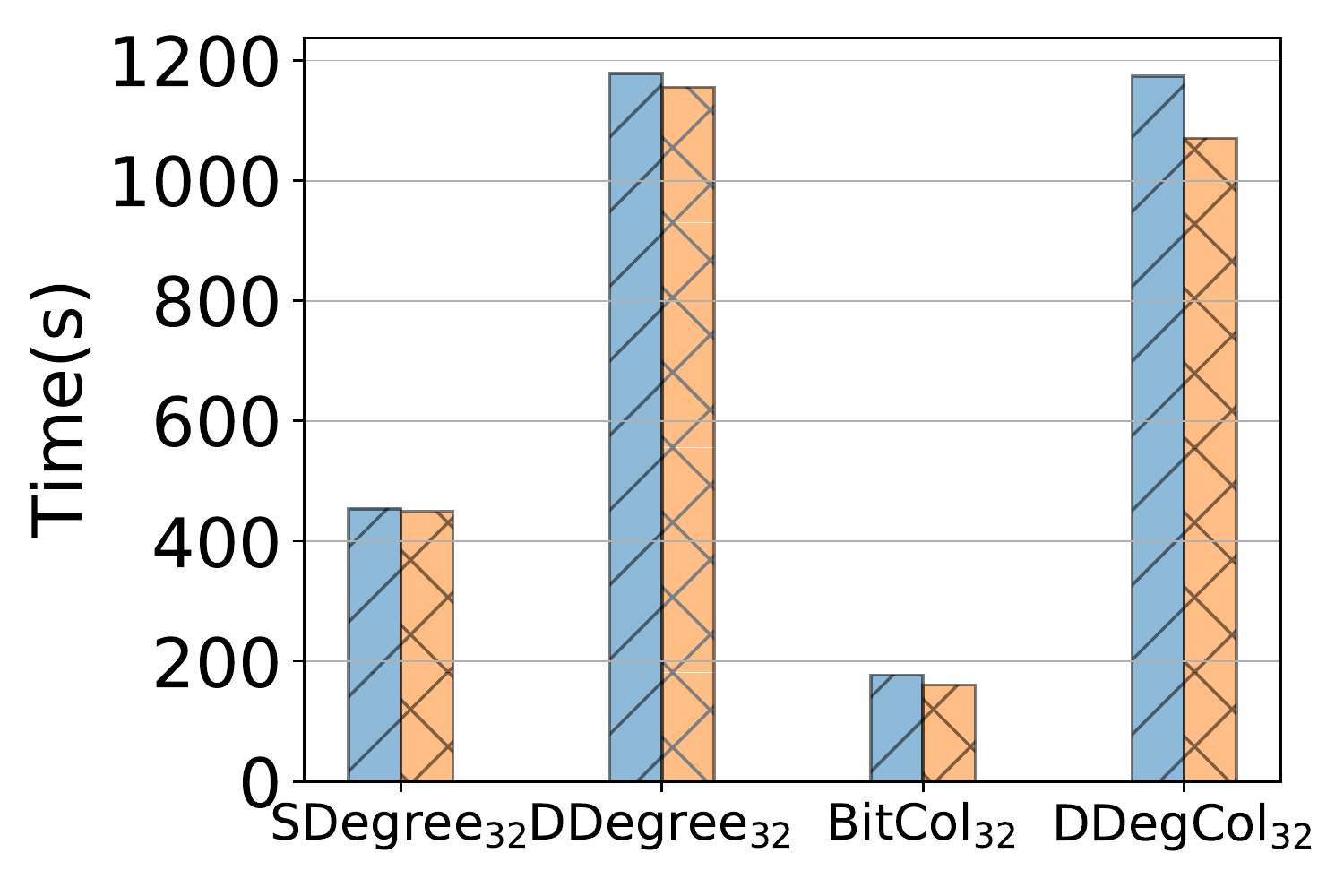}%
		\label{fig:BerkStan_EdgeParallel}
	}
	\vspace{-2mm}
	\caption{NodeParallel vs. EdgeParallel}
	\label{fig:EdgeParallel}
\end{figure}
\noindent \textbf{Exp-II: NodeParallel vs. EdgeParallel.} In this subsection, we evaluate the strategy of EdgeParallel~\cite{baseline2}. The basic idea is that each thread handles the intersection of out-neighbors of one edge's two endpoints. Compared to NodeParallel where the processing of each thread is based on the out-neighbors of one node, EdgeParallel processes ``smaller" out-neighbor sets. Therefore, EdgeParallel can achieve a higher degree of parallelism.

Fig.\ref{fig:EdgeParallel} shows the results on DBLP ($k=8$) and BerkStan ($k=7$). From Fig.\ref{fig:EdgeParallel}, the performance of NodeParallel is worse than that of EdgeParallel, indicating that EdgeParallel has a higher degree of parallelism.

As illustrated in Fig.\ref{fig:kclique-degree-32e} and Fig.\ref{fig:kclique-color-32e}, $\textit{SDegree}_{32}$ and $\textit{BitCol}_{32}$ still outperform $\textit{DDegree}_{32}$ and $\textit{DDegCol}_{32}$ with the strategy of EdgeParallel respectively, which also demonstrates the superiority of our algorithms. 

\subsection{Evaluation of Preprocessing Techniques}
\noindent \textbf{Exp-I: Preprocessing Time.}
In Table.\ref{tb:preprocess}, we evaluate the preprocessing time of  \textit{DDegree}, \textit{SDegree}, \textit{DDegCol}, and \textit{BitCol}. For all datasets, we average the preprocessing time with varying $k$ within the time limit. The preprocessing time of \textit{DDegree} and \textit{DDegCol} is mainly dependent on the reordering of the original graph $G$ by degeneracy ordering, with a complete core decomposition. Both \textit{SDegree} and \textit{BitCol} apply the preprocessing techniques of \textit{Pre-Core} and \textit{Pre-List}. 
Since \textit{BitCol} requires an additional reordering with degeneracy ordering, we merge \textit{Pre-Core} with the complete core decomposition for \textit{BitCol}.

In general, the preprocessing time of the four algorithms is almost the same, since all the techniques of preprocessing run in linear time. Meanwhile, we find that the preprocessing time of \textit{SDegree} is slightly lower. This is because \textit{SDegree} does not need to perform the complete core decomposition, where the \textit{Pre-Core} can stop earlier when each vertex has a degree of no less than $k-1$. The preprocessing time of \textit{BitCol} is a bit higher, since it exploits both the complete core decomposition and the \textit{Pre-List} preprocessing. However, the speedup of \textit{BitCol} can dominate the additional time consumption from preprocessing. 
For example, \textit{DDegCol} lists all the $7$-cliques in ClueWeb09 within $3,188$ seconds, and \textit{BitCol} is $1,943$ seconds faster, which is far more than the preprocessing time ($181$ seconds).

\noindent \textbf{Exp-II: Efficiency of Preprocessing.}
As demonstrated in Fig.\ref{fig:no_pre}, we evaluate the efficiency of our proposed preprocessing algorithms
\textit{Pre-Core}  and \textit{Pre-List} on Linkedin and Pokec. We compare the running time of the complete \textit{SDegree} and \textit{BitCol} algorithms with the no-preprocessing versions. It is shown that our preprocessing algorithms achieve around $1.5$x speedup for \textit{SDegree} and \textit{BitCol} in Pokec. Furthermore, we can achieve up to an order of magnitude acceleration for Linkedin. Our preprocessing algorithms improve the performance of
\textit{SDegree}  and \textit{BitCol} by removing invalid nodes that will not be contained in any $k$-clique.
\begin{table}
	\caption{Average Preprocessing time of all algorithms}
	\centering
	\begin{tabular}{c c c c c}
		\hline
		Dataset     & 
		SDegree  & 
		BitCol & 
		DDegCol & 
		DDegree 
		\\ \hline
		BerkStan    & \textbf{0.218s} & 0.283s & 0.269s & 0.267s 
		\\ 
		Pokec       & \textbf{1.347s} & 1.911s & 1.801s & 1.796s
		\\
		DBLP        & \textbf{0.070s} & 0.101s & 0.088s & 0.090s
		\\
		CitPatents  & \textbf{2.608s} & 3.556s & 3.152s & 3.154s 
		\\
		Linkedin    & \textbf{3.482s} & 4.698s & 4.676s & 4.659s 
		\\
		WebUK05     & \textbf{0.104s} & 0.152s & 0.143s & 0.146s
		\\
		ClueWeb09   & \textbf{138.713s} & 181.113s & 159.308s & 161.204s
		\\
		Wikipedia13 & \textbf{50.056s} & 67.318s & 65.447s & 64.361s
		\\
		AllWebUK02  & \textbf{8.742s}  & 14.063s & 13.363s & 13.401s
		\\
		Stanford    & \textbf{0.130s}  & 0.165s  & 0.157s  & 0.149s
		\\ \hline
	\end{tabular}
	\vspace{2mm}
	\label{tb:preprocess}
\end{table}

\begin{figure}[htp]
	\subfigure[Linkedin (SDegree)]{
		\includegraphics[width=0.46\linewidth]{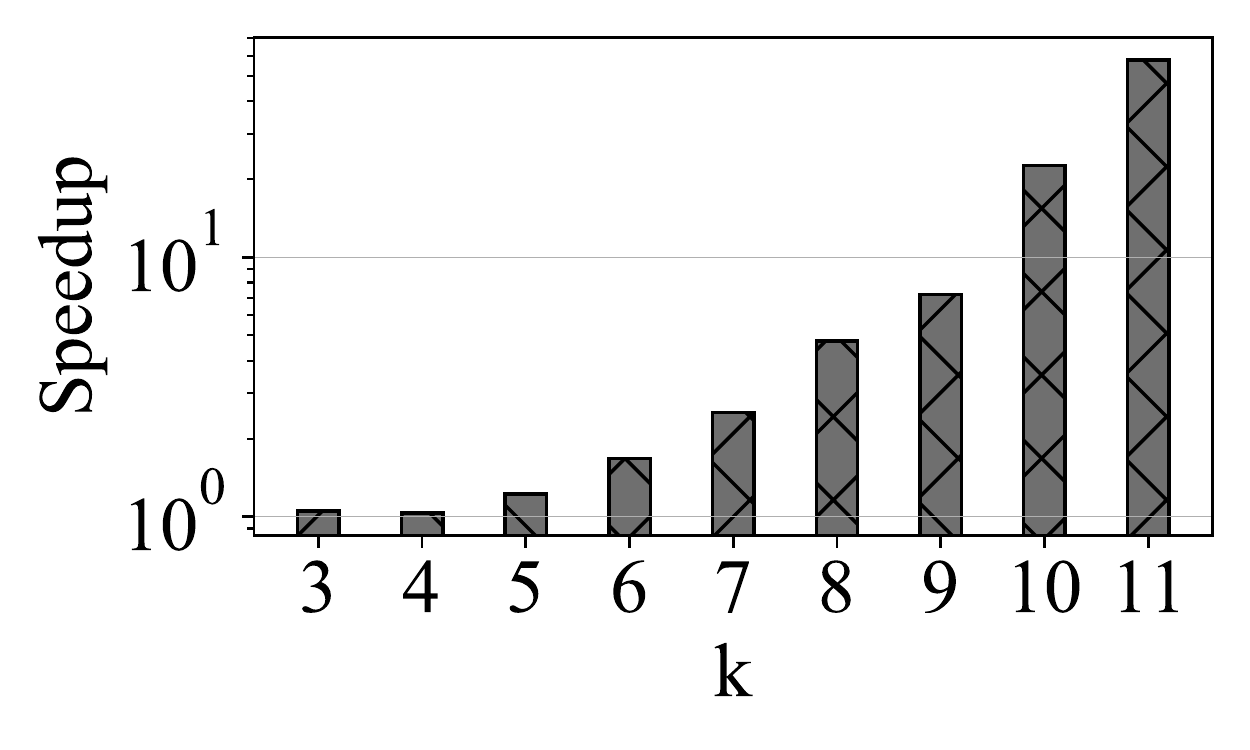}
		\label{fig:Linkedin_deg_nopre}
	}
	\subfigure[Pokec (SDegree)]{
		\includegraphics[width=0.46\linewidth]{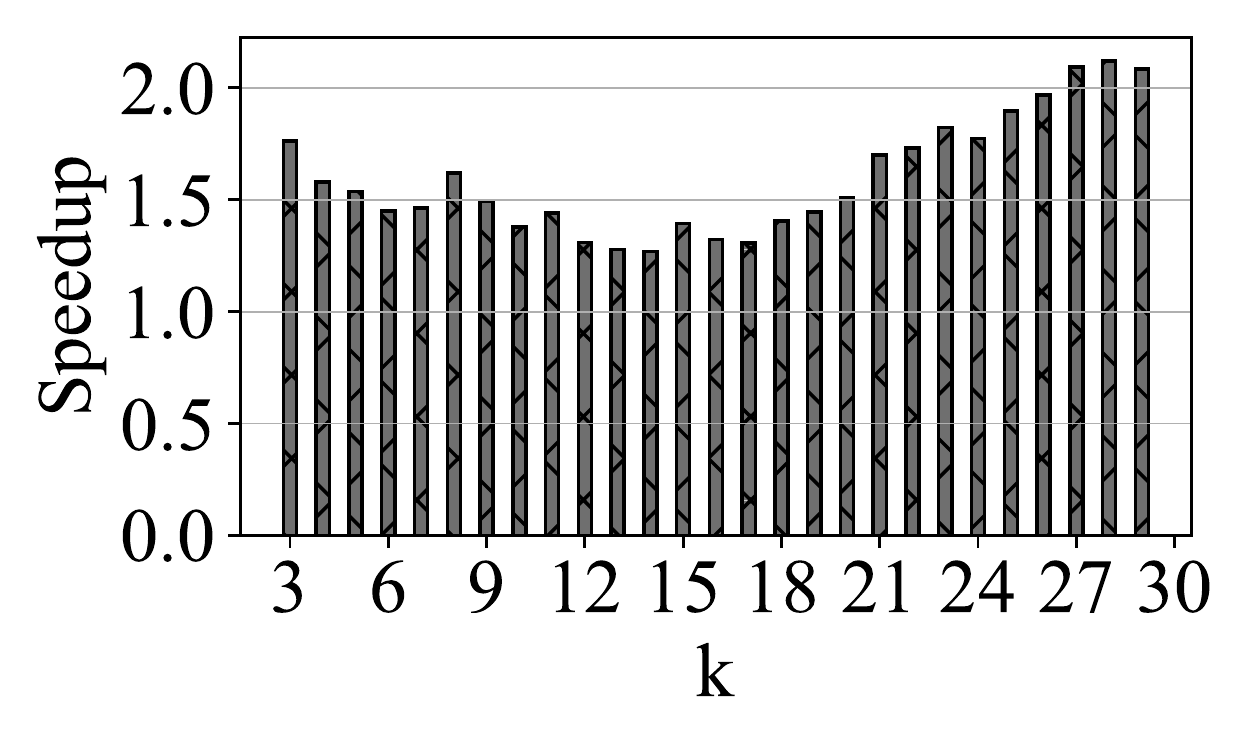}%
		\label{fig:Pokec_deg_nopre}
	}
	
	\subfigure[Linkedin (BitCol)]{
		\includegraphics[width=0.46\linewidth]{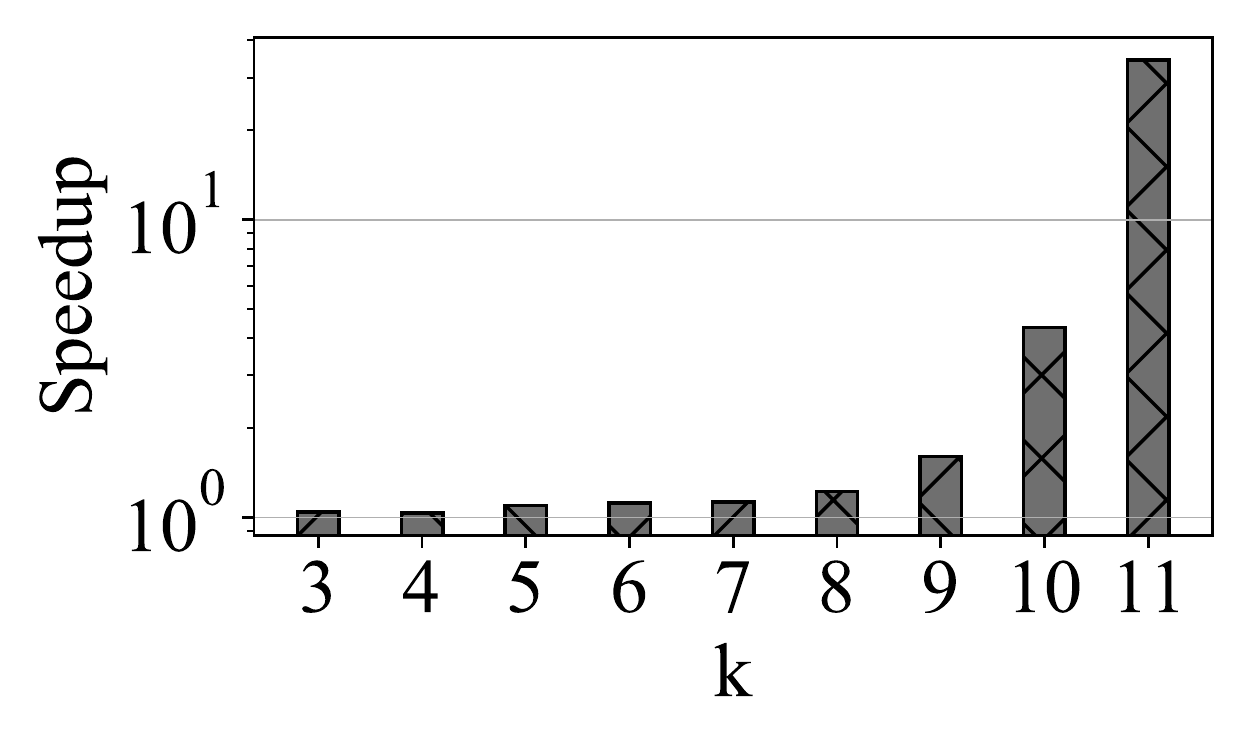}
		\label{fig:Linkedin_col_nopre}
	}
	\subfigure[Pokec (BitCol)]{
		\includegraphics[width=0.46\linewidth]{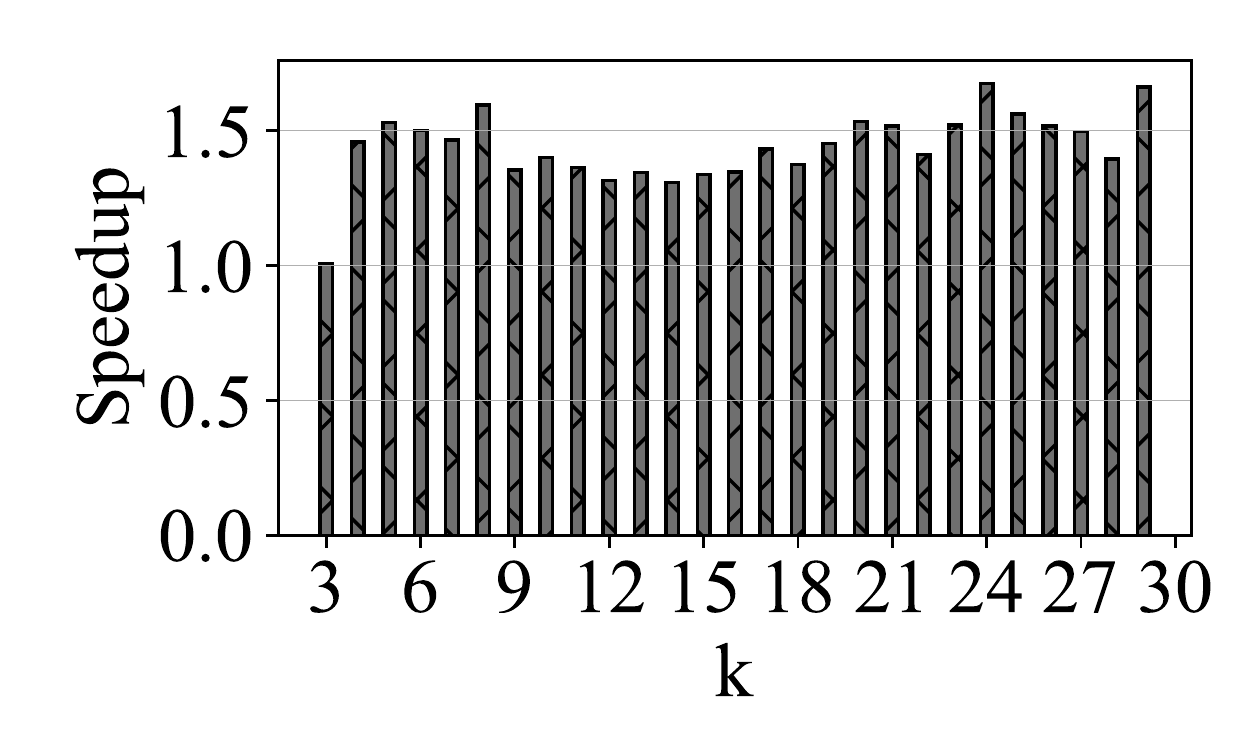}%
		\label{fig:Pokec_col_nopre}
	}
	\vspace{-2mm}
	\caption{The efficiency of preprocessing (No preprocessing = $1.0$)}
	\label{fig:no_pre}
\end{figure}

\subsection{Evaluation of Memory Consumption}
\begin{figure}[htp]
	\subfigure[Pokec ($k=18$)]{
		\includegraphics[width=0.45\linewidth]{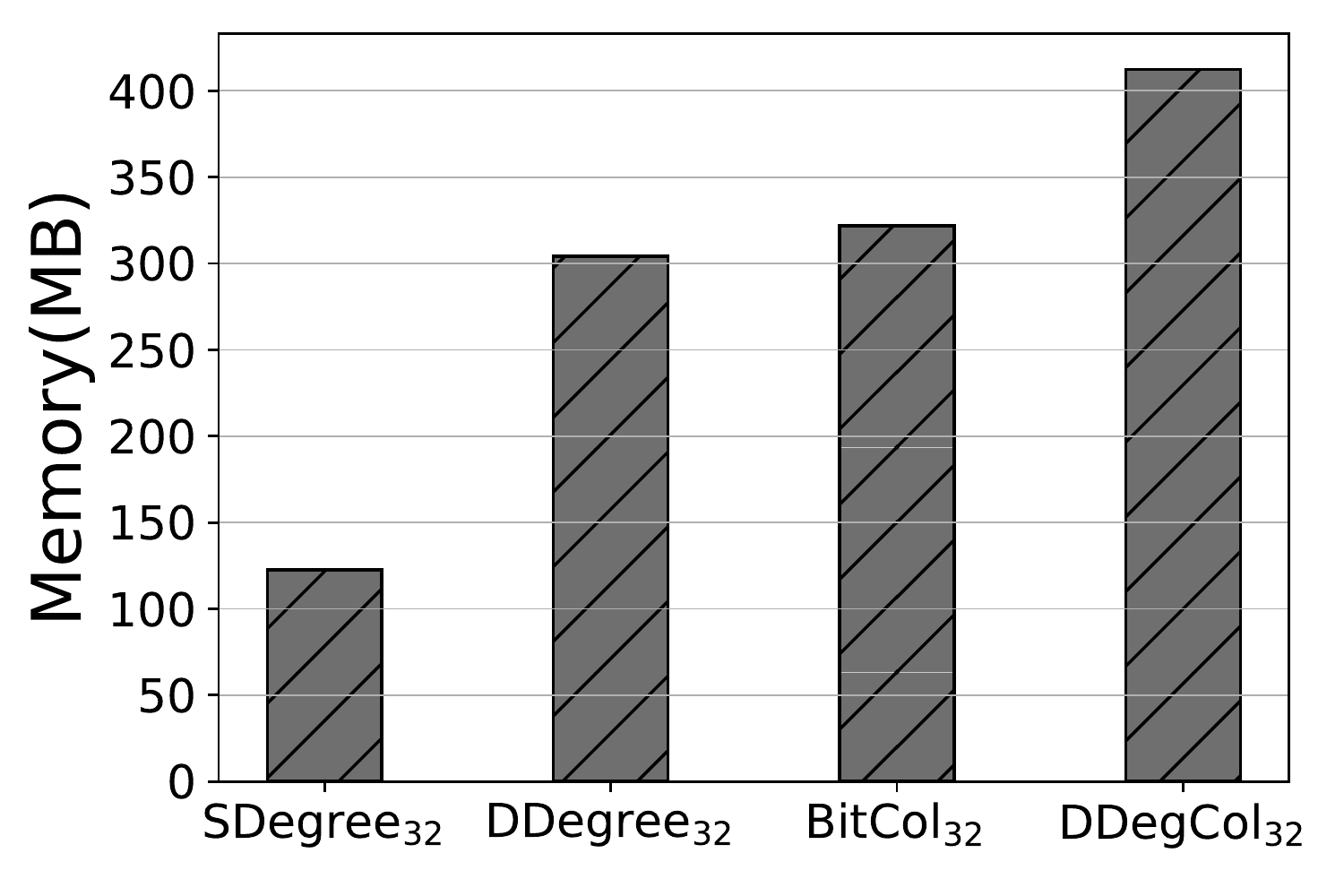}
		\label{fig:Pokec_Mem}
	}
	\subfigure[AllWebUK02 ($k=4$)]{
		\includegraphics[width=0.45\linewidth]{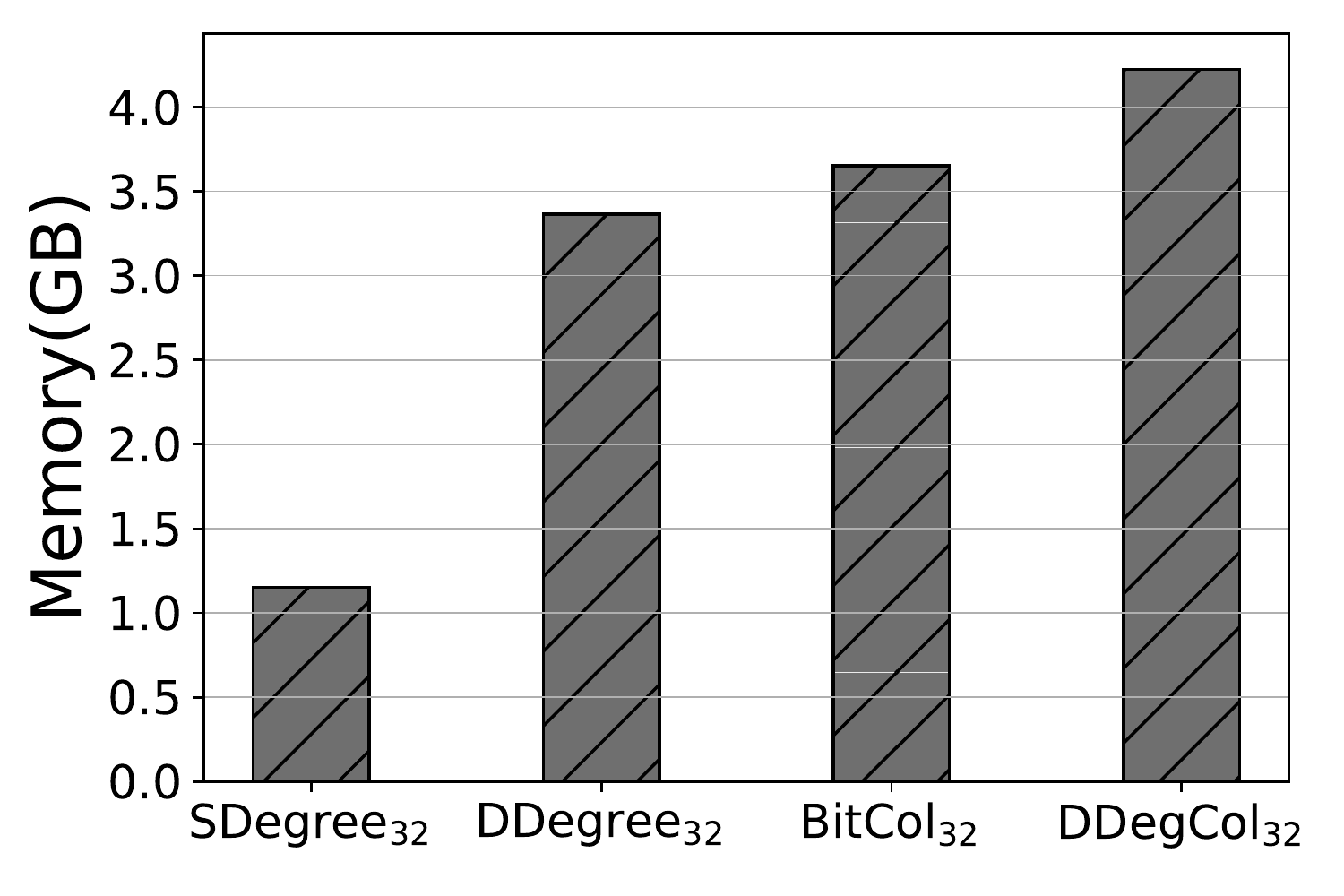}%
		\label{fig:AllWebUK02_Mem}
	}
	\vspace{-2mm}
	\caption{Memory Consumption with 32 threads (NodeParallel)}
	\label{fig:Mem}
\end{figure}
We evaluate the memory consumption on Pokec ($k=18$) and AllWebUK02 ($k=4$) with NodeParallel and $32$ threads, which is shown in Fig.\ref{fig:Mem}. We can see that the memory overhead of $\textit{SDegree}_{32}$ is minimal. As we analyzed in Section VIII, $\textit{SDegree}_{32}$ does not require extra space to construct induced subgraphs. Moreover, the memory overhead of $\textit{BitCol}_{32}$ is smaller than that of $\textit{DDegCol}_{32}$ as expected, due to the capability of bitmap vectors to compress the out-neighbor sets. However, the memory overhead of $\textit{BitCol}_{32}$ is a bit larger than that of $\textit{DDegree}_{32}$ in Fig.\ref{fig:Mem} since the color-based algorithms need to maintain additional information, such as color values.

\section{Conclusion}
In this paper, we proposed two algorithms \textit{SDegree} and \textit{BitCol} to efficiently solve the $k$-clique listing problem, based on degree ordering and color ordering, respectively. We mainly focused on accelerating the set intersection part, thereby accelerating the entire process of $k$-clique listing. Both \textit{SDegree} and \textit{BitCol} exploit the data level parallelism, which is non-trivial for the state-of-the-art algorithms.

First, two preprocessing techniques \textit{Pre-Core} and \textit{Pre-List} are developed to efficiently prune the invalid nodes that will not be contained in a $k$-clique. \textit{SDegree} is a simple but effective framework based on merge join while \textit{BitCol} improves \textit{SDegree} with bitmaps and color ordering. Our algorithms have comparable time complexity and a slightly better space complexity, compared with the state-of-the-art algorithms. We concluded from the experimental results that our algorithms outperform the state-of-the-art algorithms by $3.75$x for degree ordering and by $5.67x$ for color ordering on average.

Since a $k$-clique can be obtained by extending a vertex adjacent to all $k-1$ nodes in a $(k-1)$-clique, the existing algorithms are all based on the recursive framework to expand from a node to a $k$-clique. The state-of-the-art algorithms propose to exploit the ordering heuristics to prune invalid search space, while in this paper, we mainly focus on accelerating set intersections, which is a frequent operation in the recursive framework.

One inherent limitation for all the existing algorithms is that, when the size of the maximum clique ($\omega$) is large and $k$ is close to $\omega/2$, the problem of $k$-clique listing is often deemed infeasible. The existing algorithms take a significant amount of time to enumerate the $k$-cliques contained in the maximum cliques. Further studies can be conducted that whether we can enumerate the $k$-cliques within and outside the maximum (or near maximum) cliques separately.

\label{sec:conclusion}

\balance

\bgroup\small
\bibliographystyle{ieeetr}
\let\xxx=\bibitem\def\bibitem{\par\vspace{1mm}\xxx}
\bibliography{add}
\egroup

\end{document}